\newif\ifdraft
\draftfalse
\newif\ifprintauthors
\printauthorstrue

\ifdefined\DRAFT\drafttrue\fi
\ifdefined\PRINTAUTHORS\printauthorstrue\fi

\ifdraft
\documentclass[twocolumn,tighten,times,linenumbers,trackchanges]{aastex631}
\else
\documentclass[twocolumn,tighten,times]{aastex631}
\fi

\usepackage[T1]{fontenc}

\usepackage{amsmath}

\usepackage{wasysym}
\usepackage{xcolor}

\usepackage{appendix}

\ifdraft

\let\keeptoday\today
\def\today{\keeptoday, GitID {\normalfont\input{gitID.txt}}}
\fi

\usepackage{siunitx}
\DeclareSIUnit[quantity-product = ]\percent{\char`\%}
\DeclareSIUnit\parsec{pc}
\DeclareSIUnit\Mpc{\mega\parsec}
\DeclareSIUnit\yr{yr}
\DeclareSIUnit\day{days}
\DeclareSIUnit\GpcCubedYr{\giga\parsec\cubed\yr}
\newcommand\sifmt[2]{{\sisetup{#1}#2}}
\newcommand\sifmttwofig[1]{\sifmt{round-mode=figures,round-precision=2}{#1}}
\newcommand\sifmtonedec[1]{\sifmt{round-mode=places,round-precision=1}{#1}}
\newcommand\sifmtmonedec[1]{\sifmt{round-mode=places,round-precision=-1}{#1}}
\newcommand\sifmtthreesci[1]{\sifmt{exponent-mode=scientific,round-mode=figures,round-precision=3}{#1}}

\usepackage{txfonts}
\newlength{\capheight}
\newcommand{\Lshape}{\mbox{\rlap{\rule{0.08333em}{\capheight}}\rule{\capheight}{0.08333em}}}
\newcommand{\textcheckmark}{\ensuremath{\checkmark}}
\usepackage{textgreek}

\usepackage{acronym}
\usepackage{xspace}
\usepackage{multirow}

\usepackage{amstext}

\makeatletter
\@ifpackageloaded{cleveref}{
  \AtBeginDocument{
    \def\ltx@label#1{\cref@label{#1}}%
    \def\label@in@display@noarg#1{\cref@old@label@in@display{#1}}%
    \def\label@in@mmeasure@noarg#1{%
      \begingroup%
        \measuring@false%
        \cref@old@label@in@display{#1}%
      \endgroup}%
  }
}{}
\makeatother

\protected\def\protectedacused{\acused}

\acrodef{LIGO}[LIGO]{Laser Interferometer Gravitational-Wave Observatory}
\acrodef{LHO}[LHO]{\ac{LIGO} Hanford Observatory}
\acrodef{LLO}[LLO]{\ac{LIGO} Livingston Observatory}
\acrodef{KAGRA}[KAGRA]{KAGRA}\acused{KAGRA}
\acrodef{iKAGRA}[iKAGRA]{initial \ac{KAGRA}}
\acrodef{bKAGRA}[bKAGRA]{baseline \ac{KAGRA}}
\acrodef{GEO}[GEO\,600]{GEO\,600 \ac{GW} detector}
\acrodef{aLIGO}{Advanced \ac{LIGO}}
\acrodef{A+}{Advanced+ \ac{LIGO}}
\acrodef{Asharp}[\ensuremath{\text{A}^\sharp}]{\ac{LIGO} \acs{Asharp}}
\acrodef{AdV}{Advanced \acl{Virgo}}
\acrodef{AdV+}{Advanced \acl{Virgo}+}
\acrodef{Virgo}{Virgo}\acused{Virgo}
\acrodef{VirgoNEXT}[Virgo\_nEXT]{Virgo\_nEXT}\acused{VirgoNEXT}

\acrodef{LSC}[LSC]{\acs{LIGO} Scientific Collaboration}
\acrodef{LV}[LV]{\acs{LIGO}--\acs{Virgo} Collaboration\protect\protectedacused{LVC}}
\acrodef{LVC}[LV]{\acs{LIGO}--\acs{Virgo} Collaboration\protect\protectedacused{LV}}
\acrodef{LVK}[LVK]{\acs{LIGO}--\ac{Virgo}--\ac{KAGRA} Collaboration}
\acrodef{IGWN}[IGWN]{International \ac{GWH} Observatory Network}

\acrodef{O1}[O1]{first observing run}
\acrodef{O2}[O2]{second observing run}
\acrodef{O3}[O3]{third observing run}
\acrodef{O3a}[O3a]{first half of the third observing run}
\acrodef{O3b}[O3b]{second half of the third observing run}
\acrodef{O3GK}[O3GK]{observing run}
\acrodef{O4}[O4]{fourth observing run}
\acrodef{O4a}[O4a]{first part of the fourth observing run}
\acrodef{O4b}[O4b]{second part of the fourth observing run}
\acrodef{O4c}[O4c]{third part of the fourth observing run}
\acrodef{IR1}[IR1]{intermediate run 1}
\acrodef{O5}[O5]{fifth observing run}

\acrodef{BH}[BH]{black hole}
\acrodef{BBH}[BBH]{binary \ac{BH}}
\acrodef{BNS}[BNS]{binary \ac{NS}}
\acrodef{IMBH}[IMBH]{intermediate-mass \ac{BH}}
\acrodef{NS}[NS]{neutron star}
\acrodef{BHNS}[BHNS]{\ac{BH}--\ac{NS} binary}
\acrodefplural{BHNS}[BHNSs]{\ac{BH}--\ac{NS} binaries}
\acrodef{NSBH}[NSBH]{\ac{NS}--\ac{BH} binary}
\acrodefplural{NSBH}[NSBHs]{\ac{NS}--\ac{BH} binaries}
\acrodef{PBH}[PBH]{primordial \ac{BH}}
\acrodef{CBC}[CBC]{compact binary coalescence}

\acrodef{GW}[GW]{gravitational wave\protect\protectedacused{GWH}}
\acrodef{GWH}[GW]{gravitational-wave\protect\protectedacused{GW}}
\acrodef{IFO}[IFO]{interferometer}
\acrodef{SNR}[SNR]{signal-to-noise ratio}
\acrodef{FAR}[FAR]{false-alarm rate}
\acrodef{IFAR}[IFAR]{inverse false-alarm rate}
\acrodef{FAP}[FAP]{false alarm probability}
\acrodef{PSD}[PSD]{power spectral density}
\acrodef{GR}[GR]{general relativity}
\acrodef{NR}[NR]{numerical relativity}
\acrodef{PN}[PN]{post-Newtonian}
\acrodef{EOB}[EOB]{effective-one-body}
\acrodef{ROM}[ROM]{reduced-order model}
\acrodef{IMR}[IMR]{inspiral--merger--ringdown}
\acrodef{PDF}[pdf]{probability density function}
\acrodef{PE}[PE]{parameter estimation}
\acrodef{CI}[CI]{credible interval}
\acrodef{CL}[CL]{credible level}
\acrodef{EOS}[EoS]{equation of state}
\acrodef{KLD}[KLD]{Kullback--Leibler divergence}
\acrodef{JSD}[JSD]{Jensen--Shannon divergence}
\acrodef{GCN}[GCN]{General Coordinates Network}
\acrodef{GWTC}[GWTC]{Gravitational-Wave Transient Catalog}
\acrodef{GWOSC}[GWOSC]{Gravitational Wave Open Science Center}
\acrodef{WDM}[WDM]{Wilson--Daubechies--Meyer}
\acrodef{DQR}[DQR]{data-quality report}

\acrodef{CWB}[cWB]{coherent WaveBurst}
\acrodef{LAL}[LAL]{\ac{LIGO} algorithm library}

\acrodef{CHRoCC}{central heating radius of curvature correction}
\acrodef{NonSENS}{non-stationary estimation and noise subtraction}
\acrodef{RF}{radio frequency}
\acrodef{PNC}{phase noise cancellation}
\acrodef{ASC}{alignment sensing and control}
\acrodef{WFS}{wave-front sensing}
\acrodef{BPC}{beam position control}
\acrodef{ADS}{alignment dither systems}
\acrodef{OMC}{output mode cleaner}
\acrodef{LVDTs}{linear variable differential transformers}
\acrodef{GAS}{geometrical anti-spring}

\acrodef{PTA}{Pulsar Timing Array}

\newcommand\gwtc[1][?]{\mbox{GWTC\if#1?\else-#1\fi}}
\newcommand\thisgwtcversionmajor{5}
\newcommand\thisgwtcversionminor{0}
\newcommand\thisgwtcversionfull{\thisgwtcversionmajor.\thisgwtcversionminor}
\newcommand\thisgwtcversion\thisgwtcversionfull
\newcommand\thisgwtc{\gwtc[\thisgwtcversion]}

\newcommand\laser{\ast}

\newcommand{\OoneStartDate}{{{2015~September~12}}}

\newcommand{\OoneEndDate}{{{2016~January~19}}}

\newcommand{\OtwoStartDate}{{{2016~November~30}}}

\newcommand{\OtwoEndDate}{{{2017~August~25}}}

\newcommand{\OthreeAStartDate}{{{2019~April~1}}}

\newcommand{\OthreeAEndDate}{{{2019~October~1}}}

\newcommand{\OthreeBStartDate}{{{2019~November~1}}}

\newcommand{\OthreeBEndDate}{{{2020~March~27}}}

\newcommand{\OfourAStartDate}{{{2023~May~24}}}
\newcommand{\OfourAStartTime}{{{15:00:00}}}
\newcommand{\OfourAEndDate}{{{2024~January~16}}}
\newcommand{\OfourAEndTime}{{{16:00:00}}}
\newcommand{\OfourBStartDate}{{{2024~April~10}}}
\newcommand{\OfourBStartTime}{{{15:00:00}}}
\newcommand{\OfourBEndDate}{{{2025~January~28}}}
\newcommand{\OfourBEndTime}{{{17:00:00}}}
\newcommand{\OfourCStartDate}{{{2025~January~28}}}

\newcommand{\OtwoVStartDate}{{{2017~August~1}}}

\newcommand{\OthreeGKStartDate}{{{2020~April~7}}}

\newcommand{\OthreeGKEndDate}{{{2020~April~21}}}

\newcommand{\OoneStartIntendedDate}{{{2015~September~18}}}

\newcommand{\OoneEndIntendedDate}{{{2016~January~12}}}

\newcommand{\OtwoPreEngrunStartLhoDate}{{{2016~October~31}}}

\newcommand{\OtwoPreEngrunStartLloDate}{{{2016~November~14}}}

\newcommand{\OtwoEarthquakeMontanaDate}{{{2017~July~6}}}

\newcommand{\OthreebEndIntendedDate}{{{2020~April~30}}}

\newcommand{\OfourEarthquakeNotoPeninsulaDate}{{{2024~January~1}}}

\newcommand{\OfourcCommissioningBreakStartDate}{{{2025~April~1}}}

\newcommand{\OfourcCommissioningBreakEndDate}{{{2025~June~11}}}

\newcommand{\GWTConeENDDate}{{{2018~October~1}}}
\newcommand{\GWTConeENDTime}{{{00:00:00}}}
\newcommand{\GWTCtwoENDDate}{{{2019~October~1}}}
\newcommand{\GWTCtwoENDTime}{{{15:00:00}}}
\newcommand{\GWTCthreeENDDate}{{{2020~May~1}}}
\newcommand{\GWTCthreeENDTime}{{{00:00:00}}}
\newcommand{\GWTCfourENDDate}{{{2024~January~31}}}
\newcommand{\GWTCfourENDTime}{{{00:00:00}}}
\newcommand{\GWTCfiveENDDate}{{{2025~January~28}}}
\newcommand{\GWTCfiveENDTime}{{{17:00:00}}}

\newcommand{\OoneAsdLhoDate}{{{2015~October~24}}}

\newcommand{\OtwoAsdLhoDate}{{{2017~June~10}}}

\newcommand{\OtwoAsdLloDate}{{{2017~August~6}}}

\newcommand{\OthreeaAsdLloDate}{{{2019~April~29}}}

\newcommand{\OthreebAsdLhoDate}{{{2020~January~4}}}

\newcommand{\OthreebAsdVirgoDate}{{{2020~February~9}}}

\newcommand{\OfouraAsdLhoDate}{{{2024~January~11}}}

\newcommand{\OfouraAsdLloDate}{{{2023~November~19}}}

\newcommand{\OfourbAsdLhoDate}{{{2024~November~16}}}

\newcommand{\OfourbAsdLloDate}{{{2024~August~19}}}

\newcommand{\OfourbAsdVirgoDate}{{{2024~April~20}}}

\DeclareSIUnit\parsec{pc}
\DeclareSIUnit\Mpc{\mega\parsec}
\DeclareSIUnit\yr{yr}
\DeclareSIUnit\GpcCubedYear{\giga\parsec\cubed\yr}

\newcommand{\OoneLHORange}{{{\qty{77.0412141711}{\Mpc}}}}
\newcommand{\OoneLLORange}{{{\qty{70.442490731}{\Mpc}}}}
\newcommand{\OtwoLHORange}{{{\qty{77.8651737027}{\Mpc}}}}
\newcommand{\OtwoLLORange}{{{\qty{97.2053833026}{\Mpc}}}}
\newcommand{\OtwoVirgoRange}{{{\qty{27.162793797}{\Mpc}}}}
\newcommand{\OthreeLHORange}{{{\qty{113.493787915}{\Mpc}}}}
\newcommand{\OthreeLLORange}{{{\qty{135.464666304}{\Mpc}}}}
\newcommand{\OthreeVirgoRange}{{{\qty{57.4187395607}{\Mpc}}}}
\newcommand{\OthreeGKGEORange}{{{\qty{1.08274164216}{\Mpc}}}}
\newcommand{\OthreeGKKAGRARange}{{{\qty{0.68632871622}{\Mpc}}}}

\newcommand{\OfourbLHORange}{{{\qty{160.922259557}{\Mpc}}}}
\newcommand{\OfourbLLORange}{{{\qty{171.076606979}{\Mpc}}}}
\newcommand{\OfourbVirgoRange}{{{\qty{53.0033013035}{\Mpc}}}}

\newcommand{\OoneDuration}{{{\qty{129.66666666666666}{\day}}}}

\newcommand{\OoneDurationAnyTwo}{{{\qty{49.004895833333336}{\day}}}}

\newcommand{\OoneDurationNone}{{{\qty{36.16134259259259}{\day}}}}
\newcommand{\OoneDurationNoneFraction}{{{\qty{27.887924878606114}{\percent}}}}

\newcommand{\OoneDurationHL}{{{\qty{49.004895833333336}{\day}}}}
\newcommand{\OoneDurationHLFraction}{{{\qty{37.79297879177378}{\percent}}}}

\newcommand{\OtwoDurationAnyTwo}{{{\qty{122.19149305555555}{\day}}}}

\newcommand{\OtwoDurationHLV}{{{\qty{15.252222222222223}{\day}}}}

\newcommand{\OtwoVDurationVAnyFraction}{{{\qty{85.10010393046107}{\percent}}}}

\newcommand{\OthreeDuration}{{{\qty{361.0833333333333}{\day}}}}

\newcommand{\OthreeDurationNone}{{{\qty{42.059872685185184}{\day}}}}
\newcommand{\OthreeDurationNoneFraction}{{{\qty{11.648245377849578}{\percent}}}}

\newcommand{\OthreeDurationHLV}{{{\qty{154.28855324074075}{\day}}}}
\newcommand{\OthreeDurationHLVFraction}{{{\qty{42.729347770341306}{\percent}}}}

\newcommand{\OthreeGKDurationGAny}{{{\qty{10.881168981481482}{\day}}}}
\newcommand{\OthreeGKDurationGAnyFraction}{{{\qty{79.61830962059621}{\percent}}}}
\newcommand{\OthreeGKDurationKAny}{{{\qty{7.269155092592593}{\day}}}}

\newcommand{\OthreeGKDurationGK}{{{\qty{6.380150462962963}{\day}}}}

\newcommand{\OthreeaDurationAnyOne}{{{\qty{177.14966435185184}{\day}}}}

\newcommand{\OthreeaDurationAnyTwo}{{{\qty{149.57175925925927}{\day}}}}

\newcommand{\OthreebDurationAnyOne}{{{\qty{141.8737962962963}{\day}}}}

\newcommand{\OthreebDurationAnyTwo}{{{\qty{124.60466435185185}{\day}}}}

\newcommand{\OfouraDurationAnyOne}{{{\qty{196.8159837962963}{\day}}}}

\newcommand{\OfouraDurationAnyTwo}{{{\qty{126.47008101851851}{\day}}}}

\newcommand{\OfourbDuration}{{{\qty{293.0833333333333}{\day}}}}
\newcommand{\OfourbDurationAnyOne}{{{\qty{259.8528009259259}{\day}}}}

\newcommand{\OfourbDurationAnyTwo}{{{\qty{198.91775462962963}{\day}}}}

\newcommand{\OfourbDurationNone}{{{\qty{33.23053240740741}{\day}}}}
\newcommand{\OfourbDurationNoneFraction}{{{\qty{11.338253877989448}{\percent}}}}

\newcommand{\OfourbDurationHL}{{{\qty{21.566111111111113}{\day}}}}
\newcommand{\OfourbDurationHLFraction}{{{\qty{7.358354658326225}{\percent}}}}
\newcommand{\OfourbDurationHV}{{{\qty{22.05903935185185}{\day}}}}
\newcommand{\OfourbDurationHVFraction}{{{\qty{7.526541718004612}{\percent}}}}
\newcommand{\OfourbDurationLV}{{{\qty{64.21601851851852}{\day}}}}
\newcommand{\OfourbDurationLVFraction}{{{\qty{21.910498215019114}{\percent}}}}
\newcommand{\OfourbDurationHLV}{{{\qty{91.07658564814815}{\day}}}}
\newcommand{\OfourbDurationHLVFraction}{{{\qty{31.075320664707927}{\percent}}}}
\newcommand{\OoneVTTotal}{{{\qty{1.588786e-04}{\GpcCubedYear}}}}

\newcommand{\OtwoVTTotal}{{{\qty{3.518480e-04}{\GpcCubedYear}}}}
\newcommand{\OtwoVTHL}{{{\qty{3.268875e-04}{\GpcCubedYear}}}}

\newcommand{\OtwoVTHV}{{{\qty{3.624116e-07}{\GpcCubedYear}}}}

\newcommand{\OtwoVTLV}{{{\qty{4.801174e-07}{\GpcCubedYear}}}}

\newcommand{\OtwoVTHLV}{{{\qty{2.411796e-05}{\GpcCubedYear}}}}

\newcommand{\OthreeVTTotal}{{{\qty{3.207126e-03}{\GpcCubedYear}}}}
\newcommand{\OthreeVTH}{{{\qty{4.473158e-05}{\GpcCubedYear}}}}

\newcommand{\OthreeVTL}{{{\qty{7.468101e-05}{\GpcCubedYear}}}}

\newcommand{\OthreeVTV}{{{\qty{9.715321e-06}{\GpcCubedYear}}}}

\newcommand{\OthreeVTHL}{{{\qty{7.203780e-04}{\GpcCubedYear}}}}

\newcommand{\OthreeVTHV}{{{\qty{4.094091e-05}{\GpcCubedYear}}}}

\newcommand{\OthreeVTLV}{{{\qty{5.034511e-05}{\GpcCubedYear}}}}

\newcommand{\OthreeVTHLV}{{{\qty{2.266334e-03}{\GpcCubedYear}}}}

\newcommand{\OfourbVTTotal}{{{\qty{5.245143e-03}{\GpcCubedYear}}}}
\newcommand{\OfourbVTH}{{{\qty{9.657884e-05}{\GpcCubedYear}}}}

\newcommand{\OfourbVTL}{{{\qty{3.330493e-04}{\GpcCubedYear}}}}

\newcommand{\OfourbVTV}{{{\qty{1.308837e-05}{\GpcCubedYear}}}}

\newcommand{\OfourbVTHL}{{{\qty{9.107948e-04}{\GpcCubedYear}}}}

\newcommand{\OfourbVTHV}{{{\qty{3.139248e-05}{\GpcCubedYear}}}}

\newcommand{\OfourbVTLV}{{{\qty{9.565171e-05}{\GpcCubedYear}}}}

\newcommand{\OfourbVTHLV}{{{\qty{3.764587e-03}{\GpcCubedYear}}}}

\newcommand{\HANFORDPOWERONE}{{{\qty{21}{\watt}}}}
\newcommand{\HANFORDPOWERTWO}{{{\qty{26}{\watt}}}}
\newcommand{\HANFORDPOWERTHREEA}{{{\qty{34}{\watt}}}}
\newcommand{\HANFORDPOWERTHREEB}{{{\qty{34}{\watt}}}}
\newcommand{\HANFORDPOWERFOURA}{{{\qty{57}{\watt}}}}
\newcommand{\HANFORDPOWERFOURB}{{{\qty{57}{\watt}}}}
\newcommand{\LIVINGSTONPOWERONE}{{{\qty{22}{\watt}}}}
\newcommand{\LIVINGSTONPOWERTWO}{{{\qty{25}{\watt}}}}
\newcommand{\LIVINGSTONPOWERTHREEA}{{{\qty{44}{\watt}}}}
\newcommand{\LIVINGSTONPOWERTHREEB}{{{\qty{40}{\watt}}}}
\newcommand{\LIVINGSTONPOWERFOURA}{{{\qty{64}{\watt}}}}
\newcommand{\LIVINGSTONPOWERFOURB}{{{\qty{62}{\watt}}}}
\newcommand{\VIRGOPOWERTWO}{{{\qty{10}{\watt}}}}
\newcommand{\VIRGOPOWERTHREEA}{{{\qty{18}{\watt}}}}
\newcommand{\VIRGOPOWERTHREEB}{{{\qty{26}{\watt}}}}
\newcommand{\VIRGOPOWERFOURB}{{{\qty{17}{\watt}}}}
\newcommand{\GEOPOWERTHREEGK}{{{\qty{3}{\watt}}}}
\newcommand{\KAGRAPOWERTHREEGK}{{{\qty{5}{\watt}}}}
\newcommand{\HANFORDPRGONE}{{{\num{38}}}}
\newcommand{\HANFORDPRGTWO}{{{\num{40}}}}
\newcommand{\HANFORDPRGTHREEA}{{{\num{44}}}}
\newcommand{\HANFORDPRGTHREEB}{{{\num{44}}}}
\newcommand{\HANFORDPRGFOURA}{{{\num{50}}}}
\newcommand{\HANFORDPRGFOURB}{{{\num{50}}}}
\newcommand{\LIVINGSTONPRGONE}{{{\num{38}}}}
\newcommand{\LIVINGSTONPRGTWO}{{{\num{36}}}}
\newcommand{\LIVINGSTONPRGTHREEA}{{{\num{47}}}}
\newcommand{\LIVINGSTONPRGTHREEB}{{{\num{42}}}}
\newcommand{\LIVINGSTONPRGFOURA}{{{\num{35}}}}
\newcommand{\LIVINGSTONPRGFOURB}{{{\num{40}}}}
\newcommand{\VIRGOPRGTWO}{{{\num{38}}}}
\newcommand{\VIRGOPRGTHREEA}{{{\num{36}}}}
\newcommand{\VIRGOPRGTHREEB}{{{\num{34}}}}
\newcommand{\VIRGOPRGFOURB}{{{\num{39}}}}
\newcommand{\GEOPRGTHREEGK}{{{\num{1000}}}}
\newcommand{\KAGRAPRGTHREEGK}{{{\num{12}}}}

\newcommand{\PASTROTHRESHOLDPCT}{{{\qty{50}{\percent}}}}
\newcommand{\OtwoMontanaEarthquakeMagnitude}{{{\num{5.8}}}}
\newcommand{\OtwoMontanaEarthquakeBNSDrop}{{{\qty{10}{\Mpc}}}}
\newcommand{\OfourNotoPeninsulaEarthquakeMagnitude}{{{\num{7.5}}}}
\newcommand{\SensemonRangeCoefficient}{{{\qty{1.01618726748049498e-20}{\Mpc.\second^{-1/6}}}}}
\newcommand{\FirstGenerationVirgoLaserPower}{{{\qty{20}{\watt}}}}
\newcommand{\OoneGroundMotionMagnitudeReduction}{{{\num{10}}}}
\newcommand{\OoneGroundMotionSuppressionFrequency}{{{\qty{10}{\hertz}}}}
\newcommand{\OtwoBnsRangeImprovement}{{{\qty{100}{\Mpc}}}}
\newcommand{\OtwoDutyFactorsPercentage}{{{\qty{50}{\percent}}}}
\newcommand{\OtwoCommissioningAdvBnsRange}{{{\qty{20}{\Mpc}}}}
\newcommand{\OtwoAdvLaserPower}{{{\qty{13}{\watt}}}}

\newcommand{\OfourAdvTestMassKg}{{{\qty{42}{\kg}}}}

\newcommand{\OfourBadvBnsRangeTarget}{{{\qty{54}{\Mpc}}}}
\newcommand{\OthreeGkTemperatureTestMass}{{{\qty{250}{\K}}}}
\newcommand{\OthreeGkKagraStrainSensitivity}{{{\qty{3.0e-22}{\hertz}}}}
\newcommand{\OthreeGkKagraReferenceFrequency}{{{\qty{250}{\hertz}}}}

\newcommand{\OfourKagraNumberOfSeismicNoiseIsolatorRestored}{{{\num{10}}}}
\newcommand{\OfourKagraIncreasedLaserPower}{{{\qty{10}{\watt}}}}
\newcommand{\OfourKagraCooledMirrodTemperature}{{{\qty{100}{\kelvin}}}}
\newcommand{\MinimumMassForMicrolensing}{{{\qty{e5}{\Msun}}}}
\newcommand{\VsrFourBnsRange}{{{\qty{12}{\Mpc}}}}
\newcommand{\AdvLaserPowerDesign}{{{\qty{125}{\watt}}}}

\newcommand{\Msun}{\ensuremath{\mathit{M_\odot}}}

\newcommand{\tgeo}{\ensuremath{t_{\text{geo}}}}

\newcommand{\massone}{\ensuremath{m_1}}
\newcommand{\masstwo}{\ensuremath{m_2}}
\newcommand{\Mc}{\ensuremath{\mathcal{M}}}
\newcommand{\Mtot}{\ensuremath{M}}
\newcommand{\Mf}{\ensuremath{M_\mathrm{f}}}
\newcommand{\mratio}{\ensuremath{q}}
\newcommand{\mratiosym}{\ensuremath{\eta}}

\newcommand{\Erad}{\ensuremath{E_\mathrm{rad}}}
\newcommand{\lumpeak}{\ensuremath{\ell_{\text{peak}}}}

\newcommand{\chieff}{\ensuremath{\chi_\mathrm{eff}}}
\newcommand{\chip}{\ensuremath{\chi_\mathrm{p}}}
\newcommand{\chif}{\ensuremath{\chi_\mathrm{f}}}

\newcommand{\spinone}{\ensuremath{\chi_1}}

\newcommand{\spintwo}{\ensuremath{\chi_2}}

\newcommand{\DA}{\ensuremath{D_\mathrm{A}}}
\newcommand{\DL}{\ensuremath{D_\mathrm{L}}}

\newcommand{\DM}{\ensuremath{D_{\text{M}}}}

\newcommand{\ip}[2]{\ensuremath{\langle #1 | #2 \rangle}}

\newcommand\PEpdfp{\ensuremath{p}}
\newcommand\PEdata{\ensuremath{d}}
\newcommand{\PEparameter}{\ensuremath{\boldsymbol{\theta}}}%

\newcommand\PEpdf[2][?]{\ensuremath{\PEpdfp({#2}\ifx#1?\else | {#1}\fi)}}

\newcommand\PElikelihood[1][\PEparameter]{\PEpdf[#1]{\PEdata}}

\newcommand\PEpriorpdfpi{\ensuremath{\pi}}
\newcommand\PEpdfprior[1]{\ensuremath{\PEpriorpdfpi({#1})}}
\newcommand\PEprior[1][\PEparameter]{\PEpdfprior{#1}}
\newcommand\PEpriorpe[1][\PEparameter]{{\let\keepPEpriorpdfpi\PEpriorpdfpi\def\PEpriorpdfpi{\keepPEpriorpdfpi_{\text{PE}}}\PEprior[#1]\let\PEpriorpdfpi\keepPEpriorpdfpi}}

\newcommand{\VT}{\ensuremath{\langle VT \rangle}\xspace}

\newcommand{\soft}[1]{\textsc{#1}}

\newcommand{\ASTROPY}{\soft{Astropy}\xspace}

\newcommand{\LALSUITE}{\soft{LALSuite}\xspace}

\newcommand{\NUMPY}{\soft{NumPy}\xspace}
\newcommand{\SCIPY}{\soft{SciPy}\xspace}
\newcommand{\MATPLOTLIB}{\soft{Matplotlib}\xspace}
\newcommand{\SEABORN}{\soft{seaborn}\xspace}
\newcommand{\GWPY}{\soft{GWpy}\xspace}

\newcommand{\LVKcollaboration}{The LIGO Scientific Collaboration, the Virgo Collaboration, and the KAGRA Collaboration}

\begin{document}

\title{\thisgwtc: An Introduction to Version \thisgwtcversion{} of the Gravitational-Wave Transient Catalog}

\ifprintauthors
\suppressAffiliations

\author[0000-0003-4786-2698]{A.~G.~Abac}
\affiliation{Max Planck Institute for Gravitational Physics (Albert Einstein Institute), D-14476 Potsdam, Germany}
\author{A.~Abe}
\affiliation{Department of Physics, Graduate School of Science, Osaka Metropolitan University, 3-3-138 Sugimoto-cho, Sumiyoshi-ku, Osaka City, Osaka 558-8585, Japan  }
\author{I.~Abouelfettouh}
\affiliation{LIGO Hanford Observatory, Richland, WA 99352, USA}
\author{F.~Acernese}
\affiliation{Dipartimento di Fisica ``E.R. Caianiello'', Universit\`a di Salerno, I-84084 Fisciano, Salerno, Italy}
\affiliation{INFN, Sezione di Napoli, I-80126 Napoli, Italy}
\author[0000-0002-8648-0767]{K.~Ackley}
\affiliation{University of Warwick, Coventry CV4 7AL, United Kingdom}
\author{A.~Adam}
\affiliation{OzGrav, University of Western Australia, Crawley, Western Australia 6009, Australia}
\author[0009-0004-2101-5428]{S.~Adhicary}
\affiliation{The Pennsylvania State University, University Park, PA 16802, USA}
\author{D.~Adhikari}
\affiliation{Max Planck Institute for Gravitational Physics (Albert Einstein Institute), D-30167 Hannover, Germany}
\affiliation{Leibniz Universit\"{a}t Hannover, D-30167 Hannover, Germany}
\author[0000-0002-5731-5076]{R.~X.~Adhikari}
\affiliation{LIGO Laboratory, California Institute of Technology, Pasadena, CA 91125, USA}
\author{V.~K.~Adkins}
\affiliation{Louisiana State University, Baton Rouge, LA 70803, USA}
% ...

\collaboration{0}{\LVKcollaboration\\(See the end matter for the full list of authors)}
\else
\collaboration{0}{\LVKcollaboration}
\fi

\correspondingauthor{LSC P\&P Committee, via LVK Publications as proxy}
\email{lvc.publications@ligo.org}

\begin{abstract}
The Gravitational-Wave Transient Catalog (\gwtc{}) is a collection of short-duration (transient) gravitational-wave signals identified by the LIGO--Virgo--KAGRA Collaboration in gravitational-wave data produced by the eponymous detectors.
The catalog provides information about the identified candidates, such as the arrival time and amplitude of the signal and properties of the signal's source as inferred from the observational data.
\gwtc{} is the release of this dataset and version~\thisgwtcversionfull{} extends the catalog to include observations made during the second part of the fourth LIGO--Virgo--KAGRA observing run up until 2025~January~28.
This paper marks an introduction to a collection of articles related to this version of the catalog, \thisgwtc{}. This update significantly increases the number of detected merging binary systems of black holes and neutron stars to over 300, enabling many follow-up studies toward understanding the gravitational-wave universe.
The collection of articles accompanying the catalog provides documentation of the methods used to analyze the data, summaries of the catalog of events, observational measurements drawn from the population, and detailed discussions of selected candidates.

\end{abstract}

\keywords{\IfFileExists{gwtc-common-files__standard_keywords}{Gravitational wave astronomy (675); Gravitational wave detectors (676); Gravitational wave sources (677); Stellar mass black holes (1611); Neutron stars (1108)
}{FIXME}}

\section{Overview}\label{sec:overview}

The \acl{LIGO} \citep[\acsu{LIGO};][]{2015CQGra..32g4001L} and the \ac{Virgo} \citep{2015CQGra..32b4001A} and \ac{KAGRA} \citep{2021PTEP.2021eA101A} observatories form an international network of ground-based \ac{GWH} detectors.
This paper introduces the collection of articles describing the contents of the 
\acf{LVK} \acf{GWTC} version \thisgwtcversionfull, hereafter \thisgwtc{}. 
It also reviews the methods used in constructing the catalog, together with the astrophysical and
cosmological implications of the observations, and tests of general relativity (GR) that
are performed on the observed transients.
The fifth version of the catalog brings over 300 gravitational-wave candidates to the community, including the loudest observations ever recorded, providing opportunities for unprecedented follow-up studies. 
This paper provides details on the network of \ac{GW} detectors, the observing runs, observatory evolution, and a review of the transient signals that have been identified.
In addition, we describe conventions and notations that are used throughout the collection of papers accompanying the catalog.

\subsection{The \gwtc{} Sources and Science}
Transient \ac{GW} signals may be produced by a variety of astrophysical sources, including \acp{CBC} of compact objects such as \acp{BH} and \acp{NS}, core-collapse supernovae, and other explosive phenomena \citep{KAGRA:2013rdx}.
The first observed \ac{GW} transient, GW150914, was a \ac{BBH} coalescence \citep{2016PhRvL.116f1102A}, and we have since observed a \ac{BNS} coalescence \citep{2017PhRvL.119p1101A} that had associated electromagnetic counterparts \citep{2017ApJ...848L..12A}, and \ac{NSBH} coalescences \citep{2020ApJ...896L..44A}.

This \thisgwtc{} collection of papers describes the \ac{GW} transient candidates observed by the \ac{LVK} from the \ac{O1} through the end of the \ac{O4b} and the astrophysical implications of these observations. The paper collection includes the following:
\begin{itemize}
\item ``\thisgwtc{}: Methods for Identifying and Characterizing Gravitational-wave Transients'' \citep{GWTC:Methods} reviews the procedures used to go from the calibrated output of the detectors to a list of transient candidates that includes measurements of the statistical significance and inferences on each of the corresponding astrophysical sources.
\item ``\thisgwtc{}: Observations from the Second Part of the Fourth LIGO-Virgo-KAGRA Observing Run and Updates to the Gravitational-Wave Transient Catalog'' \citep{GWTC:Results} describes the primary observational results contained in \thisgwtc{}: the significant \ac{GW} transient candidates observed through the end of the O4b observing run and the inferred source parameters under the hypothesis that these transients arise from \acp{GW} emitted by \acp{CBC} (Section~\ref{s:cbc}).
\item ``\thisgwtc{}: Population Properties of Merging Compact Binaries'' \citep{GWTC:AstroDist} describes the underlying population of \acp{CBC} inferred using \thisgwtc{} data and related astrophysical implications. 
\item ``\thisgwtc{}: Tests of General Relativity'' \citep{GWTC:TGR} describes the tests of \ac{GR} performed on the subset of signals suitable for such tests.
\item ``\thisgwtc{}: Constraints on the Cosmic Expansion Rate and Modified Gravitational-wave Propagation'' \citep{GWTC:Cosmology} describes the methods used to determine the Hubble constant and related parameters, including parameterized deviations from \ac{GR} on cosmological scales, using \thisgwtc{} candidates.
\item ``\thisgwtc{}: Searches for Gravitational Wave Lensing Signatures'' \citep{GWTC:Lensing} describes the searches for lensed \ac{GW} signals in the geometric and wave optics regime in the \thisgwtc{} dataset. It also sets constraints on the merger rate at high redshift and the relative rate of strongly lensed signals compared to unlensed ones.
\item ``Open Data from LIGO, Virgo, and KAGRA through the Second Part of the Fourth Observing Run'' \citep{OpenData} describes the publicly accessible data and other science products that can be freely accessed through the \ac{GWOSC}.
These data sets include the raw \ac{GWH} strain time series, details of the calibration and cleaning process, efforts to remove instrumental noise artifacts, and details of the online \thisgwtc{}.
\end{itemize}

To reference the whole \thisgwtc{} collection, we encourage citing this introductory paper. 
In addition to the \thisgwtc{} collection, exceptional \ac{O4b} event papers have been published: GW241011\_233834 and GW241110\_124123 \citep{2025ApJ...993L..21A}, a particularly interesting pair of asymmetric, high-spin \ac{BBH} coalescences; and GW250114\_082203 \citep{2025PhRvL.135k1403A,2026PhRvL.136d1403A}, the loudest \ac{GW} event ever recorded with a network \ac{SNR} of nearly 80. 
Furthermore, another paper on astrophysical calibration of \ac{GW} detectors with the loud events GW240925\_005809 and GW250207\_115645 is scheduled to be published \citep{gzrj-mwv3} by the time this article is submitted. 
\thisgwtc{} adds from \ac{O4b} 104 additional \ac{GW} observations with detailed source property measurements.

\subsection{The Electronic Catalog: \ac{GWTC}}

\citet{OpenData} document the released open data, including the \ac{GWTC} dataset.
The catalog contains \emph{candidates} (sometimes called \emph{events}) identified in observational data that are deemed likely to be caused by \ac{GW} signals, as well as \emph{triggers} corresponding to times selected by searches of the data for \ac{GW} transient signals that potentially contain an identifiable signal but with lower confidence of being caused by a \ac{GW}\@.

\subsubsection{The Catalog Naming Convention}
\label{ss:gwtc_versioning}

The \ac{LVK} \ac{GWTC} is a cumulative dataset containing data on all transient candidates reported by the \ac{LVK}.
Released versions of the catalog have major and minor numbers in the format
\begin{center}
\verb+GWTC-<+\textit{major}\verb+>.<+\textit{minor}\verb+>+
\end{center}
The major number is determined by the span of time containing all candidates in the catalog as described below.

Prior to \gwtc[4.0], the minor number was routinely omitted when describing a catalog version when that minor number was 0, so \gwtc[1.0], \gwtc[2.0], and \gwtc[3.0] were referred to as \gwtc[1], \gwtc[2], and \gwtc[3] in the papers that described those catalog
 versions.  Since \gwtc[4.0] we include the .0 when referring to those catalog versions.  We also say that \gwtc-\textless\textit{major}\textgreater{} can refer to \gwtc-\textless\textit{major}\textgreater.\textless\textit{minor}\textgreater{}
 for any minor version having that major version number.

Each catalog version is a superset of the previous one (apart from retracted candidates), so that, for example, \gwtc[3.0] \citep{2023PhRvX..13d1039A} contains all the candidates in \gwtc[2.1] \citep{2024PhRvD.109b2001A}.
Since \gwtc[2.1] provided a deeper list of candidates observed over the same period as \gwtc[2.0] \citep{2021PhRvX..11b1053A}, the minor version numbers of these two releases differ while
 their major version numbers remain the same. 
In general:
\begin{itemize}
\item The major number is incremented when the span of time over which observational data were searched for transients is increased.
\item The minor version resets to 0 when the major version number is increased.
\item The minor version is incremented when there is a change in the data describing the transients (additional data, modified data, or removed data) contained in the catalog within the current time span covered.
\end{itemize}

The time span covering the transient candidates in the catalog indicated by the major number is as follows:
\begin{description}
\expandafter\item[\normalfont\gwtc-1]
Contains candidates occurring in data taken before \GWTConeENDDate{} \GWTConeENDTime{}.
The \gwtc[1.0] dataset is described in \citet{2019PhRvX...9c1040A}.
\item[\normalfont\expandafter\gwtc-2]
Contains candidates occurring in data taken before \GWTCtwoENDDate{} \GWTCtwoENDTime{}.
The \gwtc[2.0] dataset is described in \citet{2021PhRvX..11b1053A} and the \gwtc[2.1] dataset in \citet{2024PhRvD.109b2001A}.
\item[\normalfont\expandafter\gwtc-3]
Contains candidates occurring in data taken before \GWTCthreeENDDate{} \GWTCthreeENDTime{}.
The \gwtc[3.0] dataset is described in \citet{2023PhRvX..13d1039A}.
\item[\normalfont\expandafter\gwtc-4]
Contains candidates occurring in data taken before \GWTCfourENDDate{} \GWTCfourENDTime{}.
The \gwtc[4.0] dataset is described in \citet{2025arXiv250818082T}.
\item[\normalfont\expandafter\gwtc-5]
Contains candidates occurring in data taken before \GWTCfiveENDDate{} \GWTCfiveENDTime{}.
The \gwtc[5.0] dataset is described in \citet{GWTC:Results}. This catalog release also includes updated search results and parameter estimation of additional events from \ac{O4a}. To clarify that these \ac{O4a} results supersede those in \gwtc[4.0], we name them \gwtc[4.1]. However, as \gwtc{} is a cumulative catalog, all \gwtc[4.1] results are included in \gwtc[5.0] by definition.
\end{description}

In addition to \ac{GWTC}, other catalogs of \ac{GW} transients include the Open Gravitational-wave Catalog (OGC), the most recent version 4-OGC contains observations from 2015 to 2020 \citep{2023ApJ...946...59N}, as well as catalogs of candidate signals identified by the IAS pipeline \citep{2019PhRvD.100b3011V,2022PhRvD.106d3009O,2024PhRvD.110d4063W,2026PhRvD.113b3003C}.
\citet{OpenData} provide details on the \ac{GWOSC} event portal,\footnote{\ac{GWOSC} event portal \url{https://gwosc.org/eventapi}} a database of published \ac{GW} transient events, including Community Catalogs \citep{Kanner:2025} containing catalog results from communities outside of the \ac{LVK}. 

\subsubsection{Candidate Naming Conventions}

The naming of our \ac{GW} candidates follows the format
\begin{center}
\verb+GW<+\textit{YY}\verb+><+\textit{MM}\verb+><+\textit{DD}\verb+>_<+\textit{hh}\verb+><+\textit{mm}\verb+><+\textit{ss}\verb+>+
\end{center}
encoding the date and Coordinated Universal Time (UTC) of the signal.
For example, GW200105\_162426 was the transient observed on 2020~January~5 at 16:24:26~UTC.
For transient signals spanning multiple-second intervals, the time assigned to a signal is an estimate of the time of peak \ac{GW} amplitude.

\ac{GW} candidates reported prior to the release of \gwtc[2.0] were designated by the abbreviated form
\begin{center}
\verb+GW<+\textit{YY}\verb+><+\textit{MM}\verb+><+\textit{DD}\verb+>+
\end{center}
including candidates first appearing in \gwtc[1.0] \citep{2019PhRvX...9c1040A}, as well as
GW190412 \citep{2020PhRvD.102d3015A},
GW190425 \citep{2020ApJ...892L...3A},
GW190521 \citep{2020PhRvL.125j1102A},
and
GW190814 \citep{2020ApJ...896L..44A}.
These candidates retain their legacy names.

\subsection{Outline}

An outline of the remainder of this article is as follows:
We briefly describe the network of ground-based \ac{GW} detectors in Section~\ref{sec:network} and their observing runs that have contributed to the \thisgwtc{} in Section~\ref{sec:observing}.
These sections are followed by short reviews of the evolution of the various observatories in Section~\ref{sec:interferometers} and of the nature of the transient sources observed in Section~\ref{sec:sources}.
A list of common acronyms is provided in Appendix~\ref{sec:glossary}.
Mathematical conventions used throughout the articles in this compendium are described in Appendix~\ref{s:conventions}.

\section{The International \ac{GW} Observatory Network}\label{sec:network}

The international ground-based \ac{GW} observatory network currently comprises four primary observatories employing laser interferometric \ac{GW} detectors.
The four observatories are the two US-based \ac{LIGO} detectors, \ac{LHO} in Washington and \ac{LLO} in Louisiana \citep{2015CQGra..32g4001L}; the European \ac{Virgo} detector \citep{2015CQGra..32b4001A}; and the Japanese \ac{KAGRA} detector \citep{2021PTEP.2021eA101A,2013PhRvD..88d3007A,2012CQGra..29l4007S}.
All these detectors are enhanced Michelson interferometers that sense relative changes in the lengths $L_1$ and $L_2$ of their two \qtyrange{3}{4}{\km} long arms caused by passing \acp{GW} in the high-frequency band \qtyrange{\sim10}{\sim1000}{\Hz} \citep{1987thyg.book..330T}.
Other \ac{GW} frequency bands include the very low frequency band \qtyrange{\sim1}{\sim100}{\nano\Hz} observed by pulsar timing arrays such as the European \acl{PTA} \citep[EPTA;][]{2016MNRAS.458.3341D}, the North American Nanohertz Observatory for Gravitational Waves \citep[NANOGrav;][]{2019BAAS...51g.195R}, the Parkes \acl{PTA} \citep[PPTA;][]{2020PASA...37...20K}, the Indian \acl{PTA} \citep[InPTA;][]{2018JApA...39...51J}, and their combined consortium the International \acl{PTA} \citep[IPTA;][]{2016MNRAS.458.1267V}; and the low-frequency band \qtyrange{\sim0.1}{\sim10}{\milli\Hz} that will be observed by the Laser Interferometer Space Antenna \citep[LISA;][]{2024arXiv240207571C}.

The fractional change in the relative lengths of the two optical paths of interferometric detectors, $\Delta(L_1-L_2)$, induced by a \ac{GW} is known as the detector strain, $h=\Delta(L_1-L_2)/L$, where $L$ is the average arm length (Section~\ref{ss:gw}).
The sensitivity of ground-based detectors is fundamentally limited below \qty{\sim1}{\Hz} by ground motion noise \citep{1984PhRvD..30..732S} and at high frequencies by shot noise \citep{1978PhRvD..17..379F,1991PhRvD..43.2470K}, see Section~\ref{sec:interferometers} for more details.
Significant noise sources at intermediate frequencies include thermal noise in the optics and their suspensions and quantum readout noise (\citealt[][{(Original work published 1972)}]{2022GReGr..54..153W};~\citealt{2017fige.book.....S,2001PhRvD..64d2006B}).
In the frequency domain, the overall detector sensitivity is characterized by the (one-sided) noise power spectral density in strain-equivalent units, $S_n(f)$, with dimensions of time (Appendix~\ref{s:conventions}).

The \ac{GEO} is a British--German instrument with \qty{600}{\meter} arms located near Hannover, Germany \citep{2010JPhCS.228a2012L,2014CQGra..31v4002A, 2016CQGra..33g5009D}.
This instrument is a laboratory for prototyping advanced interferometry techniques, but also is operated in data-taking \emph{astrowatch} mode when not being used for instrument science research \citep{2010CQGra..27h4003G,2016CQGra..33g5009D}.
Astrowatch provides \ac{GW} observing coverage for times when the larger detectors are not observing between observing runs and when the detectors are not taking scientific data, e.g., \ac{GEO} data were used to constrain post-merger signals following the first \ac{BNS} detection \citep{2017ApJ...851L..16A}.

\section{Observing Runs}\label{sec:observing}

\begin{figure*}
\begin{center}
\includegraphics{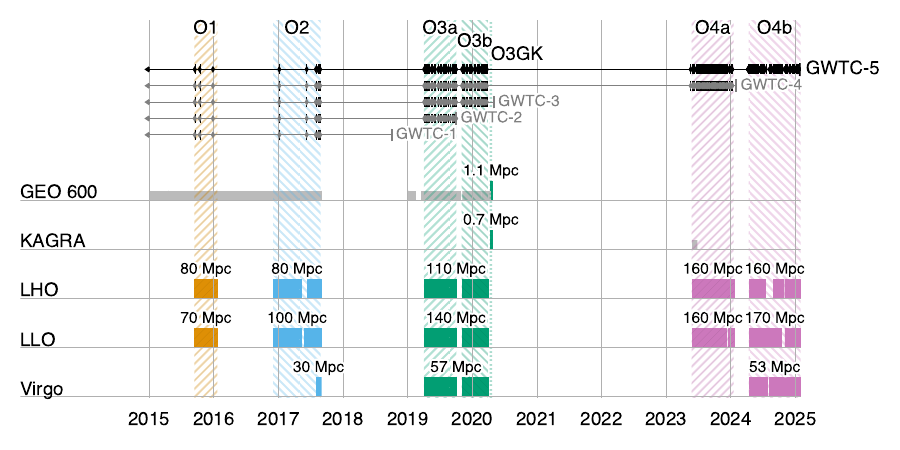}
\caption{\label{fig:obsruns}
The timeline of observing runs covering a time span starting from 2015 and lasting up to the beginning of \ac{O4c} on \OfourCStartDate{}.
The periods in which the various detectors in the network were observing are shown in this timeline, along with the typical \ac{BNS} inspiral ranges for those detectors during the observing run.
\ac{GEO} astrowatch observing periods are shown in light gray.
\ac{KAGRA} observing periods during \ac{O4a} are also shown in light gray.
In \ac{O1} and \ac{O4a}, only \ac{LHO} and \ac{LLO} were participating.
Virgo joined these two detectors for the last month of \ac{O2} and was observing alongside them throughout \ac{O3a}, \ac{O3b} and \ac{O4b}.
At the end of \ac{O3} there was a short joint observing run, \ac{O3GK}, which included \ac{GEO} and \ac{KAGRA}.
Also shown is a timeline of the observed candidates contained in \gwtc[1.0]{}, \gwtc[2.1]{}, \gwtc[3.0]{}, \gwtc[4.0]{}, and \gwtc[5.0]{} with a probability of astrophysical origin greater than or equal to \PASTROTHRESHOLDPCT{}.
The time intervals covered by the various versions of the \gwtc{} are bounded from above but not from below, as indicated by the arrows pointing left (see Section~\ref{ss:gwtc_versioning}).}
\end{center}
\end{figure*}

The \ac{GW} observing schedule is divided into observing runs, downtime for construction and commissioning, and transitional engineering runs between commissioning and observing runs \citep{KAGRA:2013rdx}.
Figure~\ref{fig:obsruns} shows a timeline of \ac{GW} observations up to the end date of the time period covered by \thisgwtc{}.
Indicated are the observing periods of each observing run and the times when each detector was in operation.
Also shown are the times when \ac{GW} transient signals were detected.

In order to quickly compare sensitivities of detectors, the \ac{GW} community uses a fiducial range, to which a typical \ac{BNS} can generally be detected.
This fiducial distance assumes that an \ac{SNR} of at least 8 is needed for a detection, and it approximates the \ac{BNS} inspiral waveform at Newtonian order (Section~\ref{s:cbc}).
The \ac{BNS} inspiral range is a volume-and-orientation-averaged measure of sensitivity to a signal from two \qty{1.4}{\Msun} bodies in a quasi-circular inspiral at a single-detector \ac{SNR} threshold of 8 \citep{1993PhRvD..47.2198F,2021CQGra..38e5010C}.
When a homogeneous \ac{BNS} population is assumed and cosmological effects are ignored, the \ac{BNS} inspiral range for a detector is determined by its noise power spectrum as
\begin{equation}\label{e:bns_range}
    R = \sifmt{round-mode=figures,round-precision=4}{\SensemonRangeCoefficient}
    \sqrt{ \int_0^\infty \frac{f^{-7/3}}{S_n(f)}\,df }\:,
\end{equation}
and the sensitive volume of the detector (also when neglecting cosmological effects) is given by $V=(4\pi/3) R^3$ (Appendix~\ref{s:conventions}). 
This measure is taken as a simple figure of merit of sensitivity to \acp{CBC}; it does not attempt to account for the true underlying astrophysical distribution describing such systems.
If the number of \ac{BNS} mergers per unit time per unit volume of space, the merger rate density of \acp{BNS}, is $\mathcal{R}$, then the expected number of \ac{BNS} signals seen with SNR greater than 8 in time $T$ would be ${\mathcal{R}}VT$.
Figure~\ref{fig:obsruns} also gives the typical \ac{BNS} inspiral range, as given in Equation~\eqref{e:bns_range}, for each detector during each observing run.

\begin{figure*} 
\begin{center}
\includegraphics{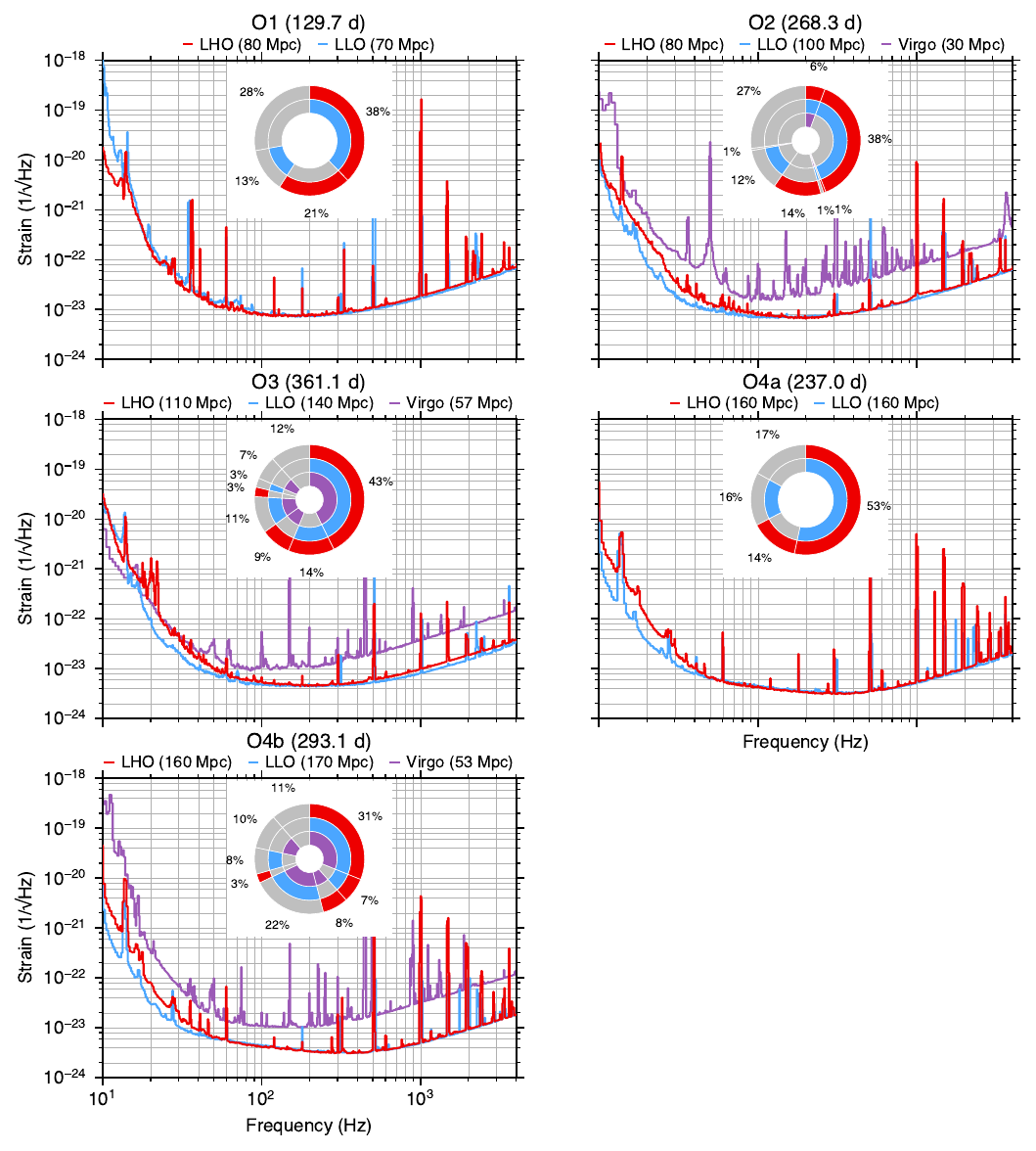}
\end{center}
\caption{
\label{fig:asd}
Representative noise amplitude spectral densities for \ac{LHO}, \ac{LLO}, and \ac{Virgo} during
\ac{O1} (\ac{LHO}, \ac{LLO}: \OoneAsdLhoDate),
\ac{O2} (\ac{LHO}: \OtwoAsdLhoDate; \ac{LLO}: \OtwoAsdLloDate; \ac{Virgo}: from \citealt{2023CQGra..40r5006A}),
\ac{O3} (\ac{LHO}: \OthreebAsdLhoDate; \ac{LLO}: \OthreeaAsdLloDate; \ac{Virgo}: \OthreebAsdVirgoDate), \ac{O4a} (\ac{LHO}: \OfouraAsdLhoDate; \ac{LLO}: \OfouraAsdLloDate),
and \ac{O4b} (\ac{LHO}: \OfourbAsdLhoDate; \ac{LLO}: \OfourbAsdLloDate; \ac{Virgo}: \OfourbAsdVirgoDate).
The \ac{BNS} inspiral ranges, defined by Equation~\eqref{e:bns_range}, for these noise curves are given in the legend.
Inset sunburst charts show the fraction of the run duration during which different combinations of detectors were observing.
Gray regions in each ring indicate portions when a detector is not operating.
The segments of the sunburst chart, clockwise from 12~o'clock, are \ac{LHO}--\ac{LLO}, \ac{LHO} alone, \ac{LLO} alone, and neither for observing runs involving only \ac{LHO} and \ac{LLO}; and \ac{LHO}--\ac{LLO}--\ac{Virgo}, \ac{LHO}--\ac{LLO}, \ac{LHO}--\ac{Virgo}, \ac{LLO}--\ac{Virgo}, \ac{LHO} alone, \ac{LLO} alone, \ac{Virgo} alone, and none for observing runs involving \ac{LHO}, \ac{LLO}, and \ac{Virgo}.}
\end{figure*}

The \emph{amplitude} strain noise spectrum is the square root of the (one-sided) noise \ac{PSD} in strain-equivalent units $S_n^{1/2}(f)$ having
dimensions of time\textsuperscript{1/2}.
The amplitude strain noise spectra of \ac{LHO}, \ac{LLO}, and \ac{Virgo} during the various observing runs are shown in Figure~\ref{fig:asd}.
There is an overall reduction in the detector noise levels with successive observing runs resulting in increased sensitivity.
Figure~\ref{fig:asd} also shows the fraction of the run duration during which different combinations of detectors were observing.

\begin{figure*}
\begin{center}
\includegraphics{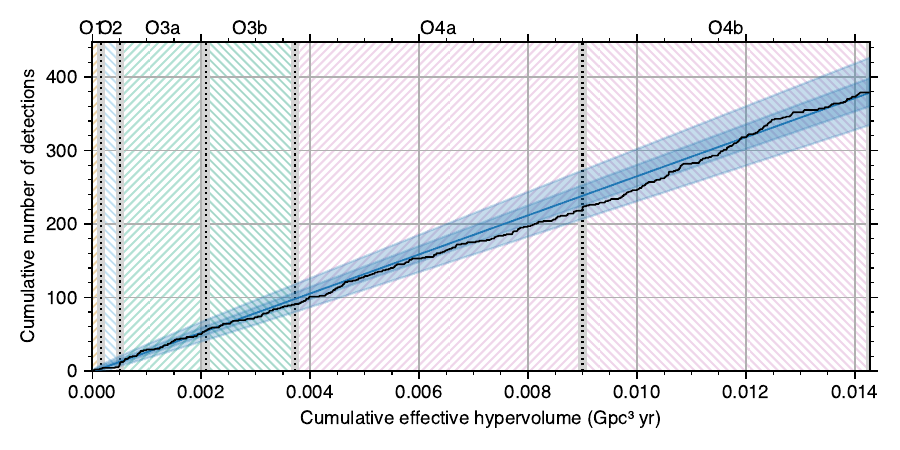}
\end{center}
\caption{\label{fig:VT-events}
The number of CBC detection candidates with a probability of astrophysical origin greater than or equal to \PASTROTHRESHOLDPCT{} versus the detector network's effective surveyed hypervolume for \ac{BNS} coalescences~\citep{2021PhRvX..11b1053A}.
The \ac{BNS} effective surveyed hypervolume is a proxy for overall sensitivity to \acp{CBC}, although its scale is set to the case of canonical \ac{BNS} signals.
The colored bands indicate the different observing runs.
The final data sets for \ac{O1}, \ac{O2}, \ac{O3a}, \ac{O3b}, \ac{O4a} and \ac{O4b} consist of \sifmtonedec{\OoneDurationAnyTwo}, \sifmtonedec{\OtwoDurationAnyTwo}, \sifmtonedec{\OthreeaDurationAnyTwo} (\sifmtonedec{\OthreeaDurationAnyOne}), \sifmtonedec{\OthreebDurationAnyTwo} (\sifmtonedec{\OthreebDurationAnyOne}), \sifmtonedec{\OfouraDurationAnyTwo} (\sifmtonedec{\OfouraDurationAnyOne}) and \sifmtonedec{\OfourbDurationAnyTwo} (\sifmtonedec{\OfourbDurationAnyOne}), respectively, with at least two detectors (one detector).
The cumulative number of probable candidates is indicated by the solid black line, while the blue line, dark-blue band and light-blue band are the median, $50\%$ confidence interval, and $90\%$ confidence interval for a Poisson distribution fit to the number of candidates at the end of \ac{O4b}, respectively. 
}
\end{figure*}

Figure~\ref{fig:VT-events} shows the cumulative number of candidates detected versus the estimated effective time--volume hypervolume $VT$ for the detector network.
For the first two observing runs (described below), only data when two detectors were operating were searched for \acp{GW}.
In this case the rate at which $VT$ is accumulated at any observing time is given by the sensitive volume $V$ for the \emph{second} most sensitive instrument observing at that time.
Beginning with the third observing run, periods during which only a single detector was observing were included in the search.
During such time, the rate at which $VT$ is accumulated is again given by the sensitive volume $V=(4\pi/3) R^3$, but where $R$ is computed from Equation~\eqref{e:bns_range} divided by 1.5, representing an effective \ac{SNR} threshold for detection of 12 rather than 8 for single-detector observation \citep{2021PhRvX..11b1053A}.
This simple estimate of $VT$, derived from the \ac{BNS} inspiral range, is an approximate one done for a quick and convenient overview.
In particular, it makes a crude approximation of whether a signal is detectable, and its numerical value is only representative of sensitivity to sources in a small region of mass space.
Actual measured sensitive hypervolume $\VT$ values for various \ac{CBC} mass regions and search methods are reported in \cite{GWTC:Results}.

\subsection{\acs{O1}\acused{O1}: The First Observing Run}
\label{ssec:O1}

O1 consists of the time period from \OoneStartDate{} to \OoneEndDate{}.
\ac{O1} includes short time periods that were originally planned to be engineering time (\OoneStartDate{} to \OoneStartIntendedDate{} and \OoneEndIntendedDate{} to \OoneEndDate{}), but which were of sufficient quality to be included in \ac{O1}.
This was the first observing run with the \ac{aLIGO} interferometers, in progress toward full \ac{aLIGO} design sensitivity~\citep{2016PhRvL.116m1103A,2016PhRvD..93k2004M}, \ac{LHO} achieving a \ac{BNS} range of \sifmtmonedec{\OoneLHORange} and \ac{LLO} a range of \sifmtmonedec{\OoneLLORange}.

Of the \sifmtonedec{\OoneDuration} duration of \ac{O1}, there were only \sifmtonedec{\OoneDurationHL} (\sifmttwofig{\OoneDurationHLFraction}) when both \ac{LHO} and \ac{LLO} were observing jointly, and there were \sifmtonedec{\OoneDurationNone} (\sifmttwofig{\OoneDurationNoneFraction}) when neither detector was observing.
The largest nonobserving periods were due to locking, the time spent bringing the interferometers from an uncontrolled state to their low-noise configuration ~\citep{2014CQGra..31x5010S}, and environmental issues such as earthquakes, wind, and microseismic noise arising from ocean storms~\citep{2015CQGra..32c5017E,2016CQGra..33m4001A}.
Wind and microseismic noise have seasonal variation, as storms are more prevalent in winter months; \ac{LLO} was more susceptible to these than \ac{LHO}, mainly due to its local geophysical environment~\citep{2004CQGra..21.2255D}. 

Overall, a total effective hypervolume $VT=\sifmtthreesci{\OoneVTTotal}$ was accumulated during joint \ac{LHO}--\ac{LLO} observing during \ac{O1}.

\subsection{\acs{O2}\acused{O2}: The Second Observing Run}\label{ssec:O2}

The \ac{O2} run was from \OtwoStartDate{} to \OtwoEndDate{}.
It was preceded by an engineering run that began on \OtwoPreEngrunStartLhoDate{} at \ac{LLO} and on \OtwoPreEngrunStartLloDate{} at \ac{LHO}.
The \ac{LHO} and \ac{LLO} detectors achieved a typical \ac{BNS} range sensitivity of \sifmtmonedec{\OtwoLHORange} and \sifmtmonedec{\OtwoLLORange} respectively~\citep{2017PhRvL.118v1101A,2019PhRvX...9c1040A}.
However, on \OtwoEarthquakeMontanaDate{} \ac{LHO} was severely affected by a \OtwoMontanaEarthquakeMagnitude{}~mag earthquake in Montana, resulting in a post-earthquake sensitivity drop of approximately \OtwoMontanaEarthquakeBNSDrop{} in \ac{BNS} range for the remainder of the run \citep{2019PhRvX...9c1040A}.

The \ac{AdV} interferometer \citep{2015CQGra..32b4001A} joined \ac{O2} on \OtwoVStartDate{}, forming a three-detector network for the last month of the run.
A vacuum contamination issue required \ac{AdV} to use steel wires rather than fused silica fibers to suspend the test masses, limiting the sensitivity of \ac{AdV} \citep{2019PhRvX...9c1040A}.
In \ac{O2}, a \sifmtmonedec{\OtwoVirgoRange} \ac{BNS} range was achieved.

The \ac{LIGO} detectors saw some improvement in duty factors during nonwinter months, with an almost \OtwoDutyFactorsPercentage{} reduction in downtime due to environmental effects at both sites, though \ac{LLO} lost over twice as much observing time as \ac{LHO} to earthquakes, microseismic noise, and wind.
\ac{O2} had a planned mid-run engineering break to effect needed repairs and to attempt improvements to the sensitivity.
The Virgo instrument operated with a duty factor of approximately \sifmttwofig{\OtwoVDurationVAnyFraction} after joining \ac{O2}.
There were \sifmttwofig{\OtwoDurationHLV} of all three detectors observing simultaneously.

Overall, a total effective hypervolume $VT=\sifmtthreesci{\OtwoVTTotal}$ was accumulated during O2; of this, \sifmtthreesci{\OtwoVTHL} was accumulated during joint \ac{LHO}--\ac{LLO} observing, \sifmtthreesci{\OtwoVTHLV} was accumulated while all three detectors were observing, and only \sifmtthreesci{\OtwoVTHV} and \sifmtthreesci{\OtwoVTLV} were accumulated during joint \ac{LHO}--\ac{Virgo} and \ac{LLO}--\ac{Virgo} observing, respectively.

\subsection{\acs{O3}\acused{O3}: The Third Observing Run}\label{ssec:O3}

\ac{O3} started on \OthreeAStartDate{}, with a commissioning break from \OthreeAEndDate{} to \OthreeBStartDate{}.
This observing run was planned to continue to \OthreebEndIntendedDate{} but the COVID-19 pandemic resulted in a suspension of observing on \OthreeBEndDate{} \citep{2023PhRvX..13d1039A}.
The period of \ac{O3} prior to the commissioning break is referred to as \ac{O3a}, while the period after the break is referred to as \ac{O3b}.
\ac{KAGRA} had intended to join \ac{LIGO} and Virgo at the end of \ac{O3}, but the early end made this impossible.
Instead, \ac{KAGRA} and \ac{GEO} jointly observed for a two-week period from \OthreeGKStartDate{} to \OthreeGKEndDate{} after \ac{LIGO} and Virgo had suspended their observing.
This joint \ac{GEO}--\ac{KAGRA} run (distinct from the \ac{O3} run described previously) is referred to as \ac{O3GK} \citep{2022PTEP.2022f3F01A}.

In \ac{O3}, the \ac{LHO} and \ac{LLO} detectors achieved a \ac{BNS} range of \sifmtmonedec{\OthreeLHORange} and \sifmtmonedec{\OthreeLLORange} respectively \citep{2020PhRvD.102f2003B}.
This increase in sensitivity resulted from a variety of improvements, chief among them an increase in the input laser power, the addition of a squeezed vacuum source at the interferometer output \citep{2019PhRvL.123w1107T}, and mitigation of noise arising from scattered light \citep{2021CQGra..38b5016S}.
In addition, end-test-mass optics with lower-loss coatings, along with new reaction masses, were installed in each \ac{LIGO} interferometer \citep{2020CQGra..37i5004G,2012CQGra..29w5004A}.

The steel wires in \ac{AdV} were replaced with fused silica fibers in preparation for \ac{O3}.
Along with other improvements, such as reduction of technical noises, an increase in laser power, and the installation of a squeezed vacuum source, 
\ac{Virgo} achieved a \ac{BNS} range of up to \sifmtmonedec{\OthreeVirgoRange} \citep{2019PhRvL.123w1108A}.

Over all of \ac{O3a} and \ac{O3b}, \sifmtonedec{\OthreeDuration} combined, there were \sifmtonedec{\OthreeDurationHLV} (\sifmttwofig{\OthreeDurationHLVFraction}) of three-detector observation and only \sifmtonedec{\OthreeDurationNone} (\sifmttwofig{\OthreeDurationNoneFraction}) during which no detector was observing.
The total effective hypervolume $VT$ accumulated was \sifmtthreesci{\OthreeVTTotal}.
Of this, \sifmtthreesci{\OthreeVTHLV} was accumulated during three-detector observations, \sifmtthreesci{\OthreeVTHL} when \ac{LHO} and \ac{LLO} were observing, \sifmtthreesci{\OthreeVTHV} when \ac{LHO} and \ac{Virgo} were observing, \sifmtthreesci{\OthreeVTLV} when \ac{LLO} and \ac{Virgo} were observing.
The amount accumulated with only a single detector observing was \sifmtthreesci{\OthreeVTH}, \sifmtthreesci{\OthreeVTL}, and \sifmtthreesci{\OthreeVTV} for \ac{LHO}, \ac{LLO}, and \ac{Virgo}, respectively.

The first operation of the \ac{KAGRA} detector in an initial configuration with a simple Michelson interferometer occurred in 2016~March \citep{2018PTEP.2018a3F01A}.
In 2019~August, the first lock of the Fabry--Perot Michelson interferometer was achieved, with power recycling accomplished in 2020~January.
By the end of 2020~March, \ac{KAGRA} obtained a \ac{BNS} range of approximately \qty{1}{\Mpc} \citep{2023PTEP.2023jA101A}, and although the \ac{LIGO} and Virgo instruments had ended their \ac{O3} run, \ac{KAGRA} was operated jointly with \ac{GEO}, which had a comparable \ac{BNS} range, in \ac{O3GK} yielding \sifmtonedec{\OthreeGKDurationGK} of joint observing time.

\subsection{\acs{O4}\acused{O4}: The Fourth Observing Run}

\acs{O4} began on \OfourAStartDate{} at \OfourAStartTime{}~UTC.
This run is again divided into parts: the \acl{O4a} \citep[\acs{O4a};][]{2025ApJ...995L..18A} ended on \OfourAEndDate{} at \OfourAEndTime{}~UTC and was followed by a commissioning break; the \acf{O4b} started on \OfourBStartDate{} at \OfourBStartTime{}~UTC.
The \ac{O4b} period continued until \OfourBEndDate{} \OfourBEndTime{}~UTC, the original intended end of \ac{O4}; however, it was decided to continue observing into a \acf{O4c}, which ended on {2025~November~18} at {16:00}~UTC.
The period covered by \thisgwtc{} contains events that occurred in \ac{O4b} and earlier observing runs only (see Section~\ref{ss:gwtc_versioning}). \ac{O4c} analyses are underway and will be included in future versions of the \gwtc{}. 
Around \ac{O4}, during transitional engineering runs between commissioning and observing, some candidates have been identified; upon careful follow-up and vetting, some of these candidates are included in the \gwtc{} \citep{GWTC:Results}.

The two \ac{LIGO} detectors were observing during \ac{O4b}, with \ac{LHO} having a \ac{BNS} range of approximately \sifmttwofig{\OfourbLHORange} and \ac{LLO} of \sifmttwofig{\OfourbLLORange}. The Virgo detector was observing with a median \ac{BNS} range of \sifmttwofig{\OfourbVirgoRange}.
During the \sifmtonedec{\OfourbDuration} there were \sifmtonedec{\OfourbDurationHLV} (\sifmttwofig{\OfourbDurationHLVFraction}) of three-detector joint observation, with \sifmtonedec{\OfourbDurationHL} (\sifmttwofig{\OfourbDurationHLFraction}) when both \ac{LIGO} detectors were observing, \sifmtonedec{\OfourbDurationHV} (\sifmttwofig{\OfourbDurationHVFraction}) when \ac{LHO} and \ac{Virgo} were observing, \sifmtonedec{\OfourbDurationLV} (\sifmttwofig{\OfourbDurationLVFraction}) when \ac{LLO} and \ac{Virgo} were observing, and \sifmtonedec{\OfourbDurationNone} (\sifmttwofig{\OfourbDurationNoneFraction}) when no detector was observing.
\ac{KAGRA} also continued commissioning to improve sensitivity, with the goal of joining \ac{O4} toward the end of the run.

During \ac{O4b}, the total effective hypervolume $VT$ accumulated was
\sifmtthreesci{\OfourbVTTotal}.
This is divided into \sifmtthreesci{\OfourbVTH} during which \ac{LHO} alone was observing, \sifmtthreesci{\OfourbVTL} during which \ac{LLO} alone was observing, and \sifmtthreesci{\OfourbVTV} during which \ac{Virgo} alone was observing. \sifmtthreesci{\OfourbVTHL} during which both \ac{LIGO} detectors were observing, \sifmtthreesci{\OfourbVTHV} during which \ac{LHO} and \ac{Virgo} detectors were observing, \sifmtthreesci{\OfourbVTLV} during which \ac{LLO} and \ac{Virgo} detectors were observing, and \sifmtthreesci{\OfourbVTHLV} during which all three detectors were observing.

A six-month observing run, designated as \ac{IR1}, is planned to begin in late October or mid-November of 2026. The timeline for \iac{O5} is being assessed in order to maximize the scientific output of the global network. Updates to the planned observing schedule will be provided once such decisions are made.\footnote{%
\ac{LVK} observing run plans \url{https://observing.docs.ligo.org/plan}%
}

\section{Observatory Evolution}
\label{sec:interferometers}

The advanced-detector era is characterized by a series of technological improvements from the initial detectors that deliver higher sensitivity and greater \ac{BNS} inspiral range, which made possible the era of \ac{GW} observation. 
Some of the key instrument science elements of the advanced-era detectors are
(i) increases in the input laser power entering the interferometer and to the circulating power in the interferometer cavities (a higher power in the arms produced a lower quantum-shot-noise-limited sensitivity above \qty{\sim200}{\hertz});
(ii) increases in test mass mirror size to accommodate larger beams, which mitigates coating thermal noise and heavier masses to reduce inertial and quantum back-action effects;
(iii) implementation of signal recycling~\citep{1988PhRvD..38.2317M} in addition to power recycling \citep{Drever:1982}, which alters the frequency band of the detectors' sensitivity (typically to give broader-band sensitivity);
(iv) implementation of monolithic test-mass suspensions, which reduces the suspension thermal noise in the detectors' sensitivity band by using the same low mechanical loss material (fused silica for \ac{LIGO} and \ac{Virgo}) for the suspension fibers as for the mirror substrate, and low loss jointing techniques and thermoelastic nulling~\citep{2012CQGra..29w5004A,2018JPhCS.957a2012T};
(v) improved passive and active seismic isolation systems and sensors to reduce ground motion coupling to the detector and to damp suspension modes~\citep{2005APh....23..557B,2015CQGra..32r5003M,2023RScI...94a4502C};
(vi) improved low thermal noise, low-absorption, high-reflectivity mirror coatings~\citep{2007CQGra..24..405H,2020CQGra..37i5004G};
(vii) injection of squeezed light to manipulate the quantum-noise-limited sensitivity of the detectors~\citep{2019PhRvL.123w1107T,2019PhRvL.123w1108A}.

Throughout the advanced-detector era of \ac{GW} observation, the \ac{LIGO} and Virgo detectors have undergone a series of performance-improving detector upgrades and commissioning activities, details of which are given in this section.
Detector upgrades include the installation of new hardware or upgrades to existing hardware in a detector. 
Commissioning activities cover a range of improvements to sensitivity and observing uptime of the instruments from targeted noise-hunting activities that remove glitches, lines, and broadband noise, to improved control schemes that mitigate instabilities and improve detector robustness. 

Alongside this has been the effort to build and commission the KAGRA detector utilizing advanced technologies such as cryogenic cooling of the test masses and an underground location~\citep{2019CQGra..36p5008A}. 
This schedule of planned upgrades and commissioning activities between observing runs ensures that the maximal science output is achieved from the network~\citep{2020LRR....23....3A}. 
In terms of valuable scientific output, a successful upgraded detector that has been offline for a period of time rapidly overtakes a non-upgraded detector in continuous observational mode in terms of number of significant detections and the resolution and sky localization of high-interest signals.

The \ac{aLIGO} and \ac{AdV} detectors are designed to be dual-recycled Fabry--Perot Michelson interferometers with orthogonal kilometer-scale arms \citep{2015CQGra..32g4001L,2015CQGra..32b4001A}.
Each arm contains a Fabry--Perot optical cavity, and a beam splitter at the corner between the arms forms a Michelson interferometer that measures the change in the relative phase of the light induced by changes in the lengths of these cavities \citep{1987thyg.book..330T,1988PhRvD..38..433V}.
Additional power-recycling and signal-recycling cavities are created by adding mirrors in the symmetric and antisymmetric ports of the interferometer.
These improve sensitivity by building up the light power on the beam splitter and beneficially modifying the response of the interferometer, respectively \citep{1988PhRvD..38.2317M}.
The input and end mirrors on each of the Fabry--Perot cavities are the test masses whose separations are affected by \acp{GW}.
The mirrors are isolated by multistage pendulums that suppress the ground motion by more than \OoneGroundMotionMagnitudeReduction{} orders of magnitude at frequencies around \OoneGroundMotionSuppressionFrequency{}.
Monolithic fused silica fibers are used on the bottom stage of the suspension system to suppress thermal noise, and the mirrors themselves are fused silica substrates with low-loss, highly reflective coatings \citep{2012CQGra..29w5004A}.

Ground-based interferometers generally have the same fundamental limiting noise sources \citep{2022GReGr..54..153W,2017fige.book.....S}, with the response of each detector and the exact extent to which each noise limits sensitivity being specific to the detailed design of each detector. 
At low observational frequency below \qty{\sim10}{\hertz} the detectors are limited by a combination of seismic noise, gravity gradient noise, suspension thermal noise, and quantum radiation pressure noise.
Thermal noise in the mirror optical coatings is a significant noise source at intermediate frequencies \qtyrange{\sim50}{\sim200}{\hertz}~\citep{2007CQGra..24..405H}, and at high frequencies, above \qty{\sim200}{\hertz}, sensitivity is limited by the quantum shot noise.

In addition to these fundamental noise sources, the detectors are also limited by technical noise.
This includes scattered-light noise, which occurs when some fraction of light is deflected from the interferometer beam path and is incident on another moving surface, varying the phase of the light; this couples noise into the interferometer readout if part of this light is reflected back into the main beam~\citep{2012AIPC.1446..150A,2012OExpr..20.8329O}.
Interferometer control-system noise is when signals couple between the multiple feedback loops that control the degrees of freedom of the interferometer and requires complicated optimization of control loop parameters to mitigate~\citep{2020PhRvD.102f2003B,2020Galax...8...85A}. 
Laser noise due to fluctuations in the frequency, intensity, and pointing of the laser beam entering the interferometer is reduced with dedicated multistage stabilization systems to a level such that it does not impact the sensitivity of the detectors; however, suboptimal tuning of these stabilization systems can lead to laser noise affecting sensitivity~\citep{2009PhRvA..79e3824A,2021OExpr..2942144C,2025APh...16403028V}.
Environmental noise is caused when environmental effects in the vicinity of the interferometer (e.g., seismic activity) couple into the measurement of the interferometer strain signal~\citep{2006JPhCS..32...80A,2015CQGra..32c5017E,2020Galax...8...82F,2021CQGra..38n5001N,2022CQGra..39w5009A,2024CQGra..41n5003H}.
Detector commissioning seeks to mitigate such nonfundamental noise sources.

The key parameters of the \ac{LIGO}, Virgo, KAGRA, and \ac{GEO} detectors across the advanced era observing runs are given in Table~\ref{tab:inst_table}. The specific evolution of each detector in terms of detector upgrades and improvements is detailed in the remainder of this section.

\begin{deluxetable*}{llccccc}
\tablecaption{\label{tab:inst_table}%
Selected Optical and Physical Parameters of the \ac{LIGO} Hanford (\acs{LHO}), \ac{LIGO} Livingston (\acs{LLO}), Virgo, KAGRA, and \ac{GEO} Interferometers throughout the Advanced-detector Era.}
\tablehead{\colhead{Observing Period} & \colhead{Interferometer} & \colhead{Input Laser Power} & \colhead{Power-recycling Gain} & \colhead{Signal Recycling} & \colhead{Squeezing} & \colhead{Suspension Type}}
\startdata
\multirow{2}{*}{\ac{O1}} & \ac{LHO} & \HANFORDPOWERONE{} & \HANFORDPRGONE{} & \textcheckmark & $\times$ & Silica \\
& \ac{LLO} & \LIVINGSTONPOWERONE{} & \LIVINGSTONPRGONE{} & \textcheckmark & $\times$ & Silica \\
\hline
\multirow{3}{*}{\ac{O2}} & \ac{LHO} & \HANFORDPOWERTWO{} & \HANFORDPRGTWO{} & \textcheckmark & $\times$ & Silica \\
& \ac{LLO} & \LIVINGSTONPOWERTWO{} & \LIVINGSTONPRGTWO{} & \textcheckmark & $\times$ & Silica \\
& Virgo & \VIRGOPOWERTWO{} & \VIRGOPRGTWO{} & $\times$ & $\times$ & Steel \\
\hline
\multirow{3}{*}{\ac{O3a}} & \ac{LHO} & \HANFORDPOWERTHREEA{} & \HANFORDPRGTHREEA{} & \textcheckmark & \textcheckmark & Silica \\
& \ac{LLO} & \LIVINGSTONPOWERTHREEA{} & \LIVINGSTONPRGTHREEA{} & \textcheckmark & \textcheckmark & Silica \\
& Virgo & \VIRGOPOWERTHREEA{} & \VIRGOPRGTHREEA{} & $\times$ & \textcheckmark & Silica \\
\hline
\multirow{3}{*}{\ac{O3b}} & \ac{LHO} & \HANFORDPOWERTHREEB{} & \HANFORDPRGTHREEB{} & \textcheckmark & \textcheckmark & Silica \\
& \ac{LLO} & \LIVINGSTONPOWERTHREEB{} & \LIVINGSTONPRGTHREEB{} & \textcheckmark & \textcheckmark & Silica \\
& Virgo & \VIRGOPOWERTHREEB{} & \VIRGOPRGTHREEB{} & $\times$ & \textcheckmark & Silica \\
\hline
\multirow{2}{*}{\ac{O3GK}} & \ac{GEO} & \GEOPOWERTHREEGK & \GEOPRGTHREEGK & \textcheckmark & \textcheckmark & Silica \\
& \ac{KAGRA} & \KAGRAPOWERTHREEGK & \KAGRAPRGTHREEGK & $\times$ & $\times$ & Sapphire \\
\hline
\multirow{2}{*}{\ac{O4a}} & \ac{LHO} & \HANFORDPOWERFOURA{} & \HANFORDPRGFOURA{} & \textcheckmark & \textcheckmark & Silica \\
& \ac{LLO} & \LIVINGSTONPOWERFOURA{} & \LIVINGSTONPRGFOURA{} & \textcheckmark & \textcheckmark & Silica \\
\hline
\multirow{3}{*}{\ac{O4b}} & \ac{LHO} & \HANFORDPOWERFOURB{} & \HANFORDPRGFOURB{} & \textcheckmark & \textcheckmark & Silica \\
& \ac{LLO} & \LIVINGSTONPOWERFOURB{} & \LIVINGSTONPRGFOURB{} & \textcheckmark & \textcheckmark & Silica \\
& Virgo & \VIRGOPOWERFOURB{} & \VIRGOPRGFOURB{} &  \textcheckmark & \textcheckmark & Silica \\ 
\enddata
\def\tablecomments#1{\vskip1pt{\small\vskip1sp\indent\vrule height 11pt depth 2pt width 0pt\currtabletypesize{\bf Note.} {#1}\vskip1pt}}

\tablecomments{The input laser power is the power that would be measured at the power-recycling mirror (after the input mode cleaner) and is an estimate of the maximum level typically achieved during an observing period. Suspension types are monolithic fused silica fibers, sapphire fibers, or steel wires.}
\end{deluxetable*}

\subsection{LIGO Hanford and Livingston Observatories}
\label{sec:LIGO-interferometers}

\ac{LIGO} is a US national facility comprising two US-based interferometric detectors in Hanford, Washington (\acs{LHO}), and Livingston, Louisiana (\acs{LLO}), each with \qty{4}{\kilo\meter} arms.
\ac{LIGO} construction began in 1994.
From 2002 to 2010, initial power-recycled Fabry--Perot Michelson interferometers were operated at these sites in a series of science runs S1--S6 \citep{2009RPPh...72g6901A,2015CQGra..32k5012A}.
During this period, \ac{LIGO} also operated a second interferometer with \qty{2}{\kilo\meter} arms at the Hanford site.
Subsequently the \ac{aLIGO} project resulted in a major overhaul of the interferometers to improve the capabilities of the detectors~\citep{2015CQGra..32g4001L}, leading up to \ac{O1} and the first observation of \acp{GW}~\citep{2016PhRvL.116m1103A}.

Across the observing runs the following areas have been the main focus of much of the detector improvement effort:
(i) increasing the arm cavity power by increasing the injected laser power and the power-recycling gain while achieving stable operation,
(ii) mitigation of scattered-light sources and coupling mechanisms, and
(iii) reduction of quantum noise with the addition of a squeezed-light system for \ac{O3} and the subsequent improvements to the quantum-enhancement factor.
 
Both \ac{aLIGO} detectors are operated with a lower injected laser power and lower power-recycling gain than the design goal~\citep{2015CQGra..32g4001L}.
The full amount of available laser power cannot be fully utilized due to issues with maintaining long-duration stable locking of the interferometer owing to angular instabilities and point absorbers in the test-mass mirrors~\citep{2021ApOpt..60.4047B}.
This issue was the focus of commissioning efforts to improve the operating power in the arm cavities by optimizing the interferometer control loops~\citep{2020PhRvD.102f2003B} and reducing the presence of point absorbers in the mirrors.
Stray-light mitigation can be achieved by using baffles to block unwanted beam paths and with active control of known scattered-light paths.
The addition of a squeezed vacuum source at the interferometer's output alters the quantum noise in the interferometer and, with the inclusion of a filter cavity can, produce frequency-dependent squeezing which can be used to surpass the standard quantum limit on sensitivity of a laser interferometer~\citep{2019PhRvL.123w1107T,2023PhRvX..13d1021G}.

\subsubsection{\ac{O1}}
The sensitivity and limiting noise sources of the \ac{LIGO} detectors during \ac{O1} are described in \citet{2016PhRvD..93k2004M}.
Figure~\ref{fig:asd} shows a representative amplitude spectral density of the strain noise and the \ac{BNS} range.
In \ac{O1}, the typical input power entering the power-recycling cavity was \HANFORDPOWERONE{} in \ac{LHO} and \LIVINGSTONPOWERONE{} in \ac{LLO}, circulation of laser light in the power-recycling cavity increases the power on the beam splitter to be a factor of \HANFORDPRGONE{} times greater (the power-recycling gain), and a further increase in circulating power by a factor of 144 is achieved in the arms by the Fabry--Perot cavities.
The laser input power and power-recycling gain during \ac{O1} and the later observing runs are given in Table~\ref{tab:inst_table} alongside other detector parameters. 
An example of commissioning improvement is the investigation at \ac{LLO} during \ac{O1} of recurring changes in the \ac{BNS} range from \qty{65}{\Mpc} to \qty{60}{\Mpc}.
By searching for correlation between the detector range and the hundreds of data channels recorded by \ac{aLIGO}, it was found that the issue was caused by a malfunctioning temperature sensor.
This sensor was replaced, resulting in a stabler increased range~\citep{2018CQGra..35v5002W}.

\subsubsection{\ac{O2}}

After \ac{O1}, several improvements were made to both \ac{LIGO} instruments \citep{2017PhRvL.118v1101A}.
Detector upgrades included installation of new mass dampers on the end-test-mass suspensions to dampen mechanical modes, improving the stabilization of laser intensity, and installing a new output Faraday isolator and higher quantum-efficiency photodiodes at the output port to improve signal detection efficiency in the readout system.
Mitigation of scattered-light sources and other improvements to the detector sensitivity throughout \ac{O2} resulted in a \ac{BNS} range improvement to \OtwoBnsRangeImprovement{} by the end of the run \citep{2021CQGra..38m5014D}.
Commissioning tests during \ac{O2} on the \ac{LHO} detector to increase the laser power to \qty{50}{\watt} did not result in an overall improvement in performance of the \qty{80}{\Mpc} \ac{BNS} range at the end of \ac{O1}, due to point absorbers on one of the input test-mass optics \citep{2021ApOpt..60.4047B}, so the detector operated with \qty{30}{\watt} input power.
After \ac{O2}, it was demonstrated that the use of witness channels to perform noise subtraction on the strain data was able to increase the \ac{BNS} range by \qty{20}{\percent} \citep{2019CQGra..36e5011D,2019PhRvD..99d2001D}.

\subsubsection{\ac{O3}}

Leading up to \ac{O3}, several upgrades were made to the \ac{LIGO} instruments~\citep{2020PhRvD.102f2003B}.
The most significant was the installation of an in-vacuum squeezed-light injection system at each site to inject squeezed vacuum into the interferometers to reduce shot noise at frequencies above \qty{50}{\hertz}~\citep{2019PhRvL.123w1107T}.
The squeezer works by optically pumping a nonlinear crystal to modify the distribution of the quantum vacuum state that enters the interferometer~\citep{1981PhRvD..23.1693C,2019RPPh...82a6905B}.

Between \ac{O3a} and \ac{O3b}, adjustments to the squeezing subsystem produced large sensitivity improvements~\citep{2023PhRvX..13d1039A}.
Among these were the installation of higher-power laser amplifiers with stable operation and output power over \qty{70}{\watt}~\citep{2020Galax...8...84B}.
A program of installation of optical baffles was completed to improve stray-light control.
The correlation of microseismic activity with scattered-light noise was determined to be primarily caused by a scattered-light path arising from large relative motion between the end test mass and the reaction mass that is immediately behind it~\citep{2021CQGra..38b5016S}.
A control loop that makes the reaction mass follow the end mass, implemented on 2020~January~07 at \ac{LLO} and 2020~January~14 at \ac{LHO}, reduced the relative motion and mitigated the scattered-light noise \citep{2021CQGra..38m5014D}.
At \ac{LHO}, wind fences were installed to mitigate ground tilt induced by wind on the buildings~\citep{2021CQGra..38n5001N}.

\subsubsection{\ac{O4a}}

Several upgrades were implemented at \ac{LHO} and \ac{LLO} to improve the quantum-limited sensitivity of the detectors via improved quantum squeezing and higher intracavity power \citep{2024ApJ...970L..34A}.
Further upgrades to the laser amplification system were implemented with stable operation and output power over \qty{140}{\watt}~\citep{2020Galax...8...84B}.
A new vacuum system to house a \qty{300}{\meter} filter cavity was built at both detectors, along with an upgraded squeezing injection system to allow the injection of frequency-dependent squeezed vacuum to achieve quantum noise reduction across the detection frequency band~\citep{2023PhRvX..13d1021G,2024Sci...385.1318J}.
Squeezing levels in \ac{O4a} reached \qty{5.8}{\decibel} at \ac{LLO} and \qty{4.6}{\decibel} at \ac{LHO}, compared to the \qtyrange{2}{3}{\decibel} achieved in O3~\citep{2025PhRvD.111f2002C}.
Test-mass mirrors were replaced at both observatories to remove point defects on the mirrors that contributed to control challenges and excess noise~\citep{2020PhRvD.102f2003B}.
This involved a replacement of both end test masses at \ac{LLO} and the input $Y$-arm test mass at \ac{LHO}.
Replacing these test masses allowed both observatories to approximately double the input power compared to \ac{O3}, further improving the quantum-limited sensitivity of the detectors, due to higher circulating power in the Fabry--Perot arm cavities~\citep{2025PhRvD.111f2002C,2020PhRvD.102f2003B}.

Other upgrades to the \ac{LIGO} detectors include improvements to the electronics in the \ac{GWH} signal readout chain, damping of baffles to mitigate scattered light, and improvements to electronics grounding~\citep{2025PhRvD.111f2002C,2025CQGra..42h5016S}.
The photodetector transimpedance amplifiers were improved ahead of \ac{O4a} using a design tested at \ac{GEO}, resulting in a factor of 10 reduction in dark noise compared to \ac{O3}~\citep{2016OExpr..2420107G}.
At both \ac{LIGO} detectors, a septum window separating two vacuum volumes housing the output optics was removed, significantly reducing the coupling of acoustic noise.
Baffles along the arm cavity and around vacuum pumps were previously identified to couple excess scattered light in \ac{O3} and were damped to reduce their motion and therefore shift the frequency of up-converted scattered light out of the sensitive band.
Finally, injections into the building electronics ground demonstrated that many spectral features in the strain at \ac{LHO} were the result of a fluctuating ground potential~\citep{2025PhRvD.111f2002C,2025CQGra..42h5016S}.
The resistance to ground was reduced for several electronics chassis around the detector.
Additionally, the voltage biases of the test-mass electrostatic drives were adjusted to minimize the electronics noise coupling further~\citep{2025PhRvD.111f2002C}.

Detector commissioning ahead of \ac{O4a} also focused on optimization of the auxiliary controls to reduce technical noise that limited the detectors at low frequency in \ac{O3}~\citep{2020PhRvD.102f2003B}.
Alignment control noise was reduced by a factor of 10 and length control noise by a factor of two at both detectors near \qty{20}{\hertz}~\citep{2025PhRvD.111f2002C,2020PhRvD.102f2003B}.
Significant improvements to the controls included the upgrade to a camera servo system that requires no line injection to sense the alignment of the main detector optics~\citep{2025PhRvD.111f2002C}.
Suspension local control loops were reoptimized to focus on noise suppression above \qty{5}{\hertz}, reducing both noise directly coupled to the strain and noise that couples indirectly through the length and alignment controls~\citep{2025PhRvD.111f2002C}.
Both detectors were also limited by unmitigated beam jitter noise that was well-witnessed by auxiliary sensors~\citep{2025PhRvD.111f2002C}.
As such, front-end infrastructure using the \ac{NonSENS} code \citep{NonSENS} was implemented to perform noise cleaning in low latency, increasing detector sensitivity by up to \qty{5}{\Mpc} in \ac{BNS} range~\citep{2020PhRvD.101d2003V,2022PhRvD.105j2005V,2025PhRvD.111f2002C}.

\subsubsection{\ac{O4b}}

During the observing break between \ac{O4a} and \ac{O4b}, upgrades were carried out at both sites.
To improve output losses that limited optical gain and squeezing levels, the \ac{LHO} output mode cleaner was swapped with a spare.
The main feedback control loop that controls the differential arm length at \ac{LHO} was recommissioned to reduce the coupling of nonlinear noise. The differential arm length control is distributed across multiple suspension stages, with the highest frequency portion of the control sent to the bottom suspension stage, which is an electrostatic drive actuator. By offloading some of the control request to the upper suspension stages, the RMS of the control request to the electrostatic drive was reduced, thereby reducing the amount of noise from 20\,Hz to 40\,Hz in the \ac{GW} band.
This low-frequency noise improvement in turn enabled an additional round of incremental noise reductions in the form of adjustments to the feedforward subtraction of noise in the longitudinal degree of freedom. Noise contributions from the power-recycling, signal-recycling, and Michelson cavities are subtracted in the feedforward architecture \citep{2025PhRvD.111f2002C}. In addition, careful beam centering adjustment resulted in reduced cross-coupling from alignment degrees of freedom.

At \ac{LLO} both end test masses were cleaned to address excess arm cavity loss, resulting in a 15\% power-recycling gain increase~\citep{2025PhRvD.111f2002C}.
Efforts to further increase the input power led to diminishing returns due to significant variation of the detector sensitivity over 24-hour periods. Lowering the input power reduced the range variation.
Quantum squeezing was improved through adjustment of alignment controls and mode matching. \ac{LLO} achieved 6.1\,dB of quantum-noise reduction during O4b~\citep{2025PhRvD.111f2002C}.

During O4b, both \ac{LHO} and \ac{LLO} faced impacts to their duty cycle due to equipment failure. At \ac{LHO}, a key component of the output optics was damaged during a normal lockloss event, requiring an emergency vacuum incursion to replace the optic. At \ac{LLO}, the fast shutter which protects the output mode cleaner during lockloss events failed, requiring an emergency incursion. Finally, the \ac{LHO} duty cycle was further reduced due to unexplained glitches in the main laser that caused frequent locklosses. The laser was replaced with a spare unit that had no glitching behavior.

\subsubsection{Beyond \ac{O4}}

Looking to the future, there is ongoing construction of LIGO-India \citep{Souradeep:2017}, a third LIGO interferometer to be built in the Hingoli district of Maharashtra, India.
This facility will be based on \ac{aLIGO} hardware and design, and its location will provide a significant improvement in the sky localization of \ac{GW} sources \citep{2012PhRvD..85j4045V,2014JPhCS.484a2007F,2020ApJ...902...71P,2022CQGra..39b5004S,2025ApJ...985L..17P}.

In parallel, there are plans underway to upgrade the existing \ac{LIGO} detectors (and eventually LIGO-India) to \ac{A+} sensitivity \citep{KAGRA:2013rdx,2023RScI...94a4502C}.
The \ac{A+} upgrade to the \ac{LIGO} detectors is a series of detector upgrades utilizing improved technology that has been developed in parallel to the observing runs. 
The inclusion of frequency-dependent squeezing was originally planned as an \Ac{A+} upgrade but was implemented ahead of \ac{O4a} at both sites \citep{2025PhRvD.111f2002C}.
Other \ac{A+} upgrades, which will be implemented for future observing runs, include new optics with lower noise and loss, improved sensors for controlling the mirrors, a new pre-mode cleaner to reduce beam jitter noise, improved output mode cleaners with lower loss, installation of a larger beamsplitter to reduce clipping losses and scattered light, and a balanced homodyne readout system that allows for better readout control of the interferometer signal. Some of those upgrades may be included for \ac{IR1}. Both LIGO detectors will be observing in \ac{IR1}, and Virgo and KAGRA will join as they are available.

A post-\ac{O5} upgrade, referred to as \ac{Asharp}, explores more transformative changes in detector design, with the goal of increasing the sensitivity to the limits of what is possible with the existing infrastructure of the \ac{LIGO} detectors \citep{PostO5Report:2022}.
Detector improvements that facilitate the achievement of the \ac{Asharp} sensitivity include the upgrade of the laser injection system to deliver more power into the interferometer and an improved system for the thermal compensation of the test-mass mirrors.
The test-mass mirrors will be replaced with heavier masses with improved optical coatings, and \ac{Asharp} targets an improved exploitation of the quantum noise reduction from the squeezed-light system. 
The \ac{Asharp} configurations are natural outgrowths of \ac{A+} configurations and will serve as pathfinders for the next-generation Cosmic Explorer concept \citep{2021arXiv210909882E}.
Additionally, it has much technological overlap with \ac{AdV+} and \ac{VirgoNEXT} (Section~\ref{sec:Virgo-interferometers}), which presents the possibility of collaborating on developing these technologies.

\subsection{Virgo Observatory}
\label{sec:Virgo-interferometers}

The \ac{Virgo} interferometer, located in Cascina (Italy), is the largest European \ac{GWH} detector, designed in its \ac{AdV} Phase~I as a \qty{3}{\km} dual-recycled Fabry--Perot Michelson interferometer \citep{2015CQGra..32b4001A}.
Construction of \ac{Virgo} started in 1997 and was completed in 2003 \citep{2005CQGra..22S.869A}.
Four science runs of the initial \ac{Virgo} interferometer, VSR1--VSR4, took place between 2007 and 2011.
These were followed by upgrades leading to the \ac{AdV} design operated during \ac{O2} and \ac{O3}.
Subsequently, further upgrades leading to \ac{AdV+} were planned to take place in two phases, the first for operation during \ac{O4} and the second for operation during \ac{O5}.
A proposed next-generation upgrade planned post-\ac{O5}, \ac{VirgoNEXT}, would provide further sensitivity by pushing current facilities to their limit and would serve as a pathfinder for future ground-based \ac{GW} detectors.

The first-generation \ac{Virgo} detector \citep{2012JInst...7.3012A} observed jointly with the initial \ac{LIGO} detector's fourth and fifth science runs.
After several years of commissioning, from 2007~May to 2007~October the first scientific data run VSR1 (along with \ac{LIGO}) took place, for which a \ac{BNS} range of \qty{4}{\Mpc} was achieved \citep{2008JPhCS.120c2007A}.
At this stage, \ac{Virgo} was a power-recycled Fabry--Perot Michelson interferometer with a \FirstGenerationVirgoLaserPower{}  laser source.
The second \ac{Virgo} science run, VSR2 (also along with \ac{LIGO}), from 2009~July to 2010~January \citep{2012AIPC.1446..150A}, was preceded by set of major improvements to mitigate scattered light and to improve the light injection system.

The replacement of the four payloads in the Fabry--Perot cavities was the major improvement in preparation for the third \ac{Virgo} science run VSR3, from 2010~July to 2010~October \citep{2012AIPC.1446..150A}.
Issues arising from thermal noise due to improperly aligned suspension wires and degraded contrast resulting from differing radii of mirror curvature were addressed leading up to VSR4, from 2011~June to 2011~October, during which \ac{Virgo} achieved a \ac{BNS} range of \VsrFourBnsRange{}.
While the three previous VSRs were aligned with initial \ac{LIGO} science runs, \ac{Virgo} took data during this run together with \ac{GEO}.
The main upgrade consisted of the installation of the \ac{CHRoCC} on both end mirrors, which allowed the radius of curvature of the mirrors to be controlled in real time \citep{2013CQGra..30e5017A}.
\ac{Virgo} stopped observing in 2011 for the \ac{AdV} upgrade.

\subsubsection{\ac{O2}}

After the four VSR1 to VSR4 science runs, major modifications were made to the optical layout to increase the broadband sensitivity by up to an order of magnitude \citep{2017PhRvL.119n1101A}.
These upgrades marked the transition from \ac{Virgo}, a first-generation interferometer, to \ac{AdV}, a second-generation \ac{GWH} detector \citep{2015CQGra..32b4001A}.
The installation of \ac{AdV} started in 2011 and was completed in August 2016.
\ac{AdV} was planned as a dual-recycled interferometer with \AdvLaserPowerDesign{} entering the interferometer, though signal recycling was not implemented until \ac{O4}.
The main improvements included a threefold increase in the arm cavity finesse (a measure of how long light stays within the cavity),
\OfourAdvTestMassKg{} fused silica test masses with ultralow absorption and high homogeneity, new stray-light control using diaphragm baffles and a vibration isolation system~\citep{2019CQGra..36g5007V}, an improved thermal compensation system with double axicon CO$_2$ laser projectors and ring heaters~\citep{2023CQGra..40e5004N}, an improved output mode cleaner with two cascaded monolithic bow-tie resonators, and a new design of payloads triggered by the need to suspend heavier mirrors, baffles, and compensation plates.

The several months of commissioning that started at the end of 2016~October achieved the target early-stage \ac{BNS} range of \OtwoCommissioningAdvBnsRange{} in 2017~July with \OtwoAdvLaserPower{} input laser power.
After an intense campaign of noise investigations, \ac{AdV} sensitivity was considered sufficient to join \ac{aLIGO} during the \ac{O2} observing run in 2017~August \citep{2018EPJWC.18202003A}.
During \ac{O2}, the \ac{AdV} \ac{BNS} range reached \sifmtmonedec{\OtwoVirgoRange}.
As noted in Section~\ref{ssec:O2}, the low-frequency Virgo sensitivity during {\ac{O2}} was limited by thermal noise from metallic suspension wires, which were implemented as a fallback option owing to the frequent failure of monolithic suspensions after the installation of the main \ac{AdV} upgrades.

\subsubsection{\ac{O3}}

The most important \ac{Virgo} upgrades for \ac{O3} were the mitigation of suspension thermal noise by installation of monolithic suspensions and the mitigation of quantum noise by increase of input laser power and by injection of frequency-independent squeezing.
An in-air optical parametric amplifier was implemented in the \ac{Virgo} interferometer before the start of \ac{O3a}, and squeezing injections were maintained during the whole of \ac{O3}, with a \qty{3}{\decibel} gain in sensitivity at high frequency~\citep{2019PhRvL.123w1108A,2020PhRvL.125m1101A}.

Throughout \ac{O3}, work was continuously carried out to improve the Virgo sensitivity in parallel with the ongoing data taking.
Dedicated tests were made during planned breaks in operation (commissioning, calibration, and maintenance), and in-depth data analysis of these tests was performed between breaks to ensure continual improvement.
In particular, the one-month commissioning break between the \ac{O3a} and \ac{O3b} observing periods was used to get a better understanding of the \ac{Virgo} sensitivity and of some of its main limiting noises \citep{2023PhRvX..13d1039A}.
This effort culminated during the last 3 months of \ac{O3b}.

The most significant change to the \ac{Virgo} configuration between \ac{O3a} and \ac{O3b} was the increase of the input power from \VIRGOPOWERTHREEA{} to \VIRGOPOWERTHREEB{}.
As with the \ac{LIGO} detectors, it was found that the optical losses of the arms increased following the increase of the input power.

New high quantum-efficiency photodiodes that had been installed at the output (detection) port of the interferometer prior to the start of \ac{O3a} were found to increase the electronics noise at low frequency.
These were improved at the end of 2020~January during a maintenance period, by replacing pre-amplifiers.
The electronic noise disappeared completely, leading to a \ac{BNS} inspiral range gain of
{{\qty{\sim 2}{\Mpc}}}.

Finally, in the period from the end of 2020~January to the beginning of 2020~February the alignment was improved for the injection of the squeezed light into the interferometer \citep{2019PhRvL.123w1108A,2020PhRvL.125m1101A}, a critical parameter of the low-frequency sensitivity.
By mitigating scattered-light noise, the \ac{BNS} range increased by
{{\qtyrange{1}{2}{\Mpc}}}.

\subsubsection{\ac{O4}}

The first step of the \ac{AdV+} detector upgrade project
\citep{VIR-0596A-19} was
implemented for \ac{O4}. The signal recycling mirror was installed to
complete the Virgo optical configuration and increase the detector
bandwidth by a factor 8. In addition, the input mode cleaner payload
was replaced to improve its stability when operated with high laser
power; the double output mode-cleaner cavities were replaced by a
single cavity with a 10 times higher finesse; the seismic sensors of
suspended optical benches were upgraded and additional baffles
installed to reduce scattered light impact during periods of high
ground motion \citep{2021CQGra..38g5020W};
and the amplitude noise of the \ac{RF} sideband
generators was reduced by an active control system.

Moreover, a new \qty{300}{m} long vacuum system was built and a
suspended filter cavity installed in order to upgrade the squeezing
injection system to frequency dependent squeezing \citep{2023PhRvL.131d1403A}.
A network of seismic and acoustic sensors was installed for Newtonian
noise monitoring; and the calibration system was complemented with a
Newtonian calibrator for improved absolute calibration accuracy \citep{2024CQGra..41w5003A}.

The addition of signal recycling to Virgo's marginally stable recycling 
configuration rendered the interferometer difficult to control, as in 
that configuration signal recycling amplifies higher order modes. 
This prolonged commissioning, and required improvements to the power 
recycling curvature adjustment by the addition of a
\ac{CHRoCC}~\citep{2013CQGra..30e5017A} and supplementing several 
longitudinal and angular control \ac{RF} error signals with mechanical dithers of
mirrors. To further reduce control issues the input power was reduced
to \qty{17}{W}.

Signal recycling was also amplifying an optical noise; misaligning the
signal recycling reduced that amplification, improving the sensitivity
from \qty{40}{\Mpc} to about \qty{55}{\Mpc}. This was at the expense of
reducing the detector bandwidth by a factor 2 and rendering the
squeezing system ineffective.

Thus, Virgo could not join \ac{O4a} and continued commissioning,
allowing it to reach stable operation and join the \ac{O4b} observing
run with a \ac{BNS} range of \OfourBadvBnsRangeTarget{}.

The end test masses have also been replaced. In the north arm the
mirror was replaced in 2023 to remove excess uniform scattering and
resulting in a 10\% increase in power-recycling gain \citep{2025ApOpt..64.4710V}. In the west arm the mirror was replaced during
O4c in April 2025, in parallel to equipment failure repairs at LIGO,
to remove a dominant point absorber. This resulted in an
additional 10\% increase in power-recycling gain.

\subsubsection{Beyond \ac{O4}}

The O4 commissioning of the Virgo detector highlighted the challenges of operating marginally stable recycling cavities in a dual-recycled interferometer. Since these issues are expected to worsen as the power stored in the interferometer increases, achieving the planned 80~W of input power for AdV+ Phase II during O5 is no longer considered realistic. Consequently, the Virgo Collaboration has decided to implement stable recycling cavities after O4, prompting a comprehensive review of the AdV+ Phase II Project.

The Virgo upgrade for O5~\citep{2026arXiv260320342A} will incorporate components from the original AdV+ Phase II plan, elements required for the installation of stable recycling cavities, and additional items identified during the O4 commissioning phase. While the optical configuration of the 3 km arm cavities will remain unchanged, the layout of the central interferometer will undergo significant modifications. The existing vacuum chambers (housing the power and signal recycling mirrors, as well as the suspended injection and detection benches) along with their seismic isolation systems, will be removed through two openings cut into the roof of the central building. Eight new vacuum chambers, equipped with suspension systems based on the AdV Multi-SAS/ET pathfinder design~\citep{2019CQGra..36g5007V} for the recycling mirrors, will be installed along with their connecting links.

All four test masses will be replaced with new ones offering improved optical and mechanical performance. Additional upgrades will include a high-power laser source, an enhanced thermal compensation system, instrumented baffles for stray-light monitoring~\citep{2025PhRvD.111d2001A}, a new timing distribution system, and updated electronics for seismic isolator control. Further improvements will target the frequency-dependent squeezing system, the auxiliary green laser system used during lock acquisition, and the calibration system, which will be upgraded to reduce systematic errors.

These enhancements will be implemented in two phases. After the first phase, the expected sensitivity range is {{\qtyrange{90}{130}{\Mpc}}}, ultimately reaching {{\qtyrange{120}{160}{\Mpc}}} upon completion of the upgrade project. 

\subsection{KAGRA Observatory}

The \ac{KAGRA} interferometer, situated in Japan’s Kamioka mine, is the only large-scale \ac{GWH} detector in East Asia.
It is designed as a cryogenic, \qty{3}{\km}, dual-recycled Fabry--Perot Michelson interferometer.
The \ac{KAGRA} project was funded in 2010, construction began in 2012 and tunnel excavation was completed in 2014 \citep{2021PTEP.2021eA101A}.
Following installation and assembly in the tunnel, two operations using temporary detector configurations served as key project milestones: the \ac{iKAGRA} operation in 2016~April \citep{2018PTEP.2018a3F01A}, and the \ac{bKAGRA} phase-1 operation in 2018~April.
During the \ac{bKAGRA} phase-1 operation, both cryogenic technology and the full-scale vibration isolation systems of \ac{KAGRA} were successfully demonstrated \citep{2019CQGra..36p5008A}.
By the summer of 2019, the primary installation of instruments was completed, allowing for the commissioning of the detector to begin immediately.
In 2019~October, a memorandum of agreement forming the \ac{LVK} was signed, and the \ac{LVK} international observation network was launched \citep{LVKMOA:2019}.
After that, the commissioning phase continued until 2020~March, marking the commencement of the detector’s scientific operation.

\subsubsection{\ac{O3GK}}

\ac{O3GK} was a joint observation conducted with the \ac{GEO} detector in 2020~April \citep{2023PTEP.2023jA101A}, just after the early termination of \ac{O3b}.
The \ac{O3GK} operation marked the first joint observation between \ac{KAGRA} and \ac{GEO}.
This collaboration aimed to improve the detection capabilities by combining data from both detectors. 
The optical configuration used during \ac{O3GK} was a power-recycled Fabry--Perot Michelson interferometer, with one room-temperature sapphire test mass and the others set around  \OthreeGkTemperatureTestMass{}.

During the \ac{O3GK} operation, \ac{KAGRA} observed for approximately \sifmtonedec{\OthreeGKDurationKAny}, with a strain sensitivity of \OthreeGkKagraStrainSensitivity{} at \OthreeGkKagraReferenceFrequency{}.
The \ac{BNS} inspiral range was about \sifmtonedec{\OthreeGKKAGRARange} \citep{2022PTEP.2022f3F01A}.
The sensitivity of \ac{KAGRA} during \ac{O3GK} was influenced by various noise sources, including sensor noise from local controls of the vibration isolation systems, acoustic noise, shot noise, and laser frequency noise \citep{2023PTEP.2023jA101A}.
Understanding these noise contributions was crucial for planning future improvements to the detector's sensitivity.
To enhance its performance, \ac{KAGRA} plans to implement hardware upgrades and refine its noise mitigation strategies. These improvements aim to extend the detection range and increase the precision of \ac{GW} observations.

\subsubsection{\ac{O4}}

After the O3GK run, KAGRA implemented extensive hardware upgrades and commissioning to enhance sensitivity and operational stability for the O4 run. The auxiliary laser system was improved by introducing \acl{PNC}, enabling robust lock acquisition under significant seismic conditions. Type-A suspensions underwent comprehensive refurbishment, including installation of low-noise \acl{LVDTs}, temperature-stabilized \acl{GAS} filters, and new accelerometers, reducing low-frequency vibration. Cryogenic payloads were stabilized through a multi-step cooling procedure, stricter vacuum leak limits, heaters on intermediate masses and radiation shields, and molecular monitoring to prevent frosting and maintain cavity finesse.
\Acl{ASC} was extensively deployed, incorporating \acl{WFS}, \acl{BPC}, and \acl{ADS}, which improved intra-cavity power stability and reduced contrast fluctuations, facilitating higher laser power injection. Additional measures included installation of baffles and optical dumps to mitigate stray light, repairs to the \acl{OMC}, and gate valves for efficient vacuum maintenance.
After those improvements, KAGRA participated in the initial O4a run for four weeks, achieving a median binary neutron star range of approximately 1.3 Mpc, up from 0.7 Mpc in O3GK, and increasing the duty cycle from 52\% to about 80\%. KAGRA then returned to the commissioning work for the further sensitivity improvement of the interferometer.

On \OfourEarthquakeNotoPeninsulaDate{} a \OfourNotoPeninsulaEarthquakeMagnitude{}~mag earthquake struck near the KAGRA site, marking the most significant seismic event in the area in the past century~\citep{2024GeoRL..5110993Y,2025EP&S...77...19Y}.
As a result, \OfourKagraNumberOfSeismicNoiseIsolatorRestored{} seismic noise isolators sustained damage but have since been restored.   
While further investigation and improvements were still needed for some vacuum and facility-related components, partial commissioning began in 2024~July.
By 2024~October, all earthquake-related repairs were completed, followed by noise reduction efforts across multiple domains.    
During the October commissioning, \ac{KAGRA} achieved a significant improvement on the \ac{BNS} range using a power-recycled Fabry--Perot Michelson interferometer configuration with DC readout.  
Further commissioning tasks have been performed, including reduction of suspension local control noise through updates to the control filters, reduction of photodiode dark noise below the shot noise level by mitigating electrical coupling from other electronic devices, reduction of quantum shot noise by increasing the laser power to above \OfourKagraIncreasedLaserPower{}, reduction of thermal noise by cooling the mirrors and their suspensions to below  \OfourKagraCooledMirrodTemperature{}, and reduction of frequency noise and acoustic noise through hardware improvements and control system updates.
Following these improvements, \ac{KAGRA} began operating in \ac{O4c} on \OfourcCommissioningBreakEndDate{}.

\subsection{\acs{GEO} Observatory}

The \ac{GEO} detector is a Michelson interferometer with two nearly orthogonal \qty{600}{\meter} arms \citep{2002CQGra..19.1377W}.
Rather than Fabry--Perot cavities, \ac{GEO} uses folding in the arms, in which the light traverses each arm twice, to give an optical length of \qty{1200}{\meter} for each arm.
\Ac{GEO} is sensitive to \acp{GW} in the \qty{100}{\hertz} to \qty{10}{\kilo\hertz} frequency range.
\ac{GEO} began operation in 2001.
From 2009 to 2014, it underwent a series of upgrades, the GEO-HF program, that resulted in a factor of 4 improvement in sensitivity at high frequencies \citep{2010CQGra..27h4003G,2016CQGra..33g5009D}.
In 2010, squeezed vacuum injection was first applied in \ac{GEO} \citep{2011NatPh...7..962L}, and the first long-term application of squeezing was demonstrated in \ac{GEO} in 2011 \citep{2013PhRvL.110r1101G}.
Subsequently, \qty{6}{\deci\bel} of squeezing (equivalent to a factor of 4 increase in light power) has been achieved \citep{2021PhRvL.126d1102L}.

\Ac{GEO} has served as an advanced development center and test bed for technologies that were subsequently incorporated in larger detectors \citep{2014CQGra..31v4002A}, such as dual-recycling \citep{2002CQGra..19.1547H}, monolithic suspension \citep{10.15488/6350}, thermal compensation \citep{2004CQGra..21S.985L}, homodyne detection \citep[DC readout;][]{2009CQGra..26e5012H}, and squeezed-light injection \citep{2011NatPh...7..962L}.

\subsubsection{Astrowatch}

Following the first-generation \ac{LIGO} and \ac{Virgo} science runs, \ac{GEO} embarked on an astrowatch program of near-continual data collection (when the detector is not being used for instrument-science research) as the sole observing detector \citep{2015JPhCS.610a2015D}.
This mode of operation has continued since 2007 and allows for searches for \acp{GW} associated with external events such as gamma-ray bursts, neutrino detections, or nearby supernovae, occurring outside of other detectors' observing periods \citep[e.g.,][]{2024ApJ...977..255A}.

\subsubsection{\ac{O3GK}}

As described in Section~\ref{ssec:O3}, a two-week-long joint observing run with the \ac{GEO} and \ac{KAGRA} detectors took place in 2020~April, during which \ac{GEO} operated with an \sifmttwofig{\OthreeGKDurationGAnyFraction} duty cycle (\sifmtonedec{\OthreeGKDurationGAny} of operation) and a \ac{BNS} range of \sifmtonedec{\OthreeGKGEORange} \citep{2022PTEP.2022f3F01A}.
The laser power injected into the power recycling cavity was about \qty{3}{\watt}, which led to about \qty{3}{\kilo\watt} of circulating power in the power-recycling cavity, or \qty{1.5}{\kilo\watt} circulating power per arm \citep{2014CQGra..31v4002A,2016CQGra..33g5009D}.
Bilinear noise subtraction resulted in modest improvement in sensitivity and data quality \citep{2020PhRvD.101j2006M}.
Since \ac{GEO} and \ac{KAGRA} had similar sensitivity during \ac{O3GK}, this joint run enabled searches for \ac{GW} transient signals occurring simultaneously in both detectors, though no significant events were observed \citep{2022PTEP.2022f3F01A}.

\section{Review of Observed Transient Sources}\label{sec:sources}

The \gwtc{} includes all candidates reported by \ac{LVK} searches targeting previously observed classes of \ac{GW} signals.

It is most likely that the significant candidates in \thisgwtc{} have an astrophysical origin and were produced by \ac{CBC} sources (the remaining less significant candidates are largely nonastrophysical).
This section provides a foundational overview of transient \ac{GWH} signals, especially those from \acp{CBC}, for use in interpreting the catalog's contents and for reference in companion papers.
We first provide a short overview of the basic physics of \acp{GW} and then provide an introduction to the \ac{CBC} sources to be used as a reference for other papers in the collection of articles.
Additional detail can also be found in \citet{2007gwte.book.....M,Maggiore:2018sht} and \citet{2011gwpa.book.....C}.

\subsection{Gravitational Waves}\label{ss:gw}

In metric theories of gravity, such as \ac{GR}, the local gravitational field can be described in terms of six independent degrees of freedom that represent the relative accelerations of a collection of nearby freely falling observers \citep{2009GReGr..41.1215P,1973grav.book.....M}.
Plane wave solutions to the linearized gravitational field equations \citep{1916SPAW.......688E} represent the weak \acp{GW} in the far-field region (where the observer is far from the source and the gravitational field is treated as a perturbation to Minkowski spacetime) that are observed by \ac{GW} detectors.
The vacuum Einstein field equations of \ac{GR} then further restrict the degrees of freedom of the plane wave solutions to two transverse polarizations that propagate at the speed of light \citep{1922RSPSA.102..268E}.
These are called the \emph{plus} ($+$) polarization and the \emph{cross} ($\times$) polarization.
In a suitably chosen set of coordinates, known as the \emph{transverse-traceless gauge} \citep{1973grav.book.....M,1987thyg.book..330T}, which is akin to the radiation gauge in classical electromagnetism, the perturbation to the Minkowski metric for these two polarizations is given by the two functions of spacetime $h_+$ and $h_\times$, respectively.
These polarizations represent two spin-2 purely transverse tensor modes \citep{1972gcpa.book.....W}.
The transverse-traceless gauge is a useful choice because world lines that are the histories of fixed points in these spatial coordinates are geodesics of the perturbed spacetime \citep{2021gaie.book.....H}.
Thus, changes in time in the metrical distance between fixed spatial coordinate locations, which is described by the time derivatives of $h_+$ and $h_\times$, represent the deviation of the geodesics at these locations.
Therefore, $h_+$ and $h_\times$ are the physical (observable) degrees of freedom of a \ac{GW}.

From an observational point of view, \ac{GW} signals are broadly classified as \emph{persistent} or \emph{transient}.
The main classes of persistent \acp{GW} include quasi-monochromatic signals, e.g., as produced by rotating \acp{NS} having a nonaxisymmetric mass distribution \citep{1979PhRvD..20..351Z}, and continuous stochastic superpositions of \acp{GW} from numerous unresolved independent sources \citep{2017LRR....20....2R}.
Here we focus on the transient signals that are cataloged in \ac{GWTC}.

\subsubsection{Transient \acs{GW} Signals}\label{sss:transient_gw_signals}

A \emph{transient} \ac{GW} is one that registers a signal of short duration (much less than the duration of the observing run) within the sensitivity band of the \ac{GW} detectors.
Such \acp{GW} can be characterized by their \emph{geocentric arrival time} \tgeo, the time at which some fiducial point in the \ac{GW}'s waveform (e.g., its peak amplitude) passes through Earth's center.
We expect that transient \acp{GW} will be observed as plane waves originating from a particular point in the sky, usually given in terms of the equatorial celestial coordinate system of R.A.\@ $\alpha$ and decl.\@ $\delta$, with a normal vector $-\boldsymbol{N}$ along this line of sight.

\begin{figure}
\begin{center}
\includegraphics[scale=0.5]{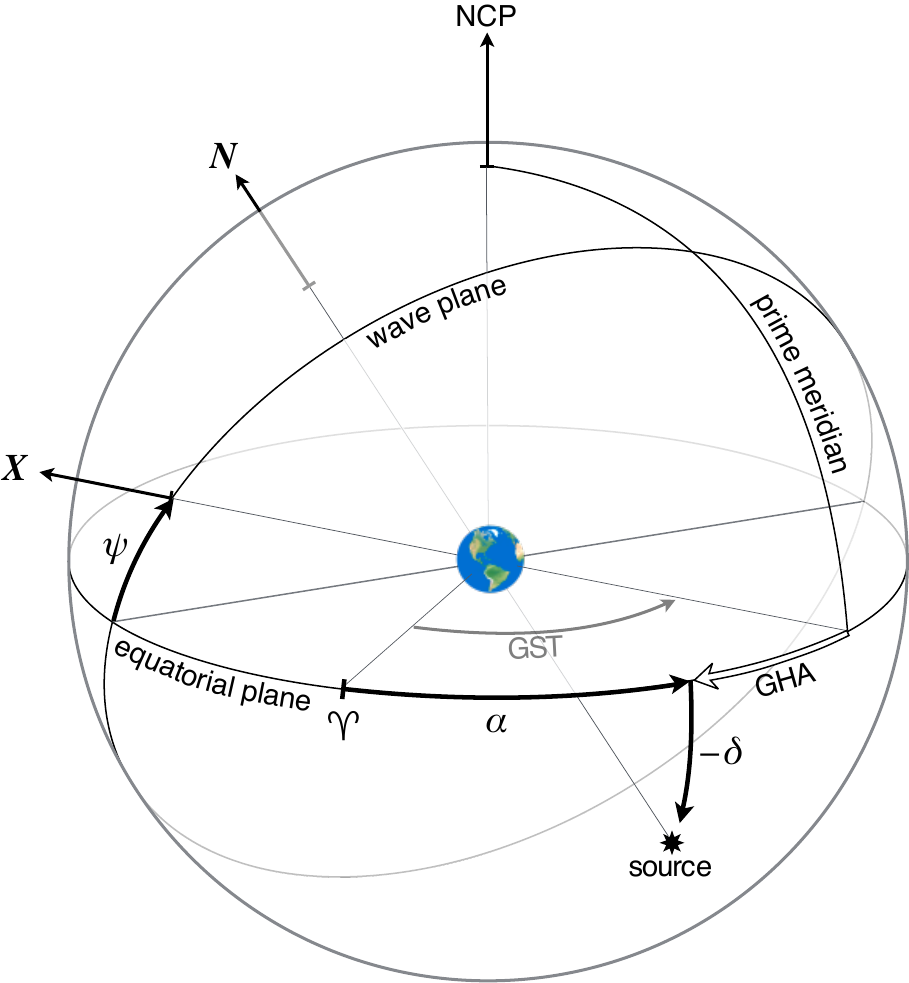}
\end{center}
\caption{\label{fig:sky}%
Relationship between the sky location in equatorial coordinates, the polarization angle, and the \ac{GWH} coordinate frame.
The direction from the source to Earth is $\boldsymbol{N}$ and the vector $\boldsymbol{X}$ defines a reference direction on the transverse plane called the wave plane.
The location of the source on the sky in the equatorial coordinate system is given by its R.A.\@ $\alpha$ and decl.\@ $\delta$.
The polarization angle $\psi$ is the angle counterclockwise about $\boldsymbol{N}$ between the equatorial plane and $\boldsymbol{X}$.
Also shown is the Greenwich sidereal time (GST), the angle between the first point of Aries \aries{} and the prime meridian, and the Greenwich hour angle (GHA) of the source, $\text{GHA}=\text{GST}-\alpha$.
$\text{NCP}$ is the north celestial pole.}
\end{figure}

\Ac{GW} detectors such as the \ac{LIGO}, Virgo, and \ac{KAGRA} detectors are designed to sense changes in the difference of the lengths of their orthogonal arms, $\Delta L = \Delta(L_1 - L_2)$, caused by \acp{GW}, via laser interferometry.
These \Lshape-shaped Michelson interferometers measure the difference in phase of coherent light, split at a beam-splitter located at the vertex of the \Lshape, after traversing the arms and recombining at the beam-splitter $\Delta\phi=2\pi\Delta L/\lambda_{\laser}$, where $\lambda_{\laser}$ is the wavelength of the laser light \citep{1971ApOpt..10.2495M,2022GReGr..54..153W,1978PhRvD..17..379F}.
For \ac{GW} transients having durations much less than a day and wavelengths much greater than the length $L$ of the detector arms, the strain induced on the arms is a linear combination of plus- and cross-polarizations of the metric perturbation \citep{1978PhRvD..17..379F,Rudenko_1980,Schutz_1987,1987thyg.book..330T}
\begin{equation}\label{e:strain}
    h = \frac{\Delta L}{L} = F_+ h_+ + F_\times h_\times.
\end{equation}
Here, $F_+$ and $F_\times$ are the detector's beam pattern functions, which depend on the position on the sky at which the \ac{GW} source is located, a polarization angle that defines the axes of the plus- and cross-polarization in the wave frame, the Earth rotation angle at the time of the signal's arrival, and the location, orientation, and geometry of the detector on Earth's surface \citep{Anderson:2001}.
Figure~\ref{fig:sky} shows the sky coordinate conventions used.
For long-duration signals, effects of Earth's rotation need to be included; for short-wavelength signals, the beam pattern functions also depend on the wavelength of the \acp{GW} \citep{2009CQGra..26o5010R,2008CQGra..25r4017R}, an effect that makes time-domain analyses more difficult \citep{2025PhRvD.111f4058V}.
Neither of these effects is significant in any of the transient signals detected to date.

Table~\ref{tab:gw_params} summarizes the parameters associated with a general transient plane \ac{GW}, a detector's response to such a \ac{GW}, and the accuracy of localization of the wave's source.

\begin{deluxetable*}{lcl}[t!]
\tablecaption{\label{tab:gw_params}%
Parameters Describing a Transient Plane \ac{GW},
a Detector's Instantaneous Antenna Response in the Long-wavelength Limit,
and Measures of Inferred Localization of the Signal on the Sky.}
\tablehead{\colhead{Parameter Name} & \colhead{Symbol} & \colhead{Notes [Dimensions]}}
\startdata
Plus- and cross-polarizations
    & $h_+$, $h_\times$
    & \parbox[t]{0.5\linewidth}{\raggedright Functions describing the plus-polarization ($h_+$) and cross-polarization ($h_\times$) of the metric perturbation [dimensionless]}
    \\
Geocentric arrival time
    & \tgeo
    & \parbox[t]{0.5\linewidth}{\raggedright Time of arrival at the center of Earth of some fiducial point in the \ac{GW}'s waveform, normally close to the peak amplitude of the waveform [time]}
    \\
Propagation direction
    & $\boldsymbol{N}$
    & \parbox[t]{0.5\linewidth}{\raggedright Direction of propagation of the \ac{GW}, the unit vector normal to the planar wavefronts; the direction to the source of the wave is $-{\boldsymbol{N}}$ [dimensionless]}
    \\
Right ascension (R.A.)
    & $\alpha$
    & \parbox[t]{0.5\linewidth}{\raggedright Azimuth of the sky location of the source of the \ac{GW} in the equatorial coordinate system (see Figure~\ref{fig:sky}) [angle]}
    \\
Declination (decl.)
    & $\delta$
    & \parbox[t]{0.5\linewidth}{\raggedright Latitude of the sky location of the source of the \ac{GW} in the equatorial coordinate system (see Figure~\ref{fig:sky}) [angle]}
    \\
Polarization angle
    & $\psi$
    & \parbox[t]{0.5\linewidth}{\raggedright Orientation of the axes defining the plus- and cross-polarization on the transverse plane of the \ac{GW} relative to the line-of-nodes of this plane and the Earth's equatorial plane (see Figure~\ref{fig:sky}) [angle]}
    \\
\hline
Plus and cross beam patterns
    & $F_+$, $F_\times$
    & \parbox[t]{0.5\linewidth}{\raggedright Antenna response of a detector to the plus-polarization ($F_+$) and the cross-polarization ($F_\times$), functions of the sky location of the source, the polarization angle, the geocentric arrival time of the signal, and the location, orientation, and geometry of the detector on Earth \citep{Anderson:2001} [dimensionless]}
    \\
Detector strain
    & $h$
    & \parbox[t]{0.5\linewidth}{\raggedright \ac{GWH}-induced strain on a detector, Equation~\eqref{e:strain}; the \ac{GW} readout of the detector is proportional to this quantity [length$/$length]}
    \\
\hline
Sky area
    & $\Delta\Omega$
    & \parbox[t]{0.5\linewidth}{\raggedright Localization area, typically taken as the 90\% credible area; if results at different \acsp*{CL} are quoted, these are indicated with a subscript, e.g., $\Delta\Omega_{50}$ is the 50\% credible area [solid angle]}
    \\
Volume localization
    & $\Delta V$
    & \parbox[t]{0.5\linewidth}{\raggedright Localization volume (for signals where the distance to the source can be estimated), typically taken as the 90\% credible volume; if results at different \acsp*{CL} are quoted, these are indicated with a subscript, e.g., $\Delta V_{50}$ is the 50\% credible volume [volume]}
    \\
\enddata
\end{deluxetable*}

\subsubsection{Sky Localization of \acsp{GW}}

A key task for multimessenger astronomy with \acp{GW} is the reconstruction of the source location, which facilitates follow-up with other astronomical facilities \citep{KAGRA:2013rdx}.

A network of detectors spaced at different locations on Earth can observe the difference in the time of arrival of the fiducial point in the waveform arising from the propagation of the plane wave across Earth and thereby reconstruct the direction of propagation $\boldsymbol{N}$ \citep{2009NJPh...11l3006F,2011CQGra..28j5021F,2011gwpa.book.....C,2016PhRvD..93b4013S}.
Such triangulation is the main way in which the sources of transient \acp{GW} are localized (but not the only one, see antenna beam patterns and strain amplitude information following Equation~\ref{e:strain}).
Hence, uncertainty in the sky location of the source, $\Delta\Omega$, partially results from the measurement uncertainty of the arrival time in each detector \citep{2011CQGra..28j5021F}.
A single detector provides no ability to determine the sky location of a source for transient signals lasting much less than a day and having wavelengths much longer than the size of the detector, as is the case for all candidates reported in \thisgwtc{}. 
However, with two detectors, the difference in times of arrival identifies a circle on the celestial sphere, centered on the axis separating the detectors, on which the wave's origin may lie.
The presence of a third detector whose location is not collinear with the other two leads to a bimodal identification of the source position.
A fourth detector, not coplanar with the other three, finally resolves the location of the source to a single area on the sky.
Additional localization information can be provided by coherently combining the observed \ac{GW} signals from an array of detectors as described below.
For some types of transient sources having known \ac{GW} emission, such as \acp{CBC}, it is also possible to estimate the distance to the source from measurements of the wave amplitude \citep{1994PhRvD..49.2658C}.
In such cases, there is a volume localization uncertainty $\Delta V$ as well \citep{2016ApJ...829L..15S,2018MNRAS.479..601D}.

The amplitudes of the strains measured in a network of detectors provide additional information about the location of the source of the \ac{GW} if the polarization of the \ac{GW} is known owing to the dependence of the beam pattern functions on the position of the source on the sky \citep[e.g.,][]{2016PhRvD..93b4013S}. The analysis of GW240925\_005809~\citep{gzrj-mwv3} includes a discussion of this effect for a real signal.
This information helps to break degeneracies in sky localization; for instance, with only two detectors, the source is typically localized to an extended arc or ring on the sky, but the amplitude response can help reduce this uncertainty to specific regions along that ring.

\paragraph{\ac{LVK} Catalog of Observed Transient \acs{GW} Signals}
In the companion paper \citet{GWTC:Results}, we describe the significant transient \ac{GW} candidates in \thisgwtc{}, highlighting those observed in \ac{O4b}.
\thisgwtc{} also provides the inferred properties of the \acp{GW}, as well as their sources, e.g., the masses and spins of the binary components under the assumption that the \acp{GW} were produced by \acp{CBC}. The exceptional \ac{O4b} event papers on GW241011\_233834 and GW241110\_124123 \citep{2025ApJ...993L..21A}, GW250114\_082203 \citep{2025PhRvL.135k1403A,2026PhRvL.136d1403A}, and GW240925\_005809 and GW250207\_115645 \citep{gzrj-mwv3}.
The \ac{GWTC} dataset, along with other open data products, is detailed in the companion paper \citet{OpenData}.

\subsubsection{Gravitational Lensing of \acsp{GW}}

Like electromagnetic waves, \acp{GW} can be gravitationally lensed by massive objects, e.g., galaxies, interposed between the \ac{GW} source and the observer.
Because of the principle of equivalence, the \ac{GW} polarization state is not affected by the gravitational interaction with a lensing non-rotating mass \citep{1973grav.book.....M}, so it is sufficient to consider scalar diffraction theory \citep{2003ApJ...595.1039T}. For rotating masses and in general for gravitomagnetic fields this requires corrections~\citep{2006PhRvD..73h4003R}.
In a thin-lens approximation, the bending of the trajectory of the \ac{GW} propagation occurs on a lens plane orthogonal to the line of sight and at the distance of the lensing body.
With $\xi_1$ and $\xi_2$ as the coordinates of the lens plane, at each point on this plane there is an observed time delay $T(\xi_1,\xi_2)$ relative to straight-line motion with no lens, corresponding to the path from the source to that point on the lens plane to the observer.
This delay accounts for the gravitational field of the lens.
\acp{GW} are deflected by a gravitational lens with the time delay field on the lens plane determining the complex phases of the interfering partial waves used to compute a frequency-dependent complex-valued magnification factor.
This factor is given by the Fresnel--Kirchhoff diffraction formula
\begin{equation}\label{e:lens-fresnel-kirchhoff}
    F(f) = i \frac{D_{\scriptscriptstyle\text{OS}}}{D_{\scriptscriptstyle\text{OL}} D_{\scriptscriptstyle\text{LS}}} \frac{(1+z_{\scriptscriptstyle\text{L}})f}{c} 
    \iint \exp\left[-2\pi if T(\xi_1,\xi_2)\right]\,d\xi_1\,d\xi_2\:,
\end{equation}
where the integral is over the lens plane, $f$ is the observed \ac{GW} frequency, $(1+z_{\scriptscriptstyle\text{L}})f$ is the blueshifted frequency of the \ac{GW} on the lens plane ($z_{\scriptscriptstyle\text{L}}$ is the redshift of the lens), and the distances $D_{\scriptscriptstyle\text{OS}}$, $D_{\scriptscriptstyle\text{OL}}$, and $D_{\scriptscriptstyle\text{LS}}$ are the distances between the observer (us) and the \ac{GW} source, between the observer and the gravitational lensing object, and between the lensing object and the source, respectively \citep{1992grle.book.....S}.
In a cosmological setting, these are \emph{angular diameter} distances \citep{1999astro.ph..5116H}.
The geometric optics limit corresponds to Fermat's principle, in which the geodesic paths taken by \acp{GW} are those passing through the lens plane at extrema of this two-dimensional time delay field $T(\xi_1,\xi_2)$, which may be local minima, which produce \emph{Type~I images}, local maxima, which produce \emph{Type~III images}, or saddle points, which produce \emph{Type~II images} \citep{1992grle.book.....S}.
Equation~\eqref{e:lens-fresnel-kirchhoff} is evaluated in this high-frequency limit by use of the stationary phase approximation to obtain
\begin{equation}\label{e:lens-magnification}
    F_j(\pm|f|) = \sqrt{\mu_j} \exp\left(\mp 2\pi i|f| t_j \pm i\pi n_j\right)\:,
\end{equation}
where $\sqrt{\mu_j}$ and $t_j$ are the magnification amplitude and observed time delay of image $j$, and $n_j$ is 0, 1/2, or 1 for Type~I, Type~II, and Type~III images, respectively.
Therefore, such images are magnified or demagnified by a factor that is positive for Type~I images and negative for Type~III images, while the gravitational waveform of Type~II images is additionally distorted, appearing as the Hilbert transform of the original waveform \citep{2017arXiv170204724D,2021PhRvD.103f4047E}.
For \ac{GW} transients, the images are a set of repeated signals from the same event observed at different times, the delays determined by the differences in the time delay field on the lens plane of the different images.
These delays are typically minutes to months for galaxy lenses \citep{2018MNRAS.476.2220L,2018PhRvD..97b3012N,2018MNRAS.480.3842O} and up to years for galaxy cluster lenses \citep{2018IAUS..338...98S,2018MNRAS.475.3823S,2020MNRAS.495.3727R,2020MNRAS.495.1666R}.
The images also appear at different points on the sky, with arcminute-scale separation, but \ac{GW} detectors have insufficient sky-localization capabilities to distinguish them in this way.
When gravitational lensing can be described in this geometric optics limit, it is referred to as strong lensing.

However, when the wavelength of the \ac{GW} is comparable to the Schwarzschild radius of the gravitational lens, the geometric optics limit of Fermat's principle is no longer valid, and the Fresnel--Kirchhoff diffraction formula of Equation~\eqref{e:lens-fresnel-kirchhoff} must be used to determine the complex-valued and frequency-dependent magnification factor.
Such lensing effects can result from objects having masses up to \MinimumMassForMicrolensing{} and searches can be done in a modeled \citep[e.g.,][]{2022ApJ...935...68W} or phenomenological~\citep{2023MNRAS.525.4149L} way.

\paragraph{Searches for Gravitational-lensing Signatures in \acs{GW} Signals}
In the companion paper \citet{GWTC:Lensing}, we present searches for gravitational lensing signatures in \thisgwtc{}.
Such signatures sought include multiple images from strong lensing, individual Type~II strongly lensed images, and waveform distortions induced by point-mass lensing.

\subsubsection{\acs{GW} Polarization and Propagation in Alternative Theories of Gravity}

In \ac{GR}, plane \ac{GW} perturbations to flat spacetime propagate at the speed of light and contain two transverse polarizations.
However, in modified theories of gravity extending beyond \ac{GR}, additional polarizations may be present, including two transverse-longitudinal spin-1 vector modes, a transverse spin-0 scalar mode, and a longitudinal spin-0 scalar mode \citep{1973PhRvD...8.3308E,1973PhRvL..30..884E,2018tegp.book.....W}.
With multiple detectors, it is possible to test for such additional polarizations \citep{1986Natur.323..310S,Schutz_1987}.
A linear combination of strain data from three detectors,  often called the \textit{null stream}, can be formed in which any \ac{GW} signal from a known sky location and containing only plus- and cross-polarizations is canceled \citep{1989PhRvD..40.3884G,2008CQGra..25k4029K,2010NJPh...12e3034S,2011gwpa.book.....C,2021arXiv210509485W}.
Any residual \ac{GW} signal found in such a null space would provide evidence for the presence of vector or scalar non-\ac{GR} polarizations.

In addition, in alternative Lorentz invariance violating theories of gravity or in which the graviton is massive, \acp{GW} are dispersive.
Certain theories of dark energy also result in dispersive \ac{GW} propagation \citep{2018PhRvL.121v1101D,2022JCAP...08..031B,2022GReGr..54..133H}.
The \ac{GW} dispersion relation between the frequency $f$ and the wavelength $\lambda$ (one where they are not inversely proportional) leads to phase speeds and/or group speeds that differ from the speed of light.
Such propagation effects can be measured for a known waveform by the anomalous arrival times of different frequency components.
A common parameterized dispersion relationship is motivated by a modified energy--momentum relationship for the graviton of the form \citep{2012PhRvD..85b4041M}
\begin{equation}
    E^2 = (pc)^2 + A_\alpha(pc)^\alpha\:,
\end{equation}
where $A_\alpha$ is a \ac{GR}-violating parameter having dimensions of $\text{(energy)}^{2-\alpha}$.
For de\,Broglie waves, $E=2\pi\hbar f$ and $p=2\pi\hbar/\lambda$, where $2\pi\hbar$ is the Planck constant.
Such a modified energy--momentum relation leads to a dispersion relation in which the phase velocity $v_\mathrm{p}$ is given by
\begin{equation}
    \left(\frac{v_\mathrm{p}}{c}\right)^2 = 1 + A_\alpha \left(\frac{2\pi\hbar c}{\lambda}\right)^{\alpha - 2}\:,
\end{equation}
where the phase velocity is related to the frequency and the wavelength of the \ac{GW}, $v_\mathrm{p}=\lambda f$.
The group velocity, $v_\mathrm{g}=v_\mathrm{p}-dv_\mathrm{p}/d\ln\lambda$, determines the difference in arrival times of different frequency components of the \ac{GW} after propagation from its source to the observer.
For small deviations from \ac{GR} ($v_\mathrm{p}\approx c$), the group velocity is frequency dependent with
\begin{equation}
     \frac{v_\mathrm{g} - c}{c} \approx \frac{1}{2} (\alpha-1) A_\alpha (2\pi\hbar f)^{\alpha-2}\:.
\end{equation}
Special cases include (i) a graviton of mass $m_g\ne0$ for which $\alpha=0$, 
$A_0=m_g^2 c^4$, and
\begin{equation}
     \frac{v_\mathrm{g} - c}{c} \approx -\frac{1}{2}\left(\frac{\lambda_g f}{c}\right)^{-2}\:,
\end{equation}
where $\lambda_g=2\pi\hbar/(m_gc)$ is the Compton wavelength of the graviton, and (ii) the case in which \acp{GW} are nondispersive but propagate at a speed different from the speed of light for which $\alpha=2$ and
\begin{equation}
    v_\mathrm{g} = c\sqrt{1+A_2}\:.
\end{equation}
Stringent bounds on the latter are provided by the close temporal association of the \ac{BNS} signal GW170817 and the gamma-ray burst GRB\,170817A (the gamma rays arriving less than \qty{2}{\second} after the \ac{BNS} \ac{GW} merger signal), resulting in $|A_2|\lesssim 10^{-14}$ \citep{2017ApJ...848L..13A}.

The Einstein--Hilbert action of \ac{GR} contains second derivatives of the spacetime metric \citep{1972gcpa.book.....W,1973grav.book.....M,1984ucp..book.....W,2019sgai.book.....C}.
Standard Model extensions having modified actions containing third derivatives of the metric can produce CPT-violating terms\footnote{Here, the CPT acronym indicates the symmetry of the Standard Model Lagrangian with respect to charge conjugation, spatial parity and time inversion. CPT-violating terms would violate this symmetry.} in the gravitational field equations, which can produce birefringence effects in which different \ac{GW} helicities propagate with different phase velocities \citep{2004PhRvD..69j5009K,2016PhLB..757..510K,2019PhRvD..99j4062M,2023PhRvD.107f4031H}.
Other theories of gravitation also have \acp{GW} with birefringent propagation \citep{2024PhRvD.110f4044Z}.
Such birefringence leads to a frequency-dependent rotation of the \ac{GW} polarization angle.

Both \ac{GW} birefringence and the modified \ac{GW} dispersion relation can potentially be anisotropic, where the magnitude of the observed effect depends on the direction to the source.

\paragraph{Tests of \acs{GR}: \acs{GW} Polarization and Propagation}
In the companion paper \citet{GWTC:TGR}, we test the \ac{GR} prediction of the polarizations of \acp{GW} by searching for evidence of vector- or scalar-polarization modes in observed \ac{GW} signals. We have also presented in the past tests of a modified dispersion relation
using \ac{GW} signals from \acp{CBC} and of anisotropic birefringence \citep{GWTC4:TGR-II}, for which it is assumed that the \ac{GW} near the source is described by \ac{GR} to a good approximation, but the waveform is affected during propagation.

\subsection{Compact Binary Coalescences}\label{s:cbc}

Binaries consisting of two \acp{BH} (\ac{BBH} systems), consisting of two \acp{NS} (\ac{BNS} systems), or in which one component is an \ac{NS} and the other a \ac{BH} (\ac{NSBH} systems), have all been observed by the \ac{LVK} \citep{2016PhRvL.116f1102A,2017PhRvL.119p1101A,2021ApJ...915L...5A}.
The detectable signal produced by such systems arises from the late stage of orbital decay, driven by \ac{GW} emission, and by the ensuing merger of the binary components and the settling of the resulting object (an \ac{NS} or \ac{BH}) to a final, stationary configuration \citep{2024arXiv240902037C}.

Table~\ref{tab:cbc_params} provides a list of parameters used to describe \acp{CBC}.

\subsubsection{Newtonian Inspiral}\label{ss:newtonian_inspiral}

\begin{figure}
\begin{center}
\includegraphics[scale=0.5]{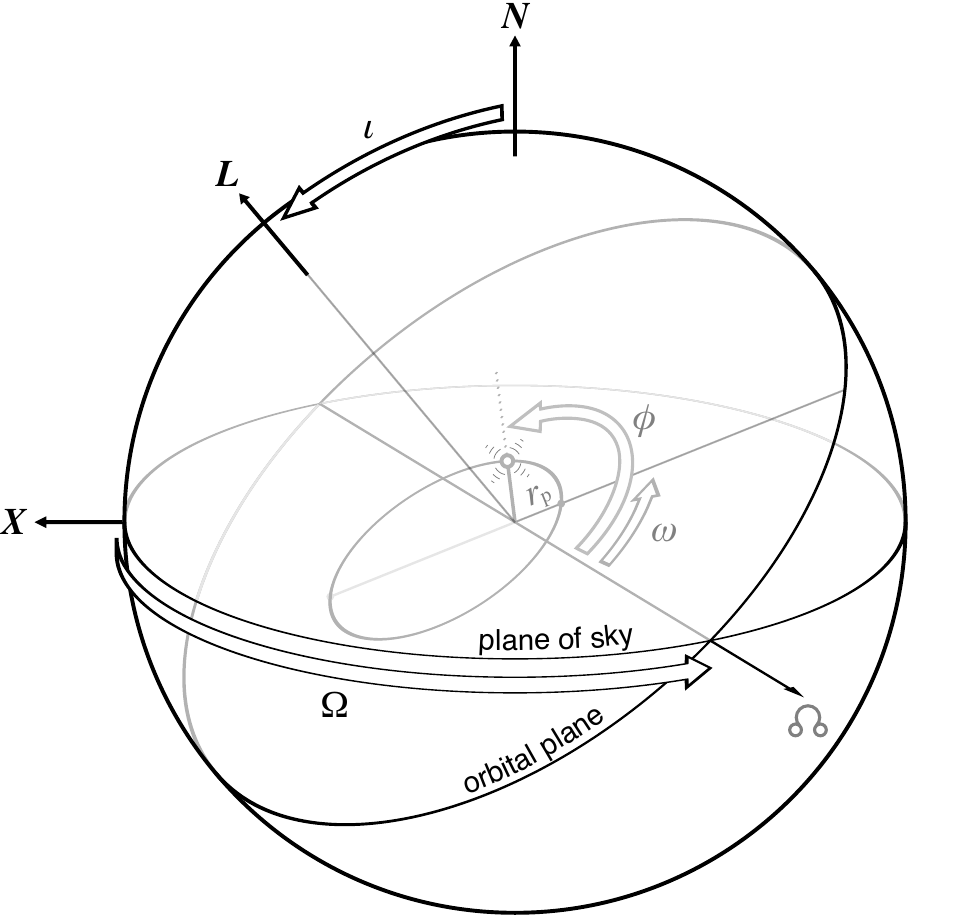}
\end{center}
\caption{\label{fig:src}%
Relationship between the orbital elements and the \ac{GWH} coordinate frame.
The direction from the source to the Earth is $\boldsymbol{N}$ and the vector $\boldsymbol{X}$ defines a reference direction on the transverse plane (the plane of the sky).
The inclination $\iota$ is the angle between $\boldsymbol{N}$ and the orbital angular momentum vector $\boldsymbol{L}$.
The longitude of the ascending node of the orbit $\Omega$ is the angle on the plane of the sky between $\boldsymbol{X}$ and the ascending node \ascnode{}, $\boldsymbol{N}\times\boldsymbol{L}$. The angle $\Omega$ is degenerate with the polarization angle $\psi$. 
The orbit of the primary about the center of mass of the system is shown.
The orbital phase $\phi$ is the angle on the orbital plane between the ascending node and position vector of the primary relative to the center of mass.
For an eccentric orbit, the distance of the primary from the center of mass at periapsis is $r_{\mathrm{p}}$ and the argument of the periapsis $\omega$ for the primary is the angle on the orbital plane between the ascending node and the position vector of the primary at periapsis.}
\end{figure}

At early stages of the inspiral, when the magnitude of the difference in velocity vectors of the two components of the binary, $v$, is much smaller than the speed of light, the orbit is determined approximately by Newtonian mechanics while the gravitational radiation is described by the quadrupole formula (\citealp{1916SPAW.......688E}, corrected by \citealp[page 279]{1922RSPSA.102..268E}).
For a quasi-circular orbit that is inclined an angle $\iota$ relative to the direction to an observer, $h_+$ and $h_\times$ are sinusoidal and are \ang{90} out of phase,
\begin{subequations}\label{e:hpc}
\begin{align}
    h_{+} &= -2(1+\cos^2\iota)\frac{GM\eta}{c^2r}\left(\frac{v}{c}\right)^2\cos2\phi \\
\intertext{and}
    h_{\times} &= -4\cos\iota\frac{GM\eta}{c^2r}\left(\frac{v}{c}\right)^2\sin2\phi\:,
\end{align}
\end{subequations}
where $r$ is the distance between the source and the observer, $\Mtot=\massone+\masstwo$ is the total mass of the system, $\mratiosym=\massone\masstwo/\Mtot^2$ is the symmetric mass ratio, $\Mtot\mratiosym$ is the reduced mass of the system, and $\phi$ is the orbital phase relative to the ascending node \citep{1963PhRv..131..435P,1987thyg.book..330T,1993PhRvD..47.2198F,1996PhRvD..54.4813W}.
The inclination angle $\iota$ as well as other orbital elements and their relation with the GW are illustrated in  Figure~\ref{fig:src}.
When $\iota=0$ or $\iota=\pi$ (face on and face off respectively), the amplitudes of the sinusoidal functions $h_+$ and $h_\times$ are equal and the \ac{GW} is circularly polarized; when $\iota=\pi/2$ (edge on), $h_\times=0$, and the \ac{GW} is linearly polarized.

The \ac{GW} luminosity of such a system, i.e., the power in gravitational radiation, is
\begin{equation}\label{e:egwdot}
   \dot{E}_{\mathrm{GW}} = \frac{32}{5} \frac{c^5}{G} \mratiosym^2 \left(\frac{v}{c}\right)^{10}.
\end{equation}
This radiation gives rise to a secular orbital decay.
Since the (Newtonian) energy of the bound system is $E_{\mathrm{orb}}=-(1/2)\eta \Mtot v^2$, and equating $\dot{E}_{\mathrm{GW}}=-\dot{E}_{\mathrm{orb}}$, we deduce that the period of the orbit, $P=2\pi G\Mtot /v^3$ by Kepler's third law, evolves according to
\begin{equation}\label{e:pdot}
    \dot{P} = - \frac{192\pi}{5} \eta \left(\frac{v}{c}\right)^5.
\end{equation}
At fixed orbital period (or orbital frequency), the orbital velocity is proportional to the cube root of the total mass, $v\propto \Mtot^{1/3}$.
It can be seen, then, that $h_+,h_\times \propto \eta \Mtot^{5/3}$, $\dot{E}_{\text{GW}} \propto (\eta \Mtot^{5/3})^2$, $E_{\mathrm{orb}} \propto \eta \Mtot^{5/3}$ and $\dot{P} \propto \eta \Mtot^{5/3}$.
At the Newtonian level of approximation, a single combination of the component masses,
\begin{equation}\label{e:chirp_mass}
    \Mc =\eta^{3/5}\Mtot
        =\frac{(\massone \masstwo)^{3/5}}{(\massone + \masstwo)^{1/5}}\:,
\end{equation}
known as the \emph{chirp mass}, solely determines both the amplitude of a \ac{GW} at fixed orbital frequency and its frequency evolution \citep{1988ESASP.283..121K,1993PhRvL..70.2984C,1993PhRvD..47.2198F}.
This chirp mass is normally the most accurately measured mass parameter for low-mass systems in which most of the signal observed arises from the pre-merger phase.

\subsubsection{\Acl{PN} Inspiral and Other Effects}

Additional terms in the \ac{GW} amplitude and frequency evolution appear at higher orders in $v/c$ in a \acf{PN} expansion in the equations of motion and in the gravitational emission~\citep{2024LRR....27....4B}.
At the Newtonian order, the frequency of the \ac{GW} is twice the frequency of the orbital motion, $f=2f_{\mathrm{orb}}=2/P$, and
\begin{equation}\label{e:orb_velocity_cubed}
    v^3 = \pi G \Mtot f\:.
\end{equation}
At $O(v/c)$ beyond this, additional components to the \ac{GW} at frequencies at one and three times the orbital frequency arise from current quadrupole and mass octupole radiation, and other components occur at $O(v^2/c^2)$ beyond Newtonian order from current octupole and mass hexadecapole radiation \citep{1980RvMP...52..299T}; the amplitudes of these \emph{higher-order multipole moments} of radiation are proportional to a different combination of component masses \citep{2008PhRvD..77d4016K}.
The frequency evolution also gains additional terms at $O(v^2/c^2)$ beyond the leading-order Newtonian term, again having a different dependence on the component masses from the leading Newtonian order \citep{1976ApJ...210..764W}. 

Spin effects from rotating binary components also appear in \ac{PN} corrections to the quadrupole waveform due to $O(v^3/c^3)$ spin--orbit and $O(v^4/c^4)$ spin--spin effects \citep{1993PhRvD..47.4183K} and to precession of the orbital plane if the spin angular momentum vectors of the bodies are not aligned (or antialigned) with the orbital angular momentum vector \citep{1994PhRvD..49.6274A}. The binary precesses around the (approximately) constant direction of total angular momentum ${\boldsymbol{J}}$, which is the vector sum of the orbital angular momentum and the spin angular momenta of the two bodies (${\boldsymbol{J}} = {\boldsymbol{L}}_{\text{PN}} + {\boldsymbol{S}}_1 + {\boldsymbol{S}}_2$, with $\boldsymbol{L}_{\text{PN}}$ the Newtonian orbital angular momentum $\boldsymbol{L}$ with \ac{PN} corrections). 
The source inclination angle $\theta_{JN}$, the angle between $\boldsymbol{J}$ and the direction to Earth $\boldsymbol{N}$ (see Figure~\ref{fig:src}) is often reported instead of the orbital inclination angle $\iota$.
The dimensionless effective inspiral spin parameter $\chieff$, defined as
\begin{equation}\label{e:chi_eff}
    \chieff = \frac{c}{G\Mtot}
    \frac{\boldsymbol{L}\cdot({\boldsymbol{S}_1/\massone + \boldsymbol{S}_2/\masstwo})}{|\boldsymbol{L}|},
\end{equation}
where $\boldsymbol{S}_1$ and $\boldsymbol{S}_2$ are the spins of the two binary components and $\boldsymbol{L}$ is the orbital angular momentum about the center of mass, is a convenient spin parameter that is conserved under the orbit-averaged precession equations of motion at $O(v^4/c^4)$ \citep{2001PhRvD..64l4013D,2008PhRvD..78d4021R,2011PhRvL.106x1101A,2010PhRvD..82f4016S,2010PhRvD..81h4054K}.
Whereas $\chieff$ depends on the spin components aligned with the orbital angular momentum, a dimensionless effective precession spin parameter that depends on in-orbital-plane components of the spins,
\begin{equation}\label{e:chi_p}
    \chip = \frac{c}{G\massone} \max\left\{
        \frac{|\boldsymbol{L}\times\boldsymbol{S}_1/\massone|}{|\boldsymbol{L}|}
        ,
        \frac{3\massone+4\masstwo}{3\masstwo+4\massone}
        \frac{|\boldsymbol{L}\times\boldsymbol{S}_2/\masstwo|}{|\boldsymbol{L}|}
     \right\}\:,
\end{equation}
captures the dominant precession effects \citep{2015PhRvD..91b4043S}. 

Deformable binary components (\acp{NS} but not \acp{BH}) suffer an induced quadrupole deformation $Q_{ij}$ under an external tidal field ${\mathcal{E}}_{ij}$, where these quadrupole tensors are those appearing in a multipole expansion of the Newtonian potential centered on the body of mass $m$ as \citep{1998PhRvD..58l4031T}
\begin{equation}
    \Phi(\boldsymbol{x}) = -\frac{Gm}{|\boldsymbol{x}|}
    - \frac{1}{2}GQ_{ij}\frac{3x^ix^j-|\boldsymbol{x}|^2\delta^{ij}}{|\boldsymbol{x}|^5} + \frac{1}{2}{\mathcal{E}}_{ij} x^ix^j + \cdots\:.
\end{equation}
The \emph{dimensionless tidal deformability}, $\Lambda$, of a body of mass $m$ is defined in terms of the ratio of the induced deformation to the external tidal field as
\begin{equation}\label{e:tidal_deformability}
     \frac{GQ_{ij}}{(Gm/c^2)^5} = -\Lambda {\mathcal{E}}_{ij}\:,
\end{equation}
where \acp{BH} have $\Lambda=0$.
Newtonian tidal interactions of deformable components appear as effective $O(v^{10}/c^{10})$ corrections to the binding energy and \ac{GW} luminosity \citep{2008PhRvD..77b1502F}.
At this order, the dimensionless combination of tidal parameters given by \citep{2014PhRvL.112j1101F}
\begin{equation}\label{e:lambda-tilde}
   \tilde\Lambda = \frac{16}{13}
   \frac{(\massone + 12\masstwo)\massone^4\Lambda_1 + (\masstwo + 12\massone)\masstwo^4\Lambda_2}{(\massone + \masstwo)^5}
\end{equation}
appears, where $\Lambda_1$ and $\Lambda_2$ are the dimensionless tidal deformabilities of the two bodies, and it is this parameter that is most measurable in the waveforms produced by binaries with deformable companions \citep{2021PhRvD.103f4023P}.
Spinning bodies experience quadrupole deformation of the form
\begin{equation}
    Q_{ij} = \mathrm{diag}\left(-\frac13Q,-\frac13Q,\frac23Q\right)\:,
\end{equation}
where $Q$ is the spin-induced mass quadrupole moment scalar and the body spins about the third coordinate axis.
The quadrupole deformations induced by an object's spin result in Newtonian quadrupole--monopole effects as an effective $O(v^4/c^4)$ correction, the same order as the spin--spin coupling relativistic effects \citep{1998PhRvD..57.5287P}.
The size of the spin-induced deformation depends on the nature of the body, where the ratio of the quadrupole scalar to the square of the body's spin magnitude is given by the dimensionless parameter $\kappa$ as
\begin{equation}\label{e:spin_induced_quadrupole}
    Q = -\kappa \frac{|\boldsymbol{S}|^2}{mc^2}\:,
\end{equation}
where $m$ is the mass of the body.
For a \ac{BH}, $\kappa=1$ \citep{1980RvMP...52..299T}.

Binaries detected by ground-based observatories are commonly assumed to have negligibly small orbital eccentricity remaining by the time the orbital period has decayed to the point that the \ac{GW} frequencies have entered the high-frequency sensitivity band of the detectors.
This decay in eccentricity happens because orbital eccentricity is efficiently reduced by \ac{GW} emission during the orbital decay \citep{1964PhRv..136.1224P}.
However, there are channels of compact binary formation that could result in nonnegligible orbital eccentricity being present even at the last stages of inspiral observed by ground-based detectors \citep[e.g.,][]{2020FrASS...7...38M}. 
Moreover, the Kozai--Lidov resonance \citep{1910AN....183..345V,2016ARA&A..54..441N} due to the hierarchical interaction of a binary with a third body, possibly in a crowded stellar environment \citep{2016MNRAS.463.2443K,2022ApJ...939...48H}, can also play an important role by sustaining high levels of eccentricity and increasing the merger rate \citep{2016ApJ...828...77V}. 
The leading-order effects of orbital eccentricity would appear at the Newtonian level \citep{1963PhRv..131..435P}.
Two additional parameters are needed to describe an eccentric binary system, the eccentricity $e$ and the argument of the periapsis $\omega$.
Although these are well-defined for Newtonian two-body systems, there are different ways to generalize their definitions for relativistic systems, and there is not yet a settled convention for these parameters \citep{2023PhRvD.108j4007S}.

\paragraph{Tests of \acs{GR} from \acs{CBC} Inspiral}

\ac{PN} theory in \ac{GR} predicts the relative amplitudes of subdominant modes of \ac{GW} radiation \citep{2024LRR....27....4B}, which depend on the binary's masses and spins \citep[e.g.,][]{2009PhRvD..79j4023A}.
Thus, allowing for freedom in these amplitudes and checking whether they are consistent with those predicted by GR provides a consistency test of the agreement of the signal with the waveform model used to analyze it \citep{2022PhRvD.106h2003P}.
In a past companion paper \citep{GWTC4:TGR-I}, this test is carried out for \ac{BBH} signals, considering deviations, $\delta A_{\ell m}$, in the amplitude of the $(\ell=2, m=\pm1)$ or $(\ell=3, m=\pm3)$ subdominant multipole moments relative to the dominant $(\ell=2, m=\pm2)$ and other multipole moments.

The \ac{PN} expansion of the orbital energy and \ac{GW} energy loss makes a prediction of how the \ac{GW} phase evolves with time as the orbit decays \citep{2024LRR....27....4B}.
The \ac{PN} formalism expresses this phase evolution with a set of coefficients in a series expansion of the \ac{GW} phase in terms of powers $(v/c)^{n-5}$ and $(v/c)^{n-5}\log(v/c)$ for integer $n$ (with $n=0$ for the leading-order Newtonian inspiral) that depend on the binary components' masses and spins for point particles.
Violations of \ac{GR} can lead to differences in the values of the \ac{PN} coefficients from those predicted by \ac{GR} \citep[e.g.,][]{2009PhRvD..80l2003Y,2018PhRvD..98h4042T}, which could be observed in a \ac{GW} signal \citep{1994CQGra..11.2807B,1995PhRvL..74.1067B,2006PhRvD..74b4006A,2010PhRvD..82f4010M,2012PhRvD..85h2003L}.
In the companion paper \citet{GWTC:TGR}, we present parameterized tests for such violations.

Effects arising from the finite size of the component masses of a binary include spin-induced multipole moments, most importantly their spin-induced quadrupole moments $Q$, which also affect the orbital evolution.
For a \ac{BH}, there is a fixed relation between its spin-induced quadrupole moment and its mass and spin \citep{1998PhRvD..57.5287P}.
Deviations from this predicted value, as observed in the phase evolution of a \ac{GW} signal, can be used to distinguish a \ac{BBH} from a compact binary containing exotic, non-\ac{BH} components.
Some examples of exotic alternatives to \acp{BH}, being compact objects capable of having masses greater than the maximum mass of an \ac{NS}, include boson stars \citep{1968PhRv..172.1331K,1969PhRv..187.1767R}, gravastars \citep{2004PNAS..101.9545M}, fuzzballs \citep{2005ForPh..53..793M}, and firewalls \citep{2013JHEP...02..062A}.
Past companion papers \citep[most recently][]{GWTC4:TGR-II} have presented such parameterized tests of the nature of the components of \acp{CBC}.

\subsubsection{Compact Binary Merger and Ringdown}

The final stages of \ac{GW} emission from \acp{CBC} that result in a \ac{BH} remnant can be modeled as a linear gravitational perturbation to a Kerr \ac{BH} spacetime \citep{1973ApJ...185..649P}.
Remarkably, the partial differential equations for the outgoing \ac{GW} content of such a perturbation decouple from the other gravitational modes, and those decoupled equations are \emph{separable} into a radial equation, an angular equation, and an exponential function of time with a complex frequency \citep{1972PhRvL..29.1114T,1973ApJ...185..635T}.
The separation results in a spectrum of complex eigenfrequencies of the \ac{GW} perturbations to the \ac{BH} spacetime indexed by integer degree $\ell$ and order $m$ numbers, $\ell\ge2$ and $|m|\le\ell$, and integer overtone $n$ with $n\ge1$ \citep{1985RSPSA.402..285L,2006PhRvD..73f4030B}.
The angular eigenfunctions, which depend on $\ell$ and $m$, also depend on the dimensionless complex frequency and the dimensionless spin parameter of the remnant \ac{BH}\@.
The complex eigenfrequencies describe the spectrum of exponentially decaying sinusoidal \ac{GW} \emph{quasi-normal modes} that make up what is called the \emph{\ac{BH} ringdown}.
\ac{GR} therefore provides a prediction for the relationship between the frequency and the decay constant for the spectrum of such quasi-normal modes that depend solely on the mass and spin of the final \ac{BH}, and thus the \ac{BH} ringdown radiation can be used to test these predictions of \ac{GR}\@.

Spanning the region between the portion of the waveform that can be computed by \ac{PN} calculations at early time and the portion that can be computed by a superposition of quasi-normal modes at late time is what is known as the \emph{merger phase} of the compact binary.
Due to the nonperturbative nature of this phase, \ac{NR} solutions to Einstein's field equations are sought \citep{2014ARA&A..52..661L,2019RPPh...82a6902D}.
Such solutions both interpolate these early and late phases and also provide the information about the quasi-normal mode amplitudes and phases excited as well as the mass and spin of the remnant \ac{BH} \citep{2016ApJ...825L..19H,2017PhRvD..95b4037H,2017PhRvD..95f4024J}.

When at least one component of the binary in the \ac{CBC} is not a \ac{BH} (i.e., an \ac{NS}), the merger and ringdown phases might be considerably more complex owing to the presence of matter in the system.
\ac{NR} is typically required to compute the entire post-inspiral phase of the \acp{GW} emitted from such systems \citep{2012LRR....15....8F,2021LRR....24....5K}.
One important piece of such simulations is to determine whether disruption of an \ac{NS} component occurs, particularly in the case of \ac{NSBH} systems in which the \ac{NS} might be swallowed whole by the \ac{BH} (typical for high-mass and low-spin \acp{BH}) or might be tidally disrupted by the \ac{BH} (typical for low-mass or high-spin \acp{BH}).
Such \ac{NS} disruption would be expected to produce electromagnetic emission that could be observed by electromagnetic astronomical observatories.
Guided by numerical simulations, one can estimate whether a system having particular parameters inferred from the inspiral phase will be electromagnetically bright and so a candidate for electromagnetic follow-up observations \citep{2018PhRvD..98h1501F,2020ApJ...896...54C,2024CQGra..41h5012B}.

Depending on the masses of the initial components, the product of the merger of two \acp{NS} might be another \ac{NS}, a supramassive \ac{NS} (a uniformly spinning \ac{NS} that is more massive than the highest allowed mass for a nonspinning \ac{NS}, which remains an \ac{NS} until its angular momentum is dissipated, resulting in its collapse to a \ac{BH}), a hypermassive \ac{NS} (an \ac{NS} more massive than would be allowed for any stationary, spinning configuration, but which is temporarily supported by differential rotation, and which will ultimately collapse to a \ac{BH}), or there might be a direct collapse on a dynamical timescale to form a \ac{BH} after the merger \citep{2000ApJ...528L..29B,2017ApJ...844L..19P}.
Both the electromagnetic emission and \ac{GW} emission from these different scenarios are expected to vary considerably \citep{2017ApJ...851L..16A}.

\paragraph{Tests of \acs{GR} from \acs{CBC} Merger}

\Ac{NR} simulations of \acp{BBH} in \ac{GR} provide predictions for the \ac{GW}  waveform spanning the inspiral, merger, and final ringdown phases of evolution.
Tests for violations of \ac{GR} can be performed by subtracting the best-fit \ac{GR} waveform from the observed data and testing whether the remaining residual is consistent with detector noise or whether there is remaining signal present.
Alternatively, since \ac{NR} predicts how a final \ac{BH} mass and spin are related to the initial \ac{BH} masses and spins for a \ac{BBH} \ac{CBC} \citep{2016ApJ...825L..19H,2017PhRvD..95b4037H,2017PhRvD..95f4024J}, a test of consistency between the initial orbital parameters and the final \ac{BH} mass and spin can be performed.
Here, the initial component \ac{BH} masses and spins can be determined from the early inspiral phase of the \ac{GW} signal, while the mass and the spin of the final \ac{BH} are found from the late-time ringdown radiation.
In practice, such a consistency test divides the \ac{GW} signal into low- and high-frequency portions (below and above a given cutoff frequency) that are independently modeled with full inspiral--merger--ringdown waveforms \citep{2016PhRvD..94b1101G,2018CQGra..35a4002G}.
Companion paper \citet{GWTC:TGR} presents results from such a residual test and past companion papers \citep[most recently][]{GWTC4:TGR-I} have also presented inspiral--merger--ringdown consistency tests.

In \ac{GR}, a \ac{BH} remnant produced by a \ac{CBC} will rapidly settle to a stationary Kerr \ac{BH} \citep{1963PhRvL..11..237K}, uniquely characterized by its mass and spin \citep{1967PhRv..164.1776I,1971PhRvL..26..331C}, through emission of ringdown radiation in a spectrum of quasi-normal modes, as described earlier.
These quasi-normal modes have a discrete spectrum of complex-valued eigenfrequencies (the imaginary part of which determines the decay timescale), so a possible non-\ac{BH} remnant \citep[e.g.,][]{2013PhRvD..88f4046M}, or modifications of the spectrum in alternative theories of \ac{GR} \citep[e.g.,][]{2024PhRvD.110l4057C}, can be tested by looking for deviations in the observed ringdown radiation from the anticipated spectrum of quasi-normal modes \citep{2025arXiv250523895B}.

Furthermore, if the remnant object does not possess an event horizon, ingoing \ac{GW} radiation can be reflected off of a surface or scattered off of an inner potential and reemerge as an echo signal observed within a few seconds after the merger \citep{2016PhRvD..94h4031C,2019LRR....22....4C,2024PhRvL.133c1401S}.
The companion paper \citet{GWTC:TGR} presents tests of the nature of the remnant resulting from \ac{CBC} through observed quasi-normal mode spectra. Past companion papers \citep[most recently][]{GWTC4:TGR-III} have also presented searches for \ac{GW} echoes.

The post-inspiral portion of a \ac{BBH} signal can be phenomenologically modeled with various parameters that are fitted to \ac{NR} simulations \citep{2020PhRvD.102f4001P}.
Past companion papers \citep[most recently][]{GWTC4:TGR-II} have explored possible deviations of these parameters from their nominal values \citep{2018PhRvD..97d4033M, 2026PhRvD.113b4016R}.

\subsubsection{Redshift and Cosmological Effects}

\Acp{GW} can be redshifted, just as electromagnetic waves are. These changes are caused by the Doppler effect due to relative motion of the emitter and the observer (often described in terms of \emph{peculiar velocities} relative to the rest frame of the cosmological microwave background radiation), due to the expansion of space between the emitter and the observer, or due to gravitational redshift if the emitter and observer have different gravitational potentials.
For sources beyond the nearby Universe (having redshifts $\gtrsim 0.1$), cosmological expansion is the dominant source of redshift \citep{2022ApJ...938..112P}.

The redshift is the fractional difference between the frequency of a wave at emission at its source $f_{\text{src}}$, and its observed frequency at a detector $f_{\text{det}}$, $z=(f_{\text{src}}-f_{\text{det}})/f_{\text{det}}$ \citep{1999astro.ph..5116H}.
Thus, the observed frequency of a wave is related to its emitted frequency by $f_{\text{det}}=f_{\text{src}}/(1+z)$.
Similarly, an interval in time in the source-frame $dt_{\text{src}}$ is related to an observed interval in time by a detector $dt_{\text{det}}$ by $dt_{\text{det}}=(1+z)dt_{\text{src}}$.
Equations~\eqref{e:hpc} and \eqref{e:pdot} are both parameterized in terms of the orbital velocity $v$, which is related to the \ac{GW} frequency $f$ in the dominant mode by Equation~\eqref{e:orb_velocity_cubed}.
At a fixed moment in a \ac{GW} waveform, where the binary has some instantaneous value of $v$, we have
\begin{equation}
    v^3 = \pi G \Mtot f_{\text{src}} = \pi G \Mtot (1+z) f_{\text{det}}.
\end{equation}
That is, a redshifted signal, observed at frequency $f_{\text{det}}$, produced by a system with intrinsic mass \Mtot{} has identical morphology to an unredshifted signal produced by a system with intrinsic mass $(1+z)\Mtot$ \citep{1987GReGr..19.1163K}.
If the redshift is unknown, then the observable mass parameters are the various combinations of $(1+z)\massone$ and $(1+z)\masstwo$, e.g., $(1+z)\Mc$ and $(1+z)\Mtot$.
These mass parameters with the $1+z$ scale factor are referred to as \emph{detector-frame masses}, $\massone^{\text{det}}=(1+z)\massone$, $\masstwo^{\text{det}}=(1+z)\masstwo$, $\Mtot^{\text{det}}=(1+z)\Mtot$, and $\Mc^{\text{det}}=(1+z)\Mc$.
\ac{PN} corrections to the waveform preserve this degeneracy for point particles (and \acp{BH}).
However, for \acp{NS}, a functional relationship between the mass of an \ac{NS} and its tidal deformability means that a measurement of $\tilde\Lambda$ can break the degeneracy between mass and redshift, allowing the two to be independently
measured \citep{2012PhRvL.108i1101M}.

The amplitudes of $h_+$ and $h_\times$ given in Equation~\eqref{e:hpc} also depend on the total mass through the factor $\Mtot/r$.
If the factor $(1+z)\Mtot$ is determinable from the rate of decay of the orbital period, Equation~\eqref{e:pdot}, then the amplitude factor can be written as $[(1+z)\Mtot]/[(1+z)r]$, suggesting that the measurable amplitude distance parameter is $(1+z)r$.

The parameter $r$ that appears in inverse proportion to the \ac{GW} amplitude in Equation~\eqref{e:hpc} represents the areal radius, i.e., spheres centered on the \ac{GW} source have area $4\pi r^2$.
Within a cosmological setting, this parameter is the \emph{transverse comoving distance} $D_{\text{M}}$ \citep{1999astro.ph..5116H}.
Then, if the redshift is entirely due to the cosmological expansion of spacetime, the combination $(1+z)D_{\text{M}}$ is equal to the \emph{luminosity distance} of the source, and this becomes the observable distance parameter from the \ac{GW} amplitude.
In this sense, then, given a known cosmology (i.e., the values of the Hubble constant, matter density, and the spatial curvature), the functional relationship between luminosity distance and redshift allows the determination of the latter from the former, and the intrinsic masses, e.g., $\Mtot$, can then be deduced from
the observed mass--redshift combined parameters, e.g., $\Mtot^{\text{det}}=(1+z)\Mtot$.
However, if other redshift effects are present, e.g., due to peculiar motion of the source or the observer relative to the Hubble flow, the combination $(1+z)D_{\text{M}}$ is no longer equal to the luminosity distance.

Nevertheless, when reporting the parameters of a \ac{CBC}, we normally assume that cosmological expansion is the only significant source of redshift, and so the observed amplitude parameter $(1+z)\DM$ is referred to as luminosity distance $\DL$, while a dimensionful intrinsic mass parameter such as the primary mass $\massone$ is derived from observed detector-frame mass parameters as $\massone=\massone^{\text{det}}/[1+z(\DL)]$, where the relationship between the redshift and the luminosity distance, $z(\DL)$, is obtained by some standard cosmological model.
The only case where this was not done was for GW170817, where the measured geocentric redshift to its host galaxy NGC\,4993 was used \citep{2019PhRvX...9a1001A}.
The main uncertainty in the chirp mass of the system comes from the unknown peculiar velocity of the system relative to its host galaxy.
Unless otherwise specified, the reference cosmology used to relate luminosity distance to redshift throughout the works is a ΛCDM model \citep{2003RvMP...75..559P} corresponding to a spatially flat Friedman--Lema\^itre--Robertson--Walker spacetime \citep{1999GReGr..31.1991F,1999GReGr..31.2001F,1931MNRAS..91..483L,1935ApJ....82..284R,1936ApJ....83..187R,1936ApJ....83..257R,1937PLMS...42...90W,1972gcpa.book.....W,1973grav.book.....M} with Hubble constant $H_0=\qty{67.9}{\kilo\meter\per\second\per\Mpc}$, matter density parameter $\Omega_{\text{m}}=\num{0.3065}$, and cosmological constant density parameter $\Omega_\Lambda=1-\Omega_{\text{m}}=\num{0.6935}$ \citep[column TT+lowP+lensing+ext of Table 4]{2016A&A...594A..13P}.

\paragraph{Constraints on Cosmic Expansion from \acs{GW} Observations}

If the redshift of a \ac{GW} source can be determined independently of its distance, then a distance--redshift relationship can be obtained and used to infer cosmological parameters \citep{1986Natur.323..310S,1987GReGr..19.1163K}.
Here the \ac{CBC} is called a \emph{standard siren} (akin to the \emph{standard candles} such as Cepheid variables and Type~Ia supernovae used to measure distances to their galaxy hosts), where the luminosity distance of the \ac{CBC} is inferred from the amplitude of the \acp{GW} \citep{2005ApJ...629...15H}.
The most direct method of determining the redshift of a \ac{GW} source is if there is an electromagnetic counterpart in which spectroscopic measurements of the redshift of its host galaxy can be made \citep{1987GReGr..19.1163K,2005ApJ...629...15H,2006PhRvD..74f3006D,2018Natur.562..545C}.
This method is known as the \emph{bright sirens} method.
For example, the \ac{BNS} coalescence GW170817 \citep{2017PhRvL.119p1101A} was associated with the optical kilonova AT\,2017gfo in the galaxy NGC\,4993 \citep{2017ApJ...848L..12A}, which allowed for a measurement of the maximum a posteriori value of the Hubble constant with \qty{68.3}{\percent} \ac{CL} highest density interval $69^{+17}_{-8}$~\unit{\kilo\meter\per\second\per\Mpc} \citep{2021ApJ...909..218A}.

If no electromagnetic counterpart to a \ac{GW} is observed, various methods are available to deduce the associated redshift (\acp{BBH} are not normally expected to produce any electromagnetic radiation, unless there is matter present in their environment).
One such method, the \emph{galaxy catalog} method, also called the \emph{dark siren} method, is to obtain statistical association of \ac{GW} sources with potential host galaxies observed in surveys \citep{1986Natur.323..310S,2008PhRvD..77d3512M}.
This is usually done simultaneously with information obtained from another method, the \emph{spectral siren} method.
In the spectral siren method, a known feature in the mass distribution of the population of \acp{CBC} is used to statistically infer the redshift of a number of sources at a given distance by how the observed detector-frame mass distribution is shifted with respect to the local (zero-redshift) distribution \citep{1993ApJ...411L...5C,2012PhRvD..85b3535T,2019ApJ...883L..42F,2021PhRvD.104f2009M}.
Finally, future observations of \ac{BNS} mergers may be capable of directly inferring the redshift of the source from the \ac{GW} signal alone through measurements of the \ac{NS} tidal deformations \citep{2012PhRvL.108i1101M,2021PhRvD.104h3528C}.

The companion paper \citet{GWTC:Cosmology} reports on constraints on the cosmic expansion history based on combined \ac{CBC} bright and dark sirens, including both the galaxy catalog method and the spectral siren approach.
If \acp{GW} propagate through cosmological backgrounds differently from electromagnetic waves in a manner that produces a different distance--amplitude relation, the modified propagation effects can be observed using standard siren methods \citep{2018PhRvD..98b3510B}.
The companion paper \citet{GWTC:Cosmology} also reports on constraints on such effects of modified \ac{GW} propagation.

\subsubsection{Populations of compact binaries}

With a multitude of observed \acp{CBC}, one can infer the underlying population of these sources.
In doing so, one needs to account for the detector selection effects, e.g., the fact that events that are farther away are less likely to be detected as compared to events that are nearby.
One key element of the population that can be measured is the local merger rate density $\mathcal{R}$, representing the number of \acp{CBC} occurring per unit time per unit volume in the local Universe \citep{1991ApJ...380L..17P,2003ApJ...584..985K,2004CQGra..21S1775B,2009CQGra..26q5009B,2015PhRvD..91b3005F,2016ApJ...833L...1A}, or its evolution with cosmic redshift $\mathcal{R}(z)$, which is the number of coalescences per unit source-frame time per unit comoving volume at a cosmological redshift of $z$ \citep{2018ApJ...863L..41F}.
Another is the population distribution of masses and spins of merging compact binaries, $\PEpdf{\massone,\masstwo,{\boldsymbol{S}}_1,{\boldsymbol{S}}_2}$, which might also evolve over cosmic history, $\PEpdf[z]{\massone,\masstwo,{\boldsymbol{S}}_1,{\boldsymbol{S}}_2}$.
The measurement uncertainty in single-event parameters and the total number of detected events dictate the measurability of features in the population. 
These inferences are important in understanding the underlying astrophysical formation channels of compact binaries \citep[e.g.,][]{2017MNRAS.471.2801S,2017Natur.548..426F,2021ApJ...910..152Z,2022LRR....25....1M,2026JPhCS3177a2077C}.

\paragraph{Inference of the Population of \acsp{CBC}}
In the companion paper \citet{GWTC:AstroDist}, we present measurements of the local rate of \ac{BNS}, \ac{NSBH}, and \ac{BBH} mergers, inference of the evolution of the \ac{CBC} rate over cosmological time, and inference of the distribution of masses and spins of \acp{CBC}.

\startlongtable
\begin{deluxetable*}{lcl}
\tablecaption{\label{tab:cbc_params}%
Parameters Describing a \ac{CBC} System with Quasi-circular Orbits.}
\tablehead{\colhead{Parameter name} & \colhead{Symbol} & \colhead{Notes [Dimensions]}}
\startdata
\parbox[t]{0.3\linewidth}{\raggedright Primary and secondary masses}
    & \massone, \masstwo
    & \parbox[t]{0.5\linewidth}{\raggedright Mass of the more massive (\massone) and less massive (\masstwo) body in system, $\massone \ge \masstwo$ [mass]}
    \\
\parbox[t]{0.3\linewidth}{\raggedright Chirp mass}
    & \Mc
    & \parbox[t]{0.5\linewidth}{\raggedright See Equation~\eqref{e:chirp_mass} [mass]}
    \\
\parbox[t]{0.3\linewidth}{\raggedright Total mass}
    & \Mtot
    & \parbox[t]{0.5\linewidth}{\raggedright $\Mtot=\massone + \masstwo$ [mass]}
    \\
\parbox[t]{0.3\linewidth}{\raggedright Final mass}
    & \Mf
    & \parbox[t]{0.5\linewidth}{\raggedright Mass of the remnant [mass]}
    \\
\parbox[t]{0.3\linewidth}{\raggedright Mass ratio}
    & \mratio
    & \parbox[t]{0.5\linewidth}{\raggedright $\mratio = \masstwo/\massone \le 1$ [dimensionless]}
    \\
\parbox[t]{0.3\linewidth}{\raggedright Symmetric mass ratio}
    & \mratiosym
    & \parbox[t]{0.5\linewidth}{\raggedright $\mratiosym = \massone \masstwo/(\massone+\masstwo)^2 \le 1/4$ [dimensionless]}
    \\
\parbox[t]{0.3\linewidth}{\raggedright Energy radiated}
    & \Erad
    & \parbox[t]{0.5\linewidth}{\raggedright $\Erad = (\Mtot - \Mf)c^2$ [energy]}
    \\
\parbox[t]{0.3\linewidth}{\raggedright Peak luminosity}
    & \lumpeak
    & \parbox[t]{0.5\linewidth}{\raggedright Peak \ac{GWH} luminosity, typically 0.1\% of the Planck luminosity ($\ell_{\text{Planck}}=c^5/G$) for \ac{BBH} coalescences [power]}
    \\
\parbox[t]{0.3\linewidth}{\raggedright Primary and secondary spin vectors}
    & ${\boldsymbol{S}}_1$, ${\boldsymbol{S}}_2$
    & \parbox[t]{0.5\linewidth}{\raggedright Spin angular momentum of the primary ($\boldsymbol{S}_1$) and secondary ($\boldsymbol{S}_2$) [angular momentum]}
    \\
\parbox[t]{0.3\linewidth}{\raggedright Primary and secondary dimensionless spin magnitudes}
    & \spinone, \spintwo
    & \parbox[t]{0.5\linewidth}{\raggedright $\chi_{1,2} = c|{\boldsymbol{S}}_{1,2}|/(Gm_{1,2}^2)$; $\chi_{1,2} \le 1$ for Kerr \acp{BH} primary/secondary [dimensionless]}
    \\
\parbox[t]{0.3\linewidth}{\raggedright Remnant dimensionless spin magnitude}
    & \chif
    & \parbox[t]{0.5\linewidth}{\raggedright $\chif = cS_{\text{f}}/(G\Mf^2)$ where $S_{\text{f}}$ is the magnitude of the remnant's spin angular momentum; $\chif \le 1$ for a Kerr \ac{BH} remnant [dimensionless]}
    \\
\parbox[t]{0.3\linewidth}{\raggedright Newtonian orbital angular momentum}
    & $\boldsymbol{L}$
    & \parbox[t]{0.5\linewidth}{\raggedright Instantaneous orbital angular momentum about the center of mass; defines $z$-direction for spin coordinates [angular momentum]}
    \\
\parbox[t]{0.3\linewidth}{\raggedright Total angular momentum}
    & $\boldsymbol{J}$
    & \parbox[t]{0.5\linewidth}{\raggedright ${\boldsymbol{J}} = {\boldsymbol{L}}_{\text{PN}} + {\boldsymbol{S}}_1 + {\boldsymbol{S}}_2$, where $\boldsymbol{L}_{\text{PN}}$ is the Newtonian orbital angular momentum $\boldsymbol{L}$ with \ac{PN} corrections [angular momentum]}
    \\
\parbox[t]{0.3\linewidth}{\raggedright Primary and secondary tilt angle}
    & $\theta_1$, $\theta_2$
    & \parbox[t]{0.5\linewidth}{\raggedright Angle between ${\boldsymbol{S}}_{1,2}$ and $\boldsymbol{L}$ [angle]}
    \\
\parbox[t]{0.3\linewidth}{\raggedright Primary and secondary spin azimuthal angle}
    & $\phi_{1}$, $\phi_{2}$
    & \parbox[t]{0.5\linewidth}{\raggedright Angle measured counterclockwise about $\mathbf{L}$ from the line of nodes $\mathbf{L} \times \mathbf{N}$ to $\mathbf{L}\times (\mathbf{S}_{1,2} \times \mathbf{L})$ [angle]}
    \\
\parbox[t]{0.3\linewidth}{\raggedright Spin azimuthal angle difference}
    & $\phi_{12}$
    & \parbox[t]{0.5\linewidth}{\raggedright $\phi_{12}=\phi_{2}-\phi_{1}$ [angle]}
    \\
\parbox[t]{0.3\linewidth}{\raggedright Effective inspiral spin parameter}
    & \chieff
    & \parbox[t]{0.5\linewidth}{\raggedright See Equation~\eqref{e:chi_eff} [dimensionless]}
    \\
\parbox[t]{0.3\linewidth}{\raggedright Effective precession spin parameter}
    & \chip
    & \parbox[t]{0.5\linewidth}{\raggedright See Equation~\eqref{e:chi_p} [dimensionless]}
    \\
\parbox[t]{0.3\linewidth}{\raggedright Orbital inclination angle}
    & $\iota$
    & \parbox[t]{0.5\linewidth}{\raggedright Angle between $\boldsymbol{L}$ and the direction to Earth $\boldsymbol{N}$ (see Figure~\ref{fig:src}) [angle]}
    \\
\parbox[t]{0.3\linewidth}{\raggedright Source inclination angle}
    & $\theta_{JN}$
    & \parbox[t]{0.5\linewidth}{\raggedright Angle between $\boldsymbol{J}$ and the direction to Earth $\boldsymbol{N}$ [angle]}
    \\
\parbox[t]{0.3\linewidth}{\raggedright Viewing angle}
    & $\Theta$
    & \parbox[t]{0.5\linewidth}{\raggedright $\Theta=\min\{\theta_{JN},\pi-\theta_{JN}\}$ [angle]}
    \\
\parbox[t]{0.3\linewidth}{\raggedright Orbital phase}
    & $\phi$
    & \parbox[t]{0.5\linewidth}{\raggedright Phase of a binary's orbit, the angle on the orbital plane between the separation vector (the position vector of the primary minus the position vector of the secondary) and the line of nodes, ${\boldsymbol{L}}\times{\boldsymbol{N}}$ (see Figure~\ref{fig:src}) [angle]}
    \\
\parbox[t]{0.3\linewidth}{\raggedright Coalescence phase}
    & $\phi_{\text{c}}$
    & \parbox[t]{0.5\linewidth}{\raggedright Orbital phase, the angle on the orbital plane between the separation vector (the position vector of the primary minus the position vector of the secondary) and the line of nodes, ${\boldsymbol{L}}\times{\boldsymbol{N}}$, at a point in the evolution corresponding to the point in the waveform used to define \tgeo{} (see Table~\ref{tab:gw_params}) [angle]}
    \\
\parbox[t]{0.3\linewidth}{\raggedright Angular diameter distance}
    & \DA
    & \parbox[t]{0.5\linewidth}{\raggedright An object of transverse length $x$ is observed to subtend an angle in radians of $x/\DA$ when both the object and observer are at rest relative to a homogeneous cosmology \citep{1999astro.ph..5116H} [length]}
    \\
\parbox[t]{0.3\linewidth}{\raggedright Transverse comoving distance}
    & \DM
    & \parbox[t]{0.5\linewidth}{\raggedright Areal radius of a sphere centered on a point in an isotropic cosmology, defined so the sphere has area $4\pi\DM^2$ \citep{1999astro.ph..5116H} [length]}
    \\
\parbox[t]{0.3\linewidth}{\raggedright Luminosity distance}
    & \DL
    & \parbox[t]{0.5\linewidth}{\raggedright A source of isotropic radiation having luminosity $\ell_{\text{iso}}$ is observed to have flux $\ell_{\text{iso}}/(4\pi\DL^2)$ when both the source and observer are at rest relative to a homogeneous cosmology \citep{1999astro.ph..5116H} [length]}
    \\
\parbox[t]{0.3\linewidth}{\raggedright Redshift}
    & $z$
    & \parbox[t]{0.5\linewidth}{\raggedright The fractional difference between the frequency of a wave at emission at its source $f_{\mathrm{src}}$, and its observed frequency at a detector $f_{\mathrm{det}}$, $z=(f_{\mathrm{src}}-f_{\mathrm{det}})/f_{\mathrm{det}}$; the reference cosmology for the relationship between distance and the \emph{cosmological} redshift is given in the text [dimensionless]}
    \\
\parbox[t]{0.3\linewidth}{\raggedright Primary and secondary dimensionless tidal deformabilities}
    & $\Lambda_1$, $\Lambda_2$
    & \parbox[t]{0.5\linewidth}{\raggedright See Equation~\eqref{e:tidal_deformability}; $\Lambda_{1,2}=0$ for a \ac{BH} primary/secondary [dimensionless]}
    \\
\parbox[t]{0.3\linewidth}{\raggedright Effective tidal deformability}
    & $\tilde{\Lambda}$
    & \parbox[t]{0.5\linewidth}{\raggedright See Equation~\eqref{e:lambda-tilde}; $\tilde{\Lambda}=0$ for a \ac{BBH} [dimensionless]}
    \\
\parbox[t]{0.3\linewidth}{\raggedright Primary and secondary dimensionless spin-induced quadrupole moments}
    & $\kappa_1$, $\kappa_2$
    & \parbox[t]{0.5\linewidth}{\raggedright See Equation~\eqref{e:spin_induced_quadrupole}; $\kappa_{1,2}=1$ for a \ac{BH} primary/secondary [dimensionless]}
    \\
\parbox[t]{0.3\linewidth}{\raggedright Primary and secondary radii}
    & $R_1$, $R_2$
    & \parbox[t]{0.5\linewidth}{\raggedright Areal radii of primary and secondary objects, defined so their surface areas are $4\pi R_{1,2}^2$; used in defining \ac{NS} compactness [length]}
    \\
\parbox[t]{0.3\linewidth}{\raggedright Primary and secondary compactness}
    & $C_1$, $C_2$
    & \parbox[t]{0.5\linewidth}{\raggedright Dimensionless mass-to-radius ratios $C_{1,2}=G\massone/(c^2R_{1,2})$ of primary/secondary; $1/C_{1,2}=1+\sqrt{1+\smash{\chi_{1,2}^2}}$ for Kerr \ac{BH} primary/secondary [dimensionless]}
    \\
\parbox[t]{0.3\linewidth}{\raggedright Merger rate density}
    & $\mathcal{R}$
    & \parbox[t]{0.5\linewidth}{\raggedright Rate of binary mergers per unit volume in the local Universe; may be expressed as a function of cosmological redshift, ${\mathcal{R}}(z)$; the rate in the local Universe ${\mathcal{R}}(z=0)$ can be notated ${\mathcal{R}}_0$; subscripts can be used if considering different populations, e.g., ${\mathcal{R}}_{\text{BNS}}$, ${\mathcal{R}}_{\text{NSBH}}$, and ${\mathcal{R}}_{\text{BBH}}$ [time\textsuperscript{\textminus1} volume\textsuperscript{\textminus1}]}
    \\
\enddata
\end{deluxetable*}

\section{Synopsis}\label{sec:synopsis}

This paper serves as an introduction to the collection of papers accompanying the \ac{LVK}'s \thisgwtc{}.
We have provided an overview of the \ac{GW} detectors and observing runs of the \ac{LVK} network and of the observed \acp{GW} from \acp{CBC}.
The primary sequels to this introduction are a description of the methods used to perform searches for \acp{GW} in \ac{LVK} data and to characterize source properties of identified signals \citep{GWTC:Methods} and a summary of the main observations of \thisgwtc{}, highlighting new \ac{CBC} candidates from \ac{O4b} and their estimated masses and spins \citep{GWTC:Results}.
Other companion papers presenting science results from the analysis of the \thisgwtc{} candidates are described in Section~\ref{sec:overview}.
\ac{GWTC} provides a prodigious census of over \num{300} merging binary systems spanning two orders of magnitude in mass from \qty{\sim1}{\Msun} \acp{NS} to remnant \acp{BH} exceeding \qty{100}{\Msun}.
Study of these observations will provide new insight into the nature of these objects, their population distribution, and their formation channels.
These \ac{GW} observations allow for sensitive tests of \ac{GR} and provide information about the cosmological expansion history.

\section*{Acknowledgements}
This material is based upon work supported by NSF's LIGO Laboratory, which is a
major facility fully funded by the National Science Foundation.
The authors also gratefully acknowledge the support of
the Science and Technology Facilities Council (STFC) of the
United Kingdom, the Max-Planck-Society (MPS), and the State of
Niedersachsen/Germany for support of the construction of Advanced LIGO 
and construction and operation of the GEO\,600 detector. 
Additional support for Advanced LIGO was provided by the Australian Research Council.
The authors gratefully acknowledge the Italian Istituto Nazionale di Fisica Nucleare (INFN),  
the French Centre National de la Recherche Scientifique (CNRS) and
the Netherlands Organization for Scientific Research (NWO)
for the construction and operation of the Virgo detector
and the creation and support  of the EGO consortium. 
The authors also gratefully acknowledge research support from these agencies as well as by 
the Council of Scientific and Industrial Research of India, 
the Department of Science and Technology, India,
the Science \& Engineering Research Board (SERB), India,
the Ministry of Human Resource Development, India,
the Spanish Agencia Estatal de Investigaci\'on (AEI),
the Spanish Ministerio de Ciencia, Innovaci\'on y Universidades,
the European Union NextGenerationEU/PRTR (PRTR-C17.I1),
the ICSC - CentroNazionale di Ricerca in High Performance Computing, Big Data
and Quantum Computing, funded by the European Union NextGenerationEU,
the Comunitat Auton\`oma de les Illes Balears through the Conselleria d'Educaci\'o i Universitats,
the Conselleria d'Innovaci\'o, Universitats, Ci\`encia i Societat Digital de la Generalitat Valenciana and
the CERCA Programme Generalitat de Catalunya, Spain,
the Polish National Agency for Academic Exchange,
the National Science Centre of Poland and the European Union - European Regional
Development Fund;
the Foundation for Polish Science (FNP),
the Polish Ministry of Science and Higher Education,
the Swiss National Science Foundation (SNSF),
the Russian Science Foundation,
the European Commission,
the European Social Funds (ESF),
the European Regional Development Funds (ERDF),
the Royal Society, 
the Scottish Funding Council, 
the Scottish Universities Physics Alliance, 
the Hungarian Scientific Research Fund (OTKA),
the French Lyon Institute of Origins (LIO),
the Belgian Fonds de la Recherche Scientifique (FRS-FNRS), 
Actions de Recherche Concert\'ees (ARC) and
Fonds Wetenschappelijk Onderzoek - Vlaanderen (FWO), Belgium,
the Paris \^{I}le-de-France Region, 
the National Research, Development and Innovation Office of Hungary (NKFIH), 
the National Research Foundation of Korea,
the Natural Sciences and Engineering Research Council of Canada (NSERC),
the Canadian Foundation for Innovation (CFI),
the Brazilian Ministry of Science, Technology, and Innovations,
the International Center for Theoretical Physics South American Institute for Fundamental Research (ICTP-SAIFR), 
the Research Grants Council of Hong Kong,
the National Natural Science Foundation of China (NSFC),
the Israel Science Foundation (ISF),
the US-Israel Binational Science Fund (BSF),
the Leverhulme Trust, 
the Research Corporation,
the National Science and Technology Council (NSTC), Taiwan,
the United States Department of Energy,
and
the Kavli Foundation.
The authors gratefully acknowledge the support of the NSF, STFC, INFN and CNRS for provision of computational resources.

This work was supported by MEXT,
the JSPS Leading-edge Research Infrastructure Program,
JSPS Grant-in-Aid for Specially Promoted Research 26000005,
JSPS Grant-in-Aid for Scientific Research on Innovative Areas 2402: 24103006,
24103005, and 2905: JP17H06358, JP17H06361 and JP17H06364,
JSPS Core-to-Core Program A.\ Advanced Research Networks,
JSPS Grants-in-Aid for Scientific Research (S) 17H06133 and 20H05639,
JSPS Grant-in-Aid for Transformative Research Areas (A) 20A203: JP20H05854,
the joint research program of the Institute for Cosmic Ray Research,
University of Tokyo,
the National Research Foundation (NRF),
the Computing Infrastructure Project of the Global Science experimental Data hub
Center (GSDC) at KISTI,
the Korea Astronomy and Space Science Institute (KASI),
the Ministry of Science and ICT (MSIT) in Korea,
Academia Sinica (AS),
the AS Grid Center (ASGC) and the National Science and Technology Council (NSTC)
in Taiwan under grants including the Science Vanguard Research Program,
the Advanced Technology Center (ATC) of NAOJ, 
the Mechanical Engineering Center of KEK
and Vietnam National Foundation for Science and Technology Development 
(NAFOSTED) 103.01-2025.147.

Additional acknowledgements for support of individual authors may be found in the following document: \\
\texttt{https://dcc.ligo.org/LIGO-M2300033/public}.
For the purpose of open access, the authors have applied a Creative Commons Attribution (CC BY)
license to any Author Accepted Manuscript version arising.
We request that citations to this article use 'A. G. Abac {\it et al.} (LIGO-Virgo-KAGRA Collaboration), ...' or similar phrasing, depending on journal convention.

\facilities{EGO:Virgo, GEO\,600, Kamioka:KAGRA, LIGO.}
\software{Plots were prepared with \MATPLOTLIB{}~\citep{2007CSE.....9...90H} and \SEABORN{}~\citep{2021JOSS....6.3021W}.
\ASTROPY~\citep{2022ApJ...935..167A},
\GWPY~\citep{gwpy-software},
\LALSUITE~\citep{lalsuite, 2020SoftX..1200634W},
\NUMPY~\citep{2020Natur.585..357H},
\SCIPY~\citep{2020NatMe..17..261V}
were used for data processing in generating the figures and quantities in the manuscript.
}

\section*{Data availability}
Event data used within this work are openly available in the \thisgwtc{} online catalog, which is hosted at \url{https://gwosc.org/GWTC-5.0} and documented further in \citet{OpenData}.
Data behind Figures~\ref{fig:obsruns}, \ref{fig:asd}, and~\ref{fig:VT-events} can be found in \citet{GWTC-5.0:Intro_data_behind_figures}.

\begin{appendices}
\onecolumngrid

\section{Acronyms and glossary}\label{sec:glossary}

This is a reference of frequently used terms and acronyms.

\begin{description}

\item[\acs{A+}] \acl{A+} refers to a configuration of \ac{LIGO} following a series of upgrades, some in advance of \ac{O4} (such as the addition of a new \qty{300}{\meter} filter cavity for frequency-dependent vacuum squeezing), and some planned in advance of \ac{O5}, such as installation of new optics with lower noise and loss \citep{KAGRA:2013rdx,2023RScI...94a4502C}.

\item[\acs{Asharp}] \acl{Asharp} (A-sharp) is a proposed upgrade of the \acl{A+} interferometers anticipated on a post-\ac{O5} timeline.
The baseline \acs{Asharp} design is a room-temperature \qty{1}{\micro\meter} laser wavelength interferometer upgrade with larger test masses having coatings with lower thermal noise, higher laser power, and increased levels of vacuum squeezing \citep{PostO5Report:2022}.

\item[\acs{AdV}] \Acl{AdV} refers to an upgraded Virgo detector \citep{2015CQGra..32b4001A} with an advanced interferometer.
Virgo operated with the \acs{AdV} configuration during \ac{O2} and \ac{O3}.

\item[\acs{AdV+}] \Acl{AdV+} is an upgrade to the \ac{AdV} detector to take place in two phases: the first phase for operation during \ac{O4}, and the second phase for operation during \ac{O5}.

\item[\acs{aLIGO}] \acl{aLIGO} refers to an upgraded \ac{LIGO} configuration with advanced interferometers installed at both \ac{LHO} and \ac{LLO}. \ac{LIGO} operated with the \acs{aLIGO} configuration during \ac{O1}, \ac{O2}, \ac{O3}, and \ac{O4} \citep{2015CQGra..32g4001L}.

\item[\acs{BBH}] Binary black hole. A binary system where both components are \acp{BH}.

\item[\acs{BH}] \Acl{BH}.

\item[\acs{BHNS}] Black hole--neutron star specifically refers to systems in which the \ac{BH} formed before the \ac{NS}. See also \acs{NSBH}.

\item[\acs{bKAGRA}] \Acl{bKAGRA} is a configuration with a cryogenic resonant side-band extraction interferometer. As a preliminary step, the bKAGRA Phase-1 run from April 28 to May 6, 2018, operated a 3-km Michelson interferometer with sapphire test masses and full suspension systems, cooling one mass to cryogenic temperature, and marked the first cryogenic operation using full suspensions. \citep{2019CQGra..36p5008A}.

\item[\acs{BNS}] Binary neutron star. A binary system where both components are \acp{NS}.

\item[\acs{CBC}] \Acl{CBC}. The gravitational-radiation-driven orbital decay resulting in merger of a binary system made of two compact objects (\acp{NS} or \acp{BH}).

\item[\acs{CI}] \Acl{CI}. See \acs{CL}.

\item[\acs{CL}] \Acl{CL}.
Given a $n=1$ univariate or $n$-dimensional multivariate random variable $\boldsymbol x$ having \ac{PDF} $p({\boldsymbol x})$ and a $n$-dimensional region $R^n_\alpha$, then the \acs{CL} $\alpha$ of the region $R^n_\alpha$ is the probability of $\boldsymbol x$ lying in $R^n_\alpha$, $\alpha = P({\boldsymbol x}\in R^n_\alpha)=\int_{R^n_\alpha}p({\boldsymbol x})\,d^nx$.
The region $R^n_\alpha$ is then known as a $100\alpha\%$ \ac{CL} \emph{credible region}, with special cases: \emph{credible interval} (\acs{CI}) if $n=1$, \emph{credible area} if $n=2$, or \emph{credible volume} if $n=3$.
When $n>1$ we normally take $R^n_\alpha$ to be the region having the smallest volume that has \ac{CL} $\alpha$ (the \emph{highest-density region}).
When $n=1$ (\acs{CI}), $R^1_\alpha$ is normally chosen to be an \emph{equal-tailed interval} (also known as a symmetric interval), from the $\alpha/2$ quantile to the $1-\alpha/2$ quantile, but sometimes the smallest \emph{highest-density interval} is used instead.

\item[\acs{EOS}] \Acl{EOS} of an \ac{NS}.
For cold \acp{NS} (having temperature below the Fermi temperature), the \ac{EOS} is of a barotropic fluid, a relationship between the energy density of the fluid and its pressure.

\item[\acs{FAR}] \Acl{FAR}.
Often used as a detection threshold, the probability of any one or more of a sequence of statistical tests performed over a duration $T$ erroneously rejecting a null hypothesis is $1-\exp(-T\times\text{FAR})$.
\acs{FAR} therefore has dimensions of time\textsuperscript{\textminus1}.
When interpreted as a measure of a detection significance of a candidate detection, this is the rate at which noise alone would produce more significant candidates.

\item[\acs{GEO}] The \acl{GEO} is a British--German \Lshape-shaped interferometric \ac{GW} detector with \qty{600}{m} arms located near Hannover, Germany \citep{2002CQGra..19.1377W}.

\item[\acs{GR}] \Acl{GR}. Einstein's theory of gravitation.

\item[\acs{GW}] \Acl{GW}. See Sections~\ref{sec:overview} and \ref{ss:gw}.

\item[\acs{GWOSC}] The \acl{GWOSC} (formerly known as the \ac{LIGO} open science center) was created to provide public access to \ac{GW} data products \citep{2021SoftX..1300658A}.
The \acs{GWOSC} online data and resources can be found at \url{https://gwosc.org}.

\item[\acs{GWTC}] The \acl{GWTC} is the electronic catalog of \ac{GW} transients
 observed by \ac{LIGO}, \ac{Virgo}, and \ac{KAGRA} detectors produced by the \ac{LVK}.

\item[\acs{IFAR}] \Acl{IFAR}. 
The reciprocal of \acs{FAR}, $\text{IFAR}=(\text{FAR})^{-1}$, having dimensions of time.
A larger \acs{IFAR} implies a more significant candidate, while a larger \acs{FAR} implies a less significant candidate.

\item[\acs{IFO}] \Acl{IFO}, a type of detector that uses laser interferometry to measure changes in the lengths of optical paths induced by \acp{GW}.

\item[\acs{IGWN}] The \acl{IGWN} is a self-governing consortium using ground-based \ac{GW} interferometers to explore the fundamental physics of gravity and to observe the Universe.
The observatory network includes the \ac{KAGRA}, \ac{LHO}, \ac{LLO}, and \ac{Virgo} detectors.
 In addition, the \ac{GEO} detector serves as a technology test bed and operates in an \emph{astrowatch} mode outside of other detectors' observing periods.

\item[\acs{iKAGRA}] \Acl{iKAGRA} is a configuration of the \ac{KAGRA} detector as a simple Michelson interferometer that consists of two end test masses and a beam splitter.
The mirrors were fused silica mirrors at room temperature suspended by simplified systems.
\acs{iKAGRA} was operated from 2016~March~25 to 2016~March~31 and from
2016~April~11 to 2016~April~25 \citep{2018PTEP.2018a3F01A}.

\item[\acs{IMBH}] Intermediate-mass black hole. A \Ac{BH} in the mass range \qtyrange{\sim e2}{\sim e5}{\Msun}.

\item[\acs{IR1}] \Acl{IR1} is a six-month observing run planned to begin in late October or mid-November of 2026. See \url{https://observing.docs.ligo.org/plan} for updates on the \ac{LVK} observing run plans.

\item[\acs{KAGRA}] \acl{KAGRA} is a Japanese \Lshape-shaped interferometric \ac{GW} detector with \qty{3}{km} arms located underground at the Kamioka Mine in Japan \citep{2021PTEP.2021eA101A}.

\item[KAGRA Collaboration] The KAGRA Collaboration is a group of more than 400 individuals that carries out science related to the KAGRA detectors and their observations.

\item[\acs{LHO}] The \acl{LHO}, one of the two LIGO observatories, located in Hanford, Washington, is an \Lshape-shaped interferometric \ac{GW} detector with \qty{4}{km} arms.

\item[\acs{LIGO}] The \acl{LIGO} consists of two widely spaced installations within the United States: one in Hanford, Washington (\acs{LHO}), and the other in Livingston, Louisiana (\acs{LLO}). LIGO is operated by the LIGO Laboratory, a consortium of the California Institute of Technology and the Massachusetts Institute of Technology funded by the US National Science Foundation. Furthermore, LIGO-India is a planned advanced \ac{GW} observatory to be located in India as part of the worldwide network. LIGO-India is planned as a collaborative project between a consortium of Indian research institutions and the LIGO Laboratory in the USA, along with its international partners Australia, Germany and the UK.

\item[\acs{LLO}] The \acl{LLO}, one of the two LIGO observatories, located in Livingston, Louisiana, is an \Lshape-shaped interferometric \ac{GW} detector with \qty{4}{km} arms.

\item[\acs{LSC}] The \acl{LSC}, founded in 1997, is a group of more than 1000 scientists that carries out science related to the \ac{LIGO} detectors and their observations.

\item[\acs{LV}] The \acl{LV}. Prior to \ac{O3b}, all observational results were published by the \acs{LV}.

\item[LVC] The \acl{LVC}. The acronym \acs{LV} is now preferred.

\item[\acs{LVK}] The \acl{LVK}.

\item[\acs{NS}] \Acl{NS}.

\item[\acs{NSBH}] The general term for a neutron star--black hole binary: a binary system in which one component is an \ac{NS} and the other is a \ac{BH}.
If used in distinction with \acs{BHNS}, it refers to such systems in which the \ac{NS} formed before the \ac{BH}.

\item[\acs{NR}] \Acl{NR}, the use of numerical methods to solve relativistic field equations in curved spacetimes.

\item[\acs{O1}] The \acl{O1} began on 2015 September 12 and ended on 2016 January 19.
The \ac{LHO} and \ac{LLO} detectors participated in this observing run.

\item[\acs{O2}] The \acl{O2} began on 2016 November 30 and ended on 2017 August 25, during which the \ac{LHO} and the \ac{LLO} detectors were operating.
On 2017 August 1, the \ac{AdV} detector joined the observing run, forming a three-detector network.

\item[\acs{O3}] The \acl{O3} began on 2019 April 1 and ended on 2020 March 27, during which the \ac{LHO}, \ac{LLO}, and Virgo detectors were operating.
A commissioning break from 2019 October 1 to 2019 November 1 divided \acs{O3} into two parts, \acs{O3a} and \acs{O3b}.
A subsequent short run, \ac{O3GK}, from 2020 April 7 to 2020 April 21 with \ac{GEO} and \ac{KAGRA} observing, followed \acs{O3b}.

\item[\acs{O3a}] The first, pre--commissioning-break part of \acs{O3}, from 2019 April 1 until 2019 October 1, during which the \ac{LHO}, \ac{LLO}, and Virgo detectors were operating.

\item[\acs{O3b}] The second, post--commissioning-break part of \acs{O3}, from 2019 November 1 until 2020 March 27, during which the \ac{LHO}, \ac{LLO}, and Virgo detectors were operating.
\acs{O3b} was planned to continue until 2020 April 30 but ended early owing to the COVID-19 pandemic.

\item[\acs{O3GK}] A short observing run after \acs{O3b} from 2020 April 7 to 2020 April 21, during which the \ac{KAGRA} and \ac{GEO} detectors were observing.
\ac{KAGRA} had intended to join \ac{LIGO} and Virgo at the end of \ac{O3}, but the early end of \acs{O3b} made this impossible.

\item[\acs{O4}] The \acl{O4} began on 2023 May 24 and ended on 2025 November 18.
It is divided into parts, the first of which, \acs{O4a}, covered the period from 2023 May 24 until a commissioning break from 2024 January 16 to 2024 April 10.
During \acs{O4a}, \ac{LHO} and \ac{LLO} were observing.
Following the break, observing continued in \ac{O4b} from 2024 April 10 until an end date of 2025 January 28, with \ac{LHO}, \ac{LLO}, and Virgo observing.
It was decided to continue \ac{O4} observing in a third period \acs{O4c}, beginning 2025 January 28 and lasting until 2025 November 18.

\item[\acs{O4a}] The \acl{O4a} including data from 2023 May 24 until a commissioning break that began on 2024 January 16.
During \acs{O4a}, \ac{LHO} and \ac{LLO} were observing.
During the first 4 weeks, KAGRA was also observing.

\item[\acs{O4b}] The \acl{O4b} starting at the end of a commissioning break on 2024 April 10 and ending on 2025 January 28.
During \acs{O4b}, \ac{LHO}, \ac{LLO}, and Virgo were observing.
It was decided to continue \acs{O4} observations with a third part, \acs{O4c}, immediately following the end of \acs{O4b} on 2025 January 28.

\item[\acs{O4c}] The \acl{O4c}, extending the run beyond its intended end date through 2025 November 18.
A commissioning break in \acs{O4c} took place between \OfourcCommissioningBreakStartDate{} and \OfourcCommissioningBreakEndDate{}.

\item[\acs{O5}] The \acl{O5} is the planned future observing run to follow \ac{O4}.

\item[\acs{PDF}] \Acl{PDF}.
Given an $n=1$ univariate or $n$-dimensional multivariate random variable $\boldsymbol x$, the probability of $\boldsymbol x$ lying in an $n$-dimensional region $R^n$ is $P({\boldsymbol x}\in R^n)=\int_{R^n} p({\boldsymbol x})\,d^nx$, where $p({\boldsymbol x})$ is the \acs{PDF}.

\item[\acs{PE}] \Acl{PE}, the process of measuring the parameters that describe the source of a signal, e.g., the masses and spins of the binary components of a \ac{CBC}, from the observational data.

\item[\acs{PN}] \Acl{PN}, a perturbative method of obtaining solutions to relativistic field equations based on slow-motion and weak-field expansion of the spacetime metric and the stress--energy source.

\item[\acs{PSD}] \Acl{PSD}.
See Appendix~\ref{s:conventions}.

\item[\acs{RF}] \Acl{RF}, the radio-frequency range, from ${\sim} 1$ to \qty{\sim 150}{\mega\Hz}

\item[\acs{SNR}] \Acl{SNR}.
See Appendix~\ref{s:conventions}.

\item[Virgo] The Virgo detector is a European \Lshape-shaped interferometric \ac{GW} detector with \qty{3}{km} arms located near Cascina, Italy (near Pisa).

\item[\acs{VirgoNEXT}] \Acl{VirgoNEXT} is a planned, post-\ac{O5}, major upgrade of \ac{Virgo} to fill the gap between the current phase, \ac{AdV+}, and next-generation detectors.

\item[Virgo Collaboration] The Virgo Collaboration manages the building, operation, and development of the Virgo detector.

\end{description}

\section{Conventions for data analysis}\label{s:conventions}

This appendix serves to define the data analysis conventions that will be used throughout the \thisgwtc{} companion papers.
For a general introduction to data analysis we refer the reader to \citet{2020CQGra..37e5002A} and references therein.

Time series $a(t)$ and frequency series $\tilde{a}(t)$ are related to each other by our conventions for the \emph{Fourier transform}
\begin{equation}
    \tilde{a}(f) = \int_{-\infty}^{+\infty} a(t) \exp\left({-2\pi ift}\right) \,dt
\end{equation}
and its inverse transform
\begin{equation}
    a(t) = \int_{-\infty}^{+\infty} \tilde{a}(f) \exp\left({+2\pi ift}\right) \,df\:.
\end{equation}
With these conventions, the dimensions of $\tilde{a}$ are $[\tilde{a}] = [a] \times \text{time}$.

Detector noise is often taken to be a stochastic Gaussian process.
If $n(t)$ is a real-valued stochastic Gaussian process, then the \emph{one-sided \ac{PSD}} $S_n(f)$ is formally defined by
\begin{equation}
    \langle \tilde{n}^\ast(f') \tilde{n}(f) \rangle = \frac12 S_n(f) \delta(f - f')\:,
\end{equation}
where $\langle \cdot \rangle$ is a statistical ensemble average of realizations of $n(t)$ and $\tilde{n}^\ast$ is the complex conjugate of $\tilde{n}$.
The one-sided \ac{PSD} is defined only for $f\ge 0$.
With these conventions, the dimensions of $S_n$ are $[S_n] = [n]^2 \times \text{time}$.
Real detector noise is neither entirely stationary nor Gaussian \citep{2020CQGra..37e5002A}.
However, it is often sufficient to assume $n(t)$ is ergodic such that
\begin{equation}
    S_n(f) = \lim_{T\to\infty} \frac{2}{T} \left|
    \int_{-T/2}^{T/2} n(t) \exp\left({-2\pi ift}\right)\,dt \right|^2\:.
\end{equation}
The factor of two in the one-sided \ac{PSD} ensures that the integrated power is
\begin{equation}
    \int_0^\infty S_n(f)\,df = \lim_{T\to\infty} \frac{1}{T} \int_{-T/2}^{T/2} n^2(t)\,dt\:.
\end{equation}
The \emph{amplitude spectral density} is defined to be the square root of the \ac{PSD}, $S_n^{1/2}(f)$.

We often use a detector-noise-weighted inner product \citep{1992PhRvD..46.5236F} between two real-valued time series, $a(t)$ and $b(t)$, which is defined as
\begin{subequations}\label{e:inner_product}
\begin{align}
    \ip{a}{b} &= 4 \operatorname{Re} \int_0^{+\infty}
    \frac{\tilde{a}^\ast(f) \tilde{b}(f)}{S_n(f)} \,df \\
    &= \int_{-\infty}^{+\infty} \frac{\tilde{a}^\ast(f) \tilde{b}(f)}{(1/2)S_n(|f|)}\, df\:, \label{e:complex_inner_product}
\end{align}
\end{subequations}
where $S_n(f)$ is the detector's one-sided \ac{PSD} for the readout noise from that detector.
The second form, Equation~\eqref{e:complex_inner_product}, is an appropriate generalization of the inner product for complex-valued time series.

Since \ac{GW} detectors are insensitive at very low frequencies, the mean of the detector readout is arbitrary, and so we take the detector noise to have zero mean, $\langle n(t)\rangle = 0$.
Gaussian noise is then entirely characterized by its \ac{PSD} and its distribution is given by the probability density
\begin{equation}\label{e:noise_pdf}
    \PEpdf{n} = \frac{1}{W} \exp\left(-\frac{1}{2}\ip{n}{n}\right)\:,
\end{equation}
where $W$ is a usually neglected normalizing constant, the path integral $W=\int \exp\left({-\ip{n}{n}/2}\right)\,\mathcal{D}n$.

Consider a template waveform $u(t)$ that is unit-normalized, $\ip{u}{u}=1$, which is expected to match a hypothetical signal in detector data $\PEdata(t)$.
The \emph{matched filter \ac{SNR}} is
\begin{equation}\label{e:rho_mf}
    \rho_{\text{mf}}=\ip{u}{\PEdata}\:.
\end{equation}
If data $\PEdata(t)=n(t)+h(t)$ contain Gaussian noise $n(t)$ plus a
signal $h(t)$ that is perfectly matched by the template waveform,
$h(t)\propto u(t)$, then $\rho_{\text{mf}}$ is a random variable having a
normal distribution with unit variance and mean equal to the
\emph{optimal \ac{SNR}}
\begin{equation}
    \rho_{\text{opt}}=\sqrt{\ip{h}{h}}\:.
\end{equation}
The \emph{likelihood} that detector data $\PEdata(t)$ contains a signal $h(t)$ is given by Equation~\eqref{e:noise_pdf} with $n(t)=\PEdata(t)-h(t)$,
\begin{subequations}
\begin{align}\label{e:likelihood}
    \PElikelihood[h] &= \frac{1}{W} \exp\left(-\frac12 \ip{\PEdata-h}{\PEdata-h}\right) \\
    &= \frac{\exp\left({-\ip{\PEdata}{\PEdata}/2}\right)}{W} \exp\left(\ip{h}{\PEdata} -\frac12\ip{h}{h}\right) \\
    &= \PElikelihood[\varnothing] \exp\left(\rho_{\text{mf}}\rho_{\text{opt}}-\frac12\rho_{\text{opt}}^2\right)\:,\label{e:likelihood_3}
\end{align}
\end{subequations}
where $\rho_{\text{mf}}$ is the matched filter \ac{SNR} with unit-normalized template $u(t)\propto h(t)$ and $\PElikelihood[\varnothing]=W^{-1}\exp\left({-\ip{\PEdata}{\PEdata}/2}\right)$ is the likelihood under the no-signal hypothesis, $\PEdata(t)=n(t)$.
The likelihood is viewed as a functional of $h(t)$ for a given realization of detector data $\PEdata(t)$.
The second factor in Equation~\eqref{e:likelihood_3} is the signal-to-noise \emph{likelihood ratio} $\PElikelihood[h]/\PElikelihood[\varnothing]$.
Note that the likelihood ratio is a monotonically increasing function of the matched filter \ac{SNR}, and so $\rho_{\text{mf}}$ is the uniformly most powerful test for a known signal in Gaussian detector noise \citep{1933RSPTA.231..289N}.
If the amplitude of the signal is unknown, $h(t)=\rho_{\text{opt}}u(t)$ with unknown $\rho_{\text{opt}}$, then the likelihood is maximized for $\rho_{\text{opt}}=\rho_{\text{mf}}$ and
\begin{equation}\label{e:max_likelihood}
    \max_{\rho_{\text{opt}}} \PElikelihood[\rho_{\text{opt}} u] = \PElikelihood[\varnothing]
    \exp\left(\frac12\rho_{\text{mf}}^2\right)\:.
\end{equation}

For the Newtonian inspiral of Section~\ref{ss:newtonian_inspiral}, the
signal observed in a detector can be obtained in the frequency domain
under the stationary phase approximation as \citep{1991PhRvD..44.3819S,1993PhRvL..70.2984C}
\begin{equation}
   \tilde{h}(f) = - \sqrt{\frac{5\pi}{24}}
   \frac{G\Mc}{c^3} \frac{G\Mc}{c^2 D_{\text{eff}}}
   \left(\frac{\pi G \Mc f}{c^3}\right)^{-7/6}
   \exp\left({-i\Psi(f)}\right)\:,
\end{equation}
where $\Psi(f)$ is the stationary phase function and
\begin{equation}
   D_{\text{eff}} = r\left[{F_+^2(\vartheta,\varphi,\psi)\left(\frac12+\frac12\cos^2\iota\right)^2 + F_\times^2(\vartheta,\varphi,\psi) \cos^2\iota}\right]^{-1/2}\:.
\end{equation}
is the \emph{effective distance} \citep{2012PhRvD..85l2006A}, which is related to the distance to the binary $r$ by a factor that accounts for the orientation angles that describe the position of the source on the sky $(\vartheta,\varphi)$, its inclination $\iota$, and polarization angle $\psi$.
Since $F_+^2+F_\times^2\le 1$ (with equality for a source on the zenith or nadir of an \Lshape-shaped interferometric detector), $D_{\text{eff}}\ge r$ (with equality only if $\iota=0$ or $\iota=\pi$).
The optimal \ac{SNR} for such a signal is
\begin{equation}
   \rho_{\text{opt}}
   = \sqrt{\frac{5}{6\pi}} \frac{G\Mc}{c^2 D_{\text{eff}}}
      \left(\frac{\pi G \Mc}{c^3}\right)^{-1/6}
      \sqrt{\int_0^{+\infty}\frac{f^{-7/3}}{S_n(f)}\,df}\:.
\end{equation}

The \emph{horizon distance} $D_{\text{hor}}$ \citep{2012PhRvD..85l2006A} of a source is the effective distance of a signal from such a source that has \ac{SNR} $\rho_{\text{opt}}$ equal to some detection threshold $\rho_{\text{th}}$.
Such sources would not be expected to be detected beyond the horizon distance, but not all nearer sources will be detected either.
The \emph{sensitive volume} \citep{1993PhRvD..47.2198F,2021CQGra..38e5010C} is a measure of the effective volume of space in which randomly isotropically oriented and homogeneously distributed identical sources will produce signals in the detector with \ac{SNR} $\rho_{\text{opt}}$ greater than the threshold $\rho_{\text{th}}$,
\begin{subequations}
\begin{align}
    V &= \frac{\int_{\rho_{\text{opt}} > \rho_{\text{th}}} r^2\,\sin\vartheta\,\sin\iota\,dr\,d\vartheta\,d\varphi\,d\iota\,d\psi}{\int \sin\iota\,d\iota\,d\psi} \\
    &= \num[round-mode=figures, round-precision=5]{0.0860841994316697677}\times \frac43 \pi D_{\text{hor}}^3\:.
\end{align}
\end{subequations}
If the merger rate density is $\mathcal{R}$ then the expected number of detections in time $T$ is ${\mathcal{R}}VT$.
For a standard measure of detector sensitivity, a binary source of two \qty{1.4}{\Msun} objects ($\Mc=2^{-1/5}\times\qty{1.4}{\Msun}\approx\qty[round-mode=figures, round-precision=3]{1.21877078861}{\Msun}$) is considered and a threshold \ac{SNR} of $\rho_{\text{th}} = 8$ is adopted \citep{2021CQGra..38e5010C}.
The sensitive volume is converted into an equivalent spherical radius as
$V=(4\pi/3) R^3$ to obtain the BNS inspiral range $R=D_{\text{hor}}/\num[round-mode=figures, round-precision=6]{2.26477737785073758}$, Equation~\eqref{e:bns_range}.

In practice a discrete Fourier transform is used rather than the continuous, and window choice will vary depending on the analysis goals and methodology. For more details on the spectrum and spectral density estimation and its use in \ac{GW} data analysis, see \url{https://dcc.ligo.org/LIGO-T1400010/public}.

\section{Changelog}
\label{s:changelog}

THIS APPENDIX WILL RECORD CHANGES MADE IN FUTURE UPDATES

\begin{description}

\item[\normalfont\gwtc-4.0]\leavevmode
\begin{itemize}
    \item Initial version of this document.
\end{itemize}

\item[\normalfont\gwtc-5.0]\leavevmode
\begin{itemize}
    \item Updated catalog version throughout from GWTC‑4.0 to GWTC‑5.0.
    \item Extended observational coverage from the \acf{O4a} ending \GWTCfourENDDate{} to the \acf{O4b} ending \GWTCfiveENDDate{}.
    \item Added explicit references to newly published standalone \ac{O4b} event papers: GW241011\_233834, GW241110\_124123, GW250114\_082203 and GW240925\_005809.
    \item Updated narrative and figures in Section 3 to extend cumulative observing timelines through \ac{O4b}.
    \item Updates to Section 4 describing the detector evolution during \ac{O4b}.
    \item Wording changes throughout for clarity and general improvements.
\end{itemize}

\end{description}

\end{appendices}

\bibliography{}

\ifprintauthors
\expandafter\gdef\csname
currCollabName\the\allentries\endcsname{\LVKcollaboration}
\allauthors
\fi

\ifdraft
\listofchanges
\fi
\end{document}